\documentclass[preprintnumbers, superscriptaddress, showpacs, nofootinbib, amsfonts, amsmath, amssymb, aps, prd, notitlepage, floatfix]{revtex4-2}

\usepackage[utf8]{inputenc}
\usepackage{amsmath,amssymb,amsfonts,mathrsfs,bbold}
\usepackage{yfonts}
\usepackage[colorlinks=true,citecolor=blue,linkcolor=blue,urlcolor=blue]{hyperref}
\usepackage[dvipsnames]{xcolor}
\usepackage{lipsum}
\usepackage{slashed}
\usepackage[normalem]{ulem}
\usepackage{url}
\usepackage{physics}

\usepackage{amssymb, bm, amsmath, graphicx}
\usepackage{enumerate}
\usepackage{amsfonts}
\usepackage{mathtools}
\usepackage{latexsym}
\usepackage{epstopdf}
\usepackage{subfigure}
\usepackage{tikz}
\usetikzlibrary{decorations.pathmorphing, decorations.markings, shapes.geometric, shapes.misc, calc, math}
\tikzstyle{gluon}=[decorate, decoration={coil,aspect=0.8, amplitude=1.5pt,  segment length=3pt}]
\usepackage{stackrel}
\usepackage[most]{tcolorbox}

\usepackage{graphics,graphicx}
\usepackage{dcolumn}
\usepackage{epsfig}

\def\eq#1{{Eq.~(\ref{#1})}}
\def\fig#1{{Fig.~\ref{#1}}}
\newcommand{\ben}{\begin{eqnarray*}}
\newcommand{\een}{\end{eqnarray*}}
\newcommand{\un}[1]{\underline{#1}}

\newcommand{\ul}[1]{\underline{#1}}

\newcommand{\llangle}{\Big\langle \!\! \Big\langle}
\newcommand{\rrangle}{\Big\rangle \!\! \Big\rangle}

\newcommand{\as}{\alpha_s}

\allowdisplaybreaks

\newcommand{\pd}{\partial}

\newcommand{\dhd}{{\textstyle d}
\lower.03ex\hbox{\kern-0.38em$^{\scriptstyle-}$}\kern-0.05em{}}
\newcommand{\dbar}{{\textstyle \delta}
\lower.03ex\hbox{\kern-0.38em$^{\scriptstyle-}$}\kern-0.05em{}}
\newcommand{\half}{{1\over 2}}

\newcommand{\tord}{\textrm{T} \:}
\newcommand{\atord}{\overline{\textrm{T}} \:}

\makeatletter
\DeclareRobustCommand{\cev}[1]{%
  {\mathpalette\do@cev{#1}}%
}
\newcommand{\do@cev}[2]{%
  \vbox{\offinterlineskip
    \sbox\z@{$\m@th#1 x$}%
    \ialign{##\cr
      \hidewidth\reflectbox{$\m@th#1\vec{}\mkern4mu$}\hidewidth\cr
      \noalign{\kern-\ht\z@}
      $\m@th#1#2$\cr
    }%
  }%
}
\makeatother

\begin{document}

\title{Gluon Double-Spin Asymmetry in the Longitudinally Polarized $p+p$ Collisions}

    \makeatletter  
\def\@fnsymbol#1{\ensuremath{\ifcase#1\or *\or \dagger\or \ddagger\or
   \mathsection\or \mathparagraph\or \|\or **\or \dagger\dagger
   \or \ddagger\ddagger \or \mathsection\mathsection \else\@ctrerr\fi}}
    \makeatother

\author{Yuri V. Kovchegov} 
         \email[Email: ]{kovchegov.1@osu.edu}
         \affiliation{Department of Physics, The Ohio State
           University, Columbus, OH 43210, USA}

\author{Ming Li} 
         \email[Email: ]{li.13499@osu.edu}
         \affiliation{Department of Physics, The Ohio State
           University, Columbus, OH 43210, USA}

\begin{abstract}
We derive the first-ever small-$x$ expression for the inclusive gluon production cross section in the central rapidity region of the longitudinally polarized proton-proton collisions. The cross section depends on the polarizations of both protons, therefore comprising the numerator of the longitudinal double-spin asymmetry $A_{LL}$ for the produced gluons. The cross section is calculated in the shock wave formalism and is expressed in terms of the polarized dipole scattering amplitudes on the projectile and target protons. We show that the small-$x$ evolution corrections are included into our cross section expression if one evolves these polarized dipole amplitudes using the double-logarithmic helicity evolution derived in \cite{Kovchegov:2015pbl, Kovchegov:2016zex, Kovchegov:2018znm, Cougoulic:2022gbk}. 
Our calculation is performed for the gluon sector only, with the quark contribution left for future work. When that work is complete, the resulting formula will be applicable to longitudinally polarized proton-proton and proton-nucleus collisions, as well as to polarized semi-inclusive deep inelastic scattering (SIDIS) on a proton or a nucleus. Our results should allow one to extend the small-$x$ helicity phenomenology analysis of \cite{Adamiak:2023yhz} to the jet/hadron production data reported for the longitudinally polarized proton-proton collisions at RHIC and to polarized SIDIS measurements at central rapidities to be performed at the EIC. 
\end{abstract}

\maketitle

\tableofcontents


\section{Introduction}

Understanding helicity distributions for quarks and gluons at small values of Bjorken $x$ is an integral part of the proton spin puzzle~\cite{EuropeanMuon:1987isl, Jaffe:1989jz, Ji:1996ek, Boer:2011fh, Aidala:2012mv, Accardi:2012qut, Leader:2013jra, Aschenauer:2013woa, Aschenauer:2015eha,  Proceedings:2020eah, Ji:2020ena, AbdulKhalek:2021gbh}. Since the current and future experiments can only probe these helicity distributions down to some small value of $x = x_\textrm{min}$, and would not be able to reach all the way down to $x=0$ due to the finite energy coverage \cite{Boer:2011fh, Accardi:2012qut, Aschenauer:2013woa, Aschenauer:2015eha, Proceedings:2020eah, AbdulKhalek:2021gbh}, there appears to be a need for a theoretically controlled extrapolation of the helicity distributions down to very small $x$ values. A promising approach to accomplishing this goal is in using the small-$x$ evolution equations, which can predict the parton distributions at small $x$ given some initial conditions at the larger $x = x_0$ and assuming that the strong coupling is sufficiently small (see \cite{Kuraev:1977fs,Balitsky:1978ic, Gribov:1981ac,  Balitsky:1995ub,Balitsky:1998ya,Kovchegov:1999yj,Kovchegov:1999ua,Jalilian-Marian:1997dw,Jalilian-Marian:1997gr,Weigert:2000gi,Iancu:2001ad,Iancu:2000hn,Ferreiro:2001qy} for spin-independent small-$x$ evolution equations and \cite{Gribov:1984tu, Iancu:2003xm, Weigert:2005us, JalilianMarian:2005jf, Gelis:2010nm, Albacete:2014fwa, Kovchegov:2012mbw, Morreale:2021pnn} for reviews).

The past decade has seen a significant advancement of our theoretical understanding of helicity evolution at small $x$~\cite{Kovchegov:2015pbl, Hatta:2016aoc, Kovchegov:2016zex, Kovchegov:2016weo, Kovchegov:2017jxc, Kovchegov:2017lsr, Kovchegov:2018znm, Kovchegov:2019rrz, Cougoulic:2019aja, Kovchegov:2020hgb, Cougoulic:2020tbc, Chirilli:2021lif, Kovchegov:2021lvz, Cougoulic:2022gbk, Borden:2023ugd, Adamiak:2023okq} in the framework of the shock wave/$s$-channel evolution formalism~\cite{Mueller:1994rr,Mueller:1994jq,Mueller:1995gb,Balitsky:1995ub,Balitsky:1998ya,Kovchegov:1999yj,Kovchegov:1999ua,Jalilian-Marian:1997dw,Jalilian-Marian:1997gr,Weigert:2000gi,Iancu:2001ad,Iancu:2000hn,Ferreiro:2001qy}. The sub-eikonal and, in some cases, sub-sub-eikonal (suppressed by one or two powers of the center-of-mass energy squared $s$) corrections to the eikonal scattering of quarks and gluons on a background field have been found in \cite{Altinoluk:2014oxa,Balitsky:2015qba,Balitsky:2016dgz, Kovchegov:2017lsr, Kovchegov:2018znm, Chirilli:2018kkw, Jalilian-Marian:2018iui, Jalilian-Marian:2019kaf, Altinoluk:2020oyd, Kovchegov:2021iyc, Altinoluk:2021lvu, Kovchegov:2022kyy, Altinoluk:2022jkk, Altinoluk:2023qfr,Altinoluk:2023dww, Li:2023tlw}. This informed the in-parallel development of the small-$x$ evolution for the sub-eikonal operators related to the quark and gluon flavor-singlet helicity distribution functions $\Delta \Sigma (x, Q^2)$ and $\Delta G (x, Q^2)$, and to the $g_1$ structure function~\cite{Kovchegov:2015pbl, Hatta:2016aoc, Kovchegov:2016zex, Kovchegov:2016weo, Kovchegov:2017jxc, Kovchegov:2017lsr, Kovchegov:2018znm, Kovchegov:2019rrz, Cougoulic:2019aja, Kovchegov:2020hgb, Cougoulic:2020tbc, Chirilli:2021lif, Kovchegov:2021lvz, Cougoulic:2022gbk, Borden:2023ugd, Adamiak:2023okq}. (Flavor non-singlet helicity evolution in the $s$-channel formalism was derived in \cite{Kovchegov:2016zex}.) The corresponding formalism, involving sub-eikonal and sub-sub-eikonal operators on the light-cone, has been referred to as the light-cone operator treatment (LCOT) \cite{Kovchegov:2015pbl, Kovchegov:2016zex, Kovchegov:2018znm, Kovchegov:2021iyc, Cougoulic:2022gbk, Kovchegov:2022kyy}.

The leading-order helicity evolution derived in \cite{Kovchegov:2015pbl, Kovchegov:2016zex, Kovchegov:2018znm, Cougoulic:2022gbk} is in the double-logarithmic approximation (DLA): it sums up powers of $\as \, \ln^2 (1/x)$, where $\as$ is the strong coupling constant. Its original version was derived in \cite{Kovchegov:2015pbl, Kovchegov:2016zex, Kovchegov:2018znm} (KPS), and has recently been modified in \cite{Cougoulic:2022gbk} (KPS-CTT). A resummation of the single-logarithmic corrections (power of $\as \, \ln (1/x)$) for helicity evolution was attempted in \cite{Kovchegov:2021lvz}, but needs to be revised in light of the recent modifications of the DLA evolution found in \cite{Cougoulic:2022gbk}.

The evolution equations \cite{Kovchegov:2015pbl, Kovchegov:2016zex, Kovchegov:2018znm, Cougoulic:2022gbk} close in the large-$N_c$ \cite{tHooft:1973alw} and large-$N_c \& N_f$ \cite{Veneziano:1976wm} limits, with $N_c$ and $N_f$ the numbers of quark colors and flavors, respectively. The evolution equations have been solved numerically and analytically \cite{Cougoulic:2022gbk, Borden:2023ugd} for large $N_c$ and numerically \cite{Adamiak:2023okq} for large $N_c \& N_f$. The resulting intercepts driving the small-$x$ asymptotics of the helicity parton distribution functions (hPDFs) $\Delta \Sigma (x, Q^2)$ and $\Delta G (x, Q^2)$, and the $g_1$ structure function appear to be close to but slightly differ (at the $< 1 \%$ level at large $N_c$ \cite{Cougoulic:2022gbk, Borden:2023ugd} and $< 3 \%$ level at large $N_c \& N_f$ \cite{Adamiak:2023okq}) from those found earlier in the pioneering work by Bartels, Ermolaev and Ryskin (BER) \cite{Bartels:1995iu,Bartels:1996wc}, in which hPDFs at small $x$ were calculated employing an entirely different approach, the infrared evolution equations (IREE) formalism from Refs.~\cite{Gorshkov:1966ht,Kirschner:1983di,Kirschner:1994rq,Kirschner:1994vc,Blumlein:1995jp,Griffiths:1999dj}. The origin of this minor disagreement is not yet clear: see the Appendices of Refs.~\cite{Kovchegov:2016zex, Borden:2023ugd} for the possible reasons for the discrepancy. (There appears to be no disagreement between the flavor non-singlet helicity PDFs asymptotics found in \cite{Bartels:1995iu} using IREE and in \cite{Kovchegov:2016zex} using LCOT at large $N_c$. A disagreement between the BER formalism for the flavor-singlet distributions and the exact calculation of the 3-loop polarized anomalous dimension was observed earlier in \cite{Moch:2014sna, Blumlein:2021ryt}, but was shown to be attributable to scheme dependence.)

Since the disagreements between the LCOT and BER approaches outlined above are numerically small, it appears that the degree of agreement between the two approaches is sufficient 
to perform phenomenological analyses of the data based on these formalisms. Such analyses based on the BER formalism had been performed in Refs.~\cite{Blumlein:1995jp,Blumlein:1996hb,Ermolaev:1999jx,Ermolaev:2000sg,Ermolaev:2003zx,Ermolaev:2009cq}, before the LCOT approach was developed. More recently, a phenomenological analysis of the polarized deep inelastic scattering (DIS) and semi-inclusive DIS (SIDIS) data was performed in Ref.~\cite{Adamiak:2023yhz} using the large-$N_c \& N_f$ KPS-CTT evolution with running coupling (see also \cite{Adamiak:2021ppq} for a similar ``proof-of-principle" analysis of the polarized DIS data only using the earlier KPS evolution). The analysis of \cite{Adamiak:2023yhz} resulted in a successful fit of the world polarized DIS and SIDIS data for $x < 0.1$ and with the photon virtuality $Q^2 > 1.69$~GeV$^2$. At the same time, the analysis in \cite{Adamiak:2023yhz} found the existing polarized DIS and SIDIS data insufficient to completely fix the initial conditions for small-$x$ helicity evolution. This resulted in a significant spread of the predictions achieved in \cite{Adamiak:2023yhz} for the hPDFs and the proton $g_1$ structure function at lower values of $x$ than probed in the present or past experiments, reducing the advantage of the small-$x$ evolution approach over the more standard $Q^2$-evolution approaches~\cite{Gluck:2000dy, Leader:2005ci, deFlorian:2009vb, Leader:2010rb, Jimenez-Delgado:2013boa, Ball:2013lla, Nocera:2014gqa, deFlorian:2014yva, Leader:2014uua, Sato:2016tuz, Ethier:2017zbq, DeFlorian:2019xxt, Borsa:2020lsz, Zhou:2022wzm, Cocuzza:2022jye} based on the (spin-dependent) Dokshitzer-Gribov-Lipatov-Altarelli-Parisi (DGLAP) evolution equations \cite{Gribov:1972ri, Altarelli:1977zs, Dokshitzer:1977sg}, which cannot predict hPDFs at very small~$x$. 

One way to address this issue is to wait for the future Electron-Ion Collider (EIC) \cite{Boer:2011fh, Accardi:2012qut, Proceedings:2020eah, AbdulKhalek:2021gbh} to generate more polarized DIS and SIDIS data. However, to better test theoretical predictions, it would be desirable to reduce the spread found in \cite{Adamiak:2023yhz} before the start of the EIC experimental program. To this end, one may try utilizing the data on the jets and pion production in polarized proton-proton collisions reported by the experiments at the Relativistic Heavy Ion Collider (RHIC) (see \cite{STAR:2014wox, PHENIX:2015fxo} along with \cite{Aschenauer:2013woa, Aschenauer:2015eha} for a summary of those results). However, to describe the RHIC Spin data in the small-$x$ formalism, one needs to derive an expression for the numerator of the double-spin asymmetry $A_{LL}$, preferably in terms of the so-called polarized dipole scattering amplitudes entering the helicity evolution equations \cite{Kovchegov:2015pbl, Kovchegov:2016zex, Kovchegov:2018znm, Cougoulic:2022gbk} and employed in the polarized DIS and SIDIS analysis of \cite{Adamiak:2023yhz}. The double-spin asymmetry is defined by
\begin{align}\label{ALL}
    A_{LL} = \frac{d\sigma (++) - d\sigma (+-)}{d\sigma (++) + d\sigma (+-)},
\end{align}
where $d\sigma (++) (d\sigma (+-))$ is the differential cross section with the colliding
protons having the same (opposite) helicities. We see that to find the numerator of $A_{LL}$ one needs to calculate the part of the particle production cross section in the $p+p$ collisions dependent on the helicities of the colliding protons. This is an inclusive particle production observable, different, for instance, from the total scattering DIS cross section needed to calculate the structure functions. While the SIDIS process analyzed in \cite{Adamiak:2023yhz} also involves inclusive hadron production, the calculation there was simplified since in most of the existing polarized SIDIS data the hadron is produced in the current fragmentation region (i.e., in the forward direction for the virtual photon). Motivated by the RHIC data \cite{STAR:2014wox, PHENIX:2015fxo, Aschenauer:2013woa, Aschenauer:2015eha}, we would like to calculate hadron production in the central rapidity region, far (in rapidity) from the fragmentation regions of both colliding protons. Similar analytic calculations have been carried out at small $x$ in the unpolarized case in \cite{Kovchegov:1998bi, Kopeliovich:1998nw, Braun:2000bh, Dumitru:2001ux, Kovchegov:2001sc, Kharzeev:2003wz, Balitsky:2004rr, Chirilli:2015tea, Li:2021zmf, Li:2021yiv, Li:2021ntt}. Specifically, for gluon production in unpolarized proton--proton collisions, the target--projectile symmetric expression for inclusive gluon production cross section in the small-$x$/saturation formalism was obtained in \cite{Kovchegov:2001sc, Braun:2000bh}. The result of \cite{Kovchegov:2001sc, Braun:2000bh} can be used in the denominator of the double-spin asymmetry $A_{LL}$ in \eq{ALL}.

Our calculation will follow the standard technique for small-$x$ inclusive hadron production cross section calculations outlined in \cite{Kovchegov:2012mbw}, with the exception that we are now interested in the sub-eikonal correction to this cross section, dependent on the helicities of the colliding protons. We will employ the operator formalism in the LCOT framework. For simplicity, in this paper we will work in the gluon sector only. Hence, the colliding polarized protons for us will be dominated by gluons, and the produced particle will be a gluon as well. Since quarks also contribute to the small-$x$ helicity evolution in the DLA, they will certainly contribute to hadron production too. We leave the inclusion of the quark contributions to future work, concentrating only on the gluons here.

We derive the inclusive gluon production cross section in the polarized-particle scattering in Sec.~\ref{sec:shock_wave}, employing the shock wave formalism \cite{Balitsky:1995ub}, treating one of the scattering particles as a fast-moving projectile and another one as a target shock wave. For simplicity, we take the incoming particle to be a (longitudinally polarized) gluon. After obtaining an expression for the inclusive gluon production cross section in Sec.~\ref{sec:cross_section} (see \eq{cross_sect_1}), we recast it in the projectile--target symmetric form in \eq{kT_fact_final} from Sec.~\ref{sec:TP-symm}, in the process expressing it in terms of the polarized dipole amplitudes for scattering on the projectile and on the target. These polarized dipole amplitudes can be evaluated in the quasi--classical approximation of the Glauber--Gribov--Mueller (GGM) scattering/McLerran--Venugopalan (MV) model \cite{Mueller:1989st,McLerran:1993ni,McLerran:1993ka,McLerran:1994vd} by using the helicity-dependent version \cite{Cougoulic:2020tbc} of the MV model: this would give a quasi-classical approximation to inclusive gluon production in the polarized $p+p$ collisions, not shown explicitly below. In Sec.~\ref{sec:fixed} we cross-check our result \eqref{kT_fact_final} at the lowest non-trivial order by comparing it to the calculations existing in the literature (which we also redo in a different gauge): we find a complete agreement between all lowest-order results.

The small-$x$ helicity evolution \cite{Kovchegov:2015pbl, Kovchegov:2016zex, Kovchegov:2018znm, Cougoulic:2022gbk} is included into the expression \eqref{kT_fact_final} for the inclusive cross section in Sec.~\ref{sec:evolution} in the large-$N_c$ approximation. Similar to the unpolarized case \cite{Kovchegov:2001sc, Kovchegov:2012mbw}, the evolution effects leave the expression \eqref{kT_fact_final} for the production cross section unchanged, and are accounted for by evolving the polarized dipole amplitudes entering this expression using the large-$N_c$ KPS-CTT evolution. The rapidity intervals between the produced gluon and the target and between the produced gluon and the projectile are different, requiring evolution up to different values of rapidity for the projectile and target polarized dipole amplitudes. We summarize our main results and conclude in Sec.~\ref{sec:conclusions}.


\section{Gluon production in the shock wave formalism}
\label{sec:shock_wave}


\subsection{The scattering amplitude}
\label{sec:amplitude}

\begin{figure}[h]
    \centering
\includegraphics[width= 0.9 \textwidth]{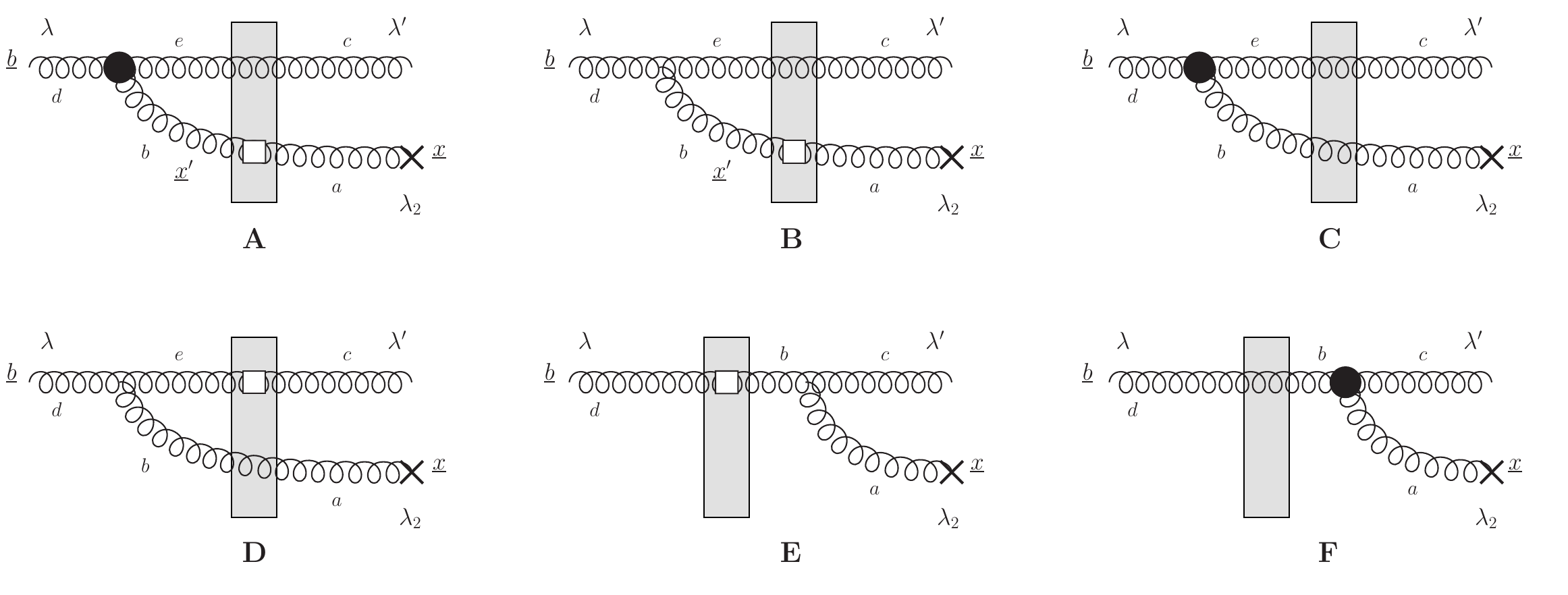}
    \caption{Diagrams contributing to gluon production in the polarized hadron--hadron scattering in the pure glue picture. The rectangle denotes the shock wave, while the produced gluon is indicated by a cross.}
    \label{fig:amplitude_gluons}
\end{figure}

The diagrams contributing to inclusive gluon production in the scattering of a longitudinally polarized gluon on a longitudinally polarized target are shown in \fig{fig:amplitude_gluons}. The produced gluon is denoted by a cross. As usual, the shaded rectangle denotes the shock wave of the target fields. The black circle denotes the sub-eikonal helicity-dependent vertex, while the small white square denotes the sub-eikonal interaction with the shock wave. This is the same notation as that used in \cite{Kovchegov:2015pbl, Hatta:2016aoc, Kovchegov:2016zex, Kovchegov:2016weo, Kovchegov:2017jxc, Kovchegov:2017lsr, Kovchegov:2018znm, Kovchegov:2019rrz, Cougoulic:2019aja, Kovchegov:2020hgb, Cougoulic:2020tbc, Kovchegov:2021lvz, Cougoulic:2022gbk}. As mentioned above, throughout this paper we will be working in the gluon sector: this is why the quark-mediated contributions are not included in \fig{fig:amplitude_gluons}.

To calculate the diagrams in \fig{fig:amplitude_gluons} we will employ the sub-eikonal expansion of the $S$-matrix for a high-energy gluon scattering in the background quark and gluon fields \cite{Balitsky:2015qba,Altinoluk:2020oyd,Kovchegov:2021iyc,Kovchegov:2018znm,Chirilli:2018kkw}. Following the notation of \cite{Cougoulic:2022gbk} we write the high-energy gluon's $S$-matrix as 
\begin{align}\label{Uxy_sub-eikonal}
(U_{{\ul x}, {\un y}; \lambda', \lambda})^{b a}   \equiv (U_{\un{x}})^{ba} \, \delta^2 (\un{x} - \un{y}) \, \delta_{\lambda, \lambda'} + \lambda \, \delta_{\lambda, \lambda'} \, (U_{{\ul x}}^{\textrm{pol} [1]})^{b a} \, \delta^2 ({\un x} - {\un y}) + \delta_{\lambda, \lambda'} \, (U_{{\ul x}, {\un y}}^{\textrm{pol} [2]})^{b a} + {\cal O} \left( \frac{1}{s^2} \right)
\end{align}
keeping terms up to and including the sub-eikonal order (that is, up to and including order-$1/s$ terms in the expansion in inverse powers of the center of mass energy squared $s$ for the projectile--target scattering). Here $b$($a$) and $\lambda'$($\lambda$) are the final(initial) gluon colors and polarizations. The transverse vectors are denoted by ${\un v} = (v^1, v^2)$. The adjoint light-cone Wilson line is defined by
\begin{align}\label{Uline}
U_{\un{x}} [x^-_f,x^-_i] = \mathcal{P} \exp \left[ ig \int\limits_{x^-_i}^{x^-_f} d{x}^- {\cal A}^+ (0^+, x^-, \un{x}) \right]
\end{align}
with the gluon field ${\cal A}^{\mu} = \sum_a A^{a \, \mu} \, T^a$, where $T^a$ are the adjoint SU($N_c$) generators, $(T^a)_{bc} = - i f^{abc}$. Here $\mathcal{P}$ is the path ordering operator and $g$ is the QCD coupling constant. The light-cone coordinates are defined by $x^\pm = (t \pm z)/\sqrt{2}$. The gluon in \eq{Uxy_sub-eikonal} is predominantly moving in the light-cone minus direction. The infinite light-cone Wilson line is $U_{\un{x}} = U_{\un{x}} [\infty, - \infty]$: this is the eikonal (order-$(1/s)^0$) contribution coming from the first term on the right of \eq{Uxy_sub-eikonal}.

The sub-eikonal (order-$1/s$) contribution is given by the second and third terms on the right of \eq{Uxy_sub-eikonal}. They are referred to as the polarized Wilson lines of the first and second kind. As in \cite{Cougoulic:2022gbk}, we separate the quark and gluon sub-eikonal operator contributions by writing
\begin{align}\label{UqG_decomp}
U_{\un x}^{\textrm{pol} [1]} = U_{\un x}^{\textrm{G} [1]} + U_{\un x}^{\textrm{q} [1]}, \ \ \  U_{{\ul x}, {\un y}}^{\textrm{pol} [2]} = U_{{\ul x}, {\un y}}^{\textrm{G} [2]} + U_{{\ul x}}^{\textrm{q} [2]} \, \delta^2 ({\un x} - {\un y}) ,
\end{align}
with
\begin{subequations}\label{UqG}
\begin{align}
& (U_{\un x}^{\textrm{G} [1]})^{ba} = \frac{2 \, i \, g \, p_1^+}{s} \int\limits_{-\infty}^{\infty} d{x}^- (U_{\un{x}} [ \infty, x^-])^{bb'} \, ({\cal F}^{12})^{b'a'} (x^-, {\un x}) \, (U_{\un{x}} [ x^-, -\infty])^{a'a}  , \label{UG1} \\
& (U_{\un x}^{\textrm{q} [1]})^{ba} = \frac{g^2 p_1^+}{2 \, s} \!\! \int\limits_{-\infty}^{\infty} \!\! d{x}_1^- \! \int\limits_{x_1^-}^\infty d x_2^- (U_{\un{x}} [ \infty, x_2^-])^{bb'} \bar{\psi} (x_2^-,\un{x}) \, t^{b'} V_{\un{x}} [x_2^-,x_1^-] \, \gamma^+ \gamma^5 \, t^{a'} \psi (x_1^-,\un{x})  (U_{\un{x}} [ x_1^-, -\infty])^{a'a} + \mbox{c.c.}  ,  \label{Uq1} \\
& (U_{{\ul x}, {\un y}}^{\textrm{G} [2]})^{ba}  = - \frac{i \, p_1^+}{s} \int\limits_{-\infty}^{\infty} d{z}^- d^2 z \ (U_{\un{x}} [ \infty, z^-])^{bb'} \, \delta^2 (\un{x} - \un{z}) \,\cev{\underline{\mathscr{D}}}^{b'c} (z^-, {\un z}) \, \underline{\mathscr{D}}^{ca'}  (z^-, {\un z}) \, (U_{\un{y}} [ z^-, -\infty])^{a'a} \, \delta^2 (\un{y} - \un{z}) , \label{UG2}  \\
& (U_{{\ul x}}^{\textrm{q} [2]} )^{ba} = - \frac{g^2 p_1^+}{2 \, s} \int\limits_{-\infty}^{\infty} \!\! d{x}_1^- \! \int\limits_{x_1^-}^\infty d x_2^- (U_{\un{x}} [ \infty, x_2^-])^{bb'} \, \bar{\psi} (x_2^-,\un{x}) \, t^{b'} \, V_{\un{x}} [x_2^-,x_1^-] \, \gamma^+ \, t^{a'} \, \psi (x_1^-,\un{x}) \,  (U_{\un{x}} [ x_1^-, -\infty])^{a'a} - \mbox{c.c.} . \label{Uq2}
\end{align}
\end{subequations}
Here $p_1^+$ is the large momentum component of the target proton, $\psi$ and $\bar \psi$ are the quark and anti-quark background fields, $\mathscr{D}^{ab}_i = \partial_i\delta^{ab} - ig (T^c)_{ab} A_i^c$ and $\cev{\mathscr{D}}^{ab}_i = \cev{\partial}_i\delta^{ab} + ig (T^c)_{ab} A_i^c$ are the right- and left-acting adjoint covariant derivatives, ${\cal F}^{12}$ is a component of the adjoint field-strength tensor, and $t^a$ are the fundamental SU($N_c$) generators. The fundamental light-cone Wilson lines are defined similarly to \eq{Uline} by
\begin{align}\label{Vline}
V_{\un{x}} [x^-_f,x^-_i] = \mathcal{P} \exp \left[ ig \int\limits_{x^-_i}^{x^-_f} d{x}^- A^+ (0^+, x^-, \un{x}) \right]
\end{align}
with the gluon field given by $A^{\mu} = \sum_a A^{a \, \mu} \, t^a$. It is useful to define the abbreviation $V_{\un{x}} = V_{\un{x}} [\infty, - \infty]$ for infinite lines.  

Since we are interested in the sub-eikonal polarization-dependent cross section, we will need to keep only the contributions to the diagrams A, C, D, E, and F in \fig{fig:amplitude_gluons} which depend on the polarization $\lambda$ of the incoming gluon. In terms of the operators, this means that the black circle in diagrams A, C and F denotes the insertion of the sub-eikonal operator containing ${\cal F}^{12}$ from \eq{UG1}, while the white box in diagrams D and E brings in the entire $U_{\un b}^{\textrm{G} [1]}$ operator. 

Using the notation outlined above, the diagrams in \fig{fig:amplitude_gluons} contribute as follows in the $A^- =0$ gauge for the minus-moving projectile and plus-moving target:
\begin{subequations}\label{ABCDEF}
\begin{align}
& i \, A = - \frac{g}{\pi} \, \lambda \, \delta_{\lambda, \lambda'} \, \frac{k^-}{p_2^-} \, \left( U_{\un b} \, T^b \right)^{cd} \, \epsilon^{ij} \, \epsilon_{\lambda_2}^{j*} \, \left\{ \frac{(x-b)^i}{|{\un x} - {\un b}|^2} \, \left[ \lambda_2 \, \left( U_{\un x}^{\textrm{pol} [1]} \right)^{ab} + \left( U_{\un x}^{\textrm{q} [2]} \right)^{ab} \right] + \int d^2 x' \, \frac{(x'-b)^i}{|{\un x}' - {\un b}|^2}  \, \left( U_{{\un x}, {\un x}'}^{\textrm{G} [2]} \right)^{ab}\right\}, \\
& i \, B = \frac{i \, g}{\pi} \, \delta_{\lambda, \lambda'} \,  \left( U_{\un b} \, T^b \right)^{cd} \, \epsilon_{\lambda_2}^{i*} \, \left\{ \frac{(x-b)^i}{|{\un x} - {\un b}|^2} \, \left[ \lambda_2 \, \left( U_{\un x}^{\textrm{pol} [1]} \right)^{ab} + \left( U_{\un x}^{\textrm{q} [2]} \right)^{ab} \right] + \int d^2 x' \, \frac{(x'-b)^i}{|{\un x}' - {\un b}|^2}  \, \left( U_{{\un x}, {\un x}'}^{\textrm{G} [2]} \right)^{ab}\right\}, \\
& i \,  C = - \frac{g}{\pi} \, \lambda \, \delta_{\lambda, \lambda'} \, \frac{k^-}{p_2^-} \, \left( U_{\un b} \, T^b \right)^{cd} \, \epsilon^{ij} \, \epsilon_{\lambda_2}^{j*} \, \frac{(x-b)^i}{|{\un x} - {\un b}|^2} \, \left( U_{\un x} \right)^{ab} , \\
& i \, D = \frac{i \, g}{\pi} \, \lambda \, \delta_{\lambda, \lambda'} \, \frac{{\un \epsilon}_{\lambda_2}^* \cdot ({\un x} - {\un b})}{|{\un x} - {\un b}|^2} \, \left( U_{\un x} \right)^{ab} \,  \left( U^{\textrm{pol} [1]}_{\un b} \, T^b \right)^{cd} , \\
& i \, E = - \frac{i \, g}{\pi} \, \lambda \, \delta_{\lambda, \lambda'} \, \frac{{\un \epsilon}_{\lambda_2}^* \cdot ({\un x} - {\un b})}{|{\un x} - {\un b}|^2} \, \left( T^a \, U^{\textrm{pol} [1]}_{\un b} \right)^{cd} , \\
& i \, F = \frac{g}{\pi} \, \lambda \, \delta_{\lambda, \lambda'} \, \frac{k^-}{p_2^-} \, \left(  T^a \, U_{\un b} \right)^{cd} \, \epsilon^{ij} \, \epsilon_{\lambda_2}^{j*} \, \frac{(x-b)^i}{|{\un x} - {\un b}|^2} = \frac{g}{\pi} \, \lambda \, \delta_{\lambda, \lambda'} \, \frac{k^-}{p_2^-} \, \left( U_{\un b} \, T^b \right)^{cd} \, \epsilon^{ij} \, \epsilon_{\lambda_2}^{j*} \, \frac{(x-b)^i}{|{\un x} - {\un b}|^2} \, \left( U_{\un b} \right)^{ab}.
\end{align}
\end{subequations}
The diagram contributions are given in the mixed representation: they are in the transverse position space while in the longitudinal momentum space. We are keeping only the sub-eikonal contributions, and only the projectile helicity-dependent terms in diagrams A, C, D, E, F. For the future reference, we are keeping the quark contributions in the polarized Wilson lines: they will be discarded shortly below. Our amplitudes are normalized as $A = M/(2s)$ \cite{Kovchegov:2012mbw} compared to the standard normalization for the scattering amplitudes $M$: this makes the eikonal contribution to the amplitudes energy-independent. 

Here $k^-$ is the minus momentum of the produced gluon, while $p_2^-$ is the momentum of the incoming (projectile) one. We assume that $k^- \ll p_2^-$. The gluon polarization four-vector is $\epsilon_\lambda^\mu = ({\un \epsilon}_\lambda \cdot {\un k}/k^-, 0^-, {\un \epsilon}_\lambda)$ with ${\un \epsilon}_\lambda = - (1/\sqrt{2}) (\lambda, i)$ \cite{Lepage:1980fj} for a gluon with momentum $k^\mu$ and polarization $\lambda$. The incoming gluon is at the transverse position $\un b$, while the produced one is at $\un x$ to the right of the shock wave and at ${\un x}'$ to the left (in diagrams A and B only). The color indices and polarizations are shown in \fig{fig:amplitude_gluons}. Latin indices denote the transverse directions, $i,j = 1,2$, while $\epsilon^{ij}$ is the transverse Levi-Civita symbol. In simplifying the expression for the diagram F we have used the following Wilson-line identity:
\begin{align}
    \left( U_{\un x} \right)^{ab} T^b = U^\dagger_{\un x} \, T^a \, U_{\un x}.
\end{align}

As one can show, emissions of the produced gluon from inside the shock wave are suppressed by one power of the logarithm of its transverse momentum \cite{Li:2023tlw}. Therefore, they are not included in \fig{fig:amplitude_gluons} and in our analysis here.

The relevant contributions to the net scattering amplitude are
\begin{align}
A ({\un x}, {\un b}) = A_\textrm{eik} + A + B + \ldots + F.
\end{align}
where the eikonal unpolarized gluon emission at the same order-$g$ in the coupling is \cite{Kovchegov:1998bi}
\begin{align}
i \, A_\textrm{eik} ({\un x}, {\un b}) = \frac{i \, g}{\pi} \, \delta_{\lambda, \lambda'} \, \frac{{\un \epsilon}_{\lambda_2}^* \cdot ({\un x} - {\un b})}{|{\un x} - {\un b}|^2} \, \left[ \left( U_{\un x} \right)^{ab} - \left( U_{\un b} \right)^{ab} \right] \,  \left( U_{\un b} \, T^b \right)^{cd} .
\end{align}


\subsection{Inclusive gluon production cross section}
\label{sec:cross_section}

\begin{figure}[t]
    \centering
\includegraphics[width= 0.9 \textwidth]{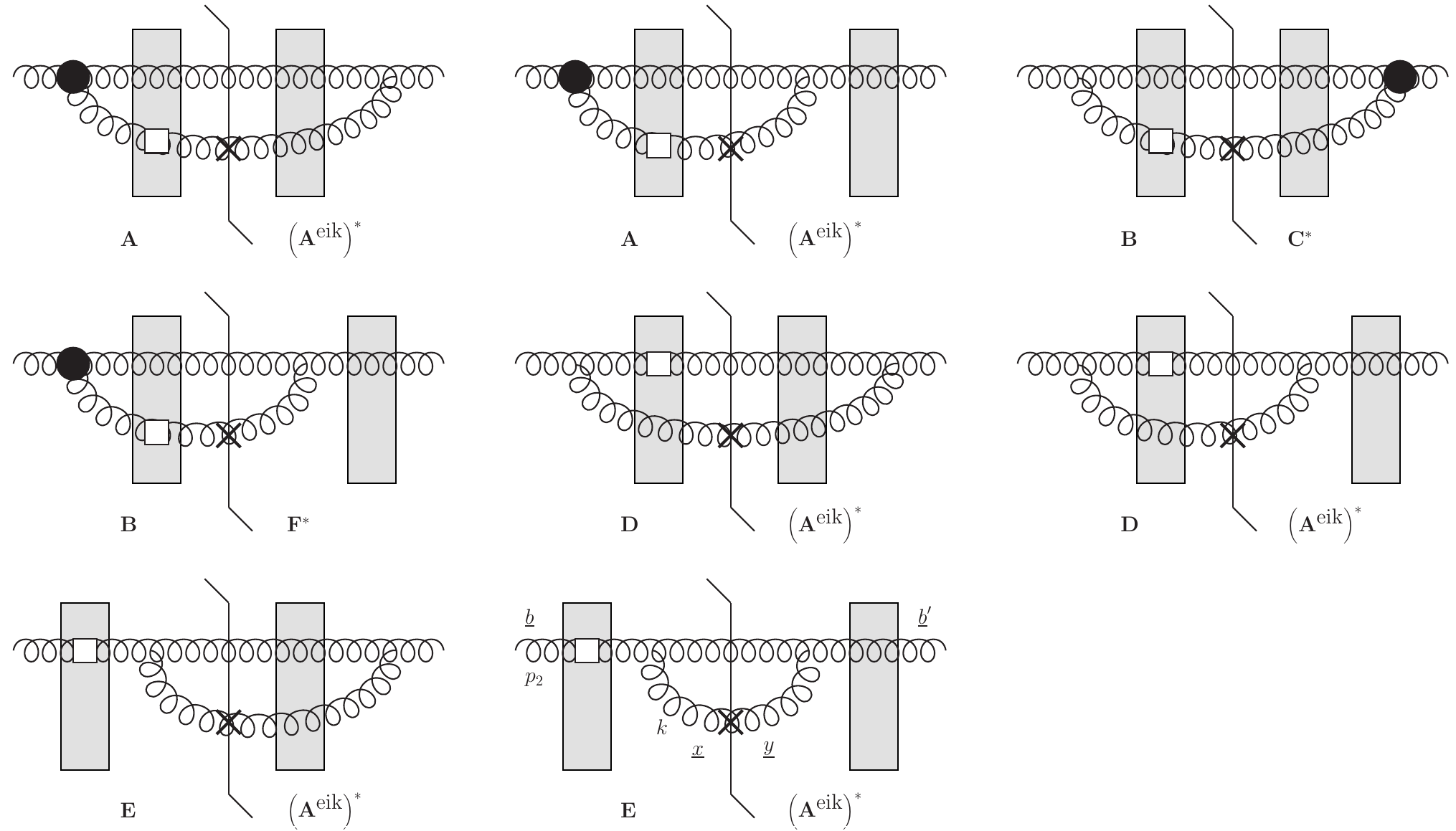}
    \caption{Diagrams contributing to the gluon production cross section in the polarized hadron--hadron scattering in the pure glue picture. The thin straight vertical line indicates the final state cut. Complex conjugate contributions are implied but not shown.}
    \label{fig:cross_section}
\end{figure}

The inclusive gluon production cross section is \cite{Kovchegov:2012mbw}
\begin{align}\label{xsect}
\frac{d \sigma (\lambda)}{d^2 k_T \, dy} = \frac{1}{2 (2 \pi)^3} \int d^2 x \, d^2 y \, d^2 b \, e^{- i {\un k} \cdot ({\un x} - {\un y})} \, \left\langle  A ({\un x}, {\un b})  \, A^*  ({\un y}, {\un b}) \right\rangle 
\end{align}
in terms of the mixed-representation scattering amplitudes we employ normalized as described above. Here $k^\mu$ is the produced gluon's 4-momentum with $k_T = |{\un k}|$, while $y= (1/2) \ln (k^-/k^+)$ is its rapidity. We are only interested in the part which depends on the polarization $\lambda$ of the incoming gluon and on the polarization of the target. Restricting the calculation to the sub-eikonal order we write the relevant contributions schematically as
\begin{align}\label{AA*}
A ({\un x}, {\un b})  \, A^* ({\un y}, {\un b}) = (A + D + E) \, A_\textrm{eik}^{\ast} + B \, (C^* + F^*) + \mbox{c.c.} .
\end{align}
This is illustrated diagrammatically in \fig{fig:cross_section}.

Dropping the quark contributions in Eqs.~\eqref{ABCDEF} we rewrite them as
\begin{subequations}\label{ABCDEF2}
\begin{align}
& A = \frac{i \, g}{\pi} \, \lambda \, \delta_{\lambda, \lambda'} \, \frac{k^-}{p_2^-} \, \left( U_{\un b} \, T^b \right)^{cd} \, \epsilon^{ij} \, \epsilon_{\lambda_2}^{j*} \, \left[ \frac{(x-b)^i}{|{\un x} - {\un b}|^2} \, \lambda_2 \, \left( U_{\un x}^{\textrm{G} [1]} \right)^{ab}  + \int d^2 x' \, \frac{(x'-b)^i}{|{\un x}' - {\un b}|^2}  \, \left( U_{{\un x}, {\un x}'}^{\textrm{G} [2]} \right)^{ab}\right], \\
& B = \frac{g}{\pi} \, \delta_{\lambda, \lambda'} \,  \left( U_{\un b} \, T^b \right)^{cd} \, \epsilon_{\lambda_2}^{i*} \, \left[ \frac{(x-b)^i}{|{\un x} - {\un b}|^2} \, \lambda_2 \, \left( U_{\un x}^{\textrm{G} [1]} \right)^{ab} + \int d^2 x' \, \frac{(x'-b)^i}{|{\un x}' - {\un b}|^2}  \, \left( U_{{\un x}, {\un x}'}^{\textrm{G} [2]} \right)^{ab}\right], \\
& C = \frac{i \, g}{\pi} \, \lambda \, \delta_{\lambda, \lambda'} \, \frac{k^-}{p_2^-} \, \left( U_{\un b} \, T^b \right)^{cd} \, \epsilon^{ij} \, \epsilon_{\lambda_2}^{j*} \, \frac{(x-b)^i}{|{\un x} - {\un b}|^2} \, \left( U_{\un x} \right)^{ab} , \\
& D = \frac{g}{\pi} \, \lambda \, \delta_{\lambda, \lambda'} \, \frac{{\un \epsilon}_{\lambda_2}^* \cdot ({\un x} - {\un b})}{|{\un x} - {\un b}|^2} \, \left( U_{\un x} \right)^{ab} \,  \left( U^{\textrm{G} [1]}_{\un b} \, T^b \right)^{cd} , \\
& E = - \frac{g}{\pi} \, \lambda \, \delta_{\lambda, \lambda'} \, \frac{{\un \epsilon}_{\lambda_2}^* \cdot ({\un x} - {\un b})}{|{\un x} - {\un b}|^2} \, \left( T^a \, U^{\textrm{G} [1]}_{\un b} \right)^{cd} , \\
& F 
= - \frac{i \, g}{\pi} \, \lambda \, \delta_{\lambda, \lambda'} \, \frac{k^-}{p_2^-} \, \left( U_{\un b} \, T^b \right)^{cd} \, \epsilon^{ij} \, \epsilon_{\lambda_2}^{j*} \, \frac{(x-b)^i}{|{\un x} - {\un b}|^2} \, \left( U_{\un b} \right)^{ab}.
\end{align}
\end{subequations}

Concentrating on individual contributions to the production cross section from \eq{AA*} we begin with the diagram A and get, after a considerable algebra similar to that in \cite{Cougoulic:2022gbk},
\begin{align}\label{AAeik}
& \frac{d \sigma^{A \, A^*_\textrm{eik} + \mbox{c.c.}} (\lambda)}{d^2 k_T \, dy} = \lambda \, \frac{4 \, \as}{(2 \pi)^4}  \frac{k^-}{p_2^-} \, \frac{N_c}{N_c^2 - 1} \, \int d^2 x \, d^2 y \, d^2 b \, e^{- i {\un k} \cdot ({\un x} - {\un y})} \, \left\{  \frac{{\un x} - {\un b}}{|{\un x} - {\un b}|^2} \, \cdot \, \frac{{\un y} - {\un b}}{|{\un y} - {\un b}|^2}  \left\langle  \Tr \left[ U_{\un y}^\dagger \, U_{{\un x}}^{\textrm{G} [1]} \right] - \Tr \left[ U_{\un b}^\dagger \, U_{{\un x}}^{\textrm{G} [1]} \right]  \right\rangle  \right. \notag \\ 
& \hspace*{3cm} \left. + \, i \int d^2 x' \, \frac{{\un x}'- {\un b}}{|{\un x}' - {\un b}|^2}  \, \times \, \frac{{\un y} - {\un b}}{|{\un y} - {\un b}|^2}   \left\langle  \Tr \left[ U_{\un y}^\dagger \, U_{{\un x}, {\un x}'}^{\textrm{G} [2]} \right] - \Tr \left[ U_{\un b}^\dagger \, U_{{\un x}, {\un x}'}^{\textrm{G} [2]} \right]  \right\rangle \right\} + \mbox{c.c.} \notag \\
& =  \lambda \, \frac{\as}{2 \pi^4}  \frac{1}{s} \, N_c \, \int d^2 x \, d^2 y \, d^2 b \, e^{- i {\un k} \cdot ({\un x} - {\un y})} \, \left\{  \frac{{\un x} - {\un b}}{|{\un x} - {\un b}|^2} \, \cdot \, \frac{{\un y} - {\un b}}{|{\un y} - {\un b}|^2} \left[ G^\textrm{adj}_{{\un x}, {\un y}} (2 k^- p_1^+) - G^\textrm{adj}_{{\un x}, {\un b}} (2 k^- p_1^+) \right] \right. \notag \\ 
& \hspace*{3cm}\left. - 2 i \, k^i \, \frac{{\un x} - {\un b}}{|{\un x} - {\un b}|^2}  \, \times \, \frac{{\un y} - {\un b}}{|{\un y} - {\un b}|^2}  \, G^{i \, \textrm{adj}}_{{\un x}, {\un b}} (2 k^- p_1^+)   \right\}  .
\end{align}
In arriving at \eq{AAeik} we have employed $\sum_{\lambda_2} \lambda_2 \, \epsilon_{\lambda_2}^{j*} \, \epsilon_{\lambda_2}^{k} = i \, \epsilon^{jk}$ and $\sum_{\lambda_2}  \epsilon_{\lambda_2}^{j*} \, \epsilon_{\lambda_2}^{k} = \delta^{jk}$, along with
\begin{align}
\int d^2 b \, \frac{{\un x}'- {\un b}}{|{\un x}' - {\un b}|^2}  \, \times \, \frac{{\un y} - {\un b}}{|{\un y} - {\un b}|^2}  = 0 . 
\end{align}
The cross product is defined by ${\un u} \times {\un v} = \epsilon^{ij} u^i v^j = u^1 v^2 - u^2 v^1$, the QCD coupling is $\as = g^2/4 \pi$, c.c. stands for complex conjugate, while $\Tr$ denotes a trace of adjoint matrices. 

We have also defined the following adjoint {\sl polarized dipole amplitudes} in the pure-gluon sector \cite{Cougoulic:2022gbk,Kovchegov:2015pbl,Kovchegov:2018znm}:
\begin{subequations}\label{G_adj_defs}
\begin{align}\label{G_adj_def}
& G^\textrm{adj}_{{\un x}, {\un y}} (\beta s) \equiv \frac{1}{2 (N_c^2 -1)} \, \mbox{Re} \, \llangle \mbox{T} \, \mbox{Tr} \left[ U_{\ul y} \, U_{{\un x}}^{\textrm{G} [1] \, \dagger} \right] + \mbox{T} \, \mbox{Tr} \left[ U_{{\un x}}^{\textrm{G} [1]} \, U_{\ul y}^\dagger \right] \rrangle (\beta s), \\ 
& G^{i \, \textrm{adj}}_{{\un x}, {\un y}} (\beta s) \equiv \frac{1}{2 (N_c^2 -1)} \, \mbox{Re} \, \llangle \mbox{T} \, \mbox{Tr} \left[ U_{\ul y} \, U_{{\un x}}^{i \, \textrm{G} [2] \, \dagger} \right] + \mbox{T} \, \mbox{Tr} \left[ U_{{\un x}}^{i \, \textrm{G} [2]} \, U_{\ul y}^\dagger \right] \rrangle (\beta s) ,
\end{align}
\end{subequations}
with the new polarized Wilson line of the second kind \cite{Cougoulic:2022gbk, Kovchegov:2017lsr}
\begin{align}\label{Ui}
U_{\un{z}}^{i \, \textrm{G} [2]} \equiv \frac{p_1^+}{2 s} \, \int\limits_{-\infty}^{\infty} d {z}^- \, U_{\un{z}} [ \infty, z^-] \, \left[ \mathscr{D}^i (z^-, \un{z}) - \cev{\mathscr{D}}^i (z^-, \un{z}) \right]  \, U_{\un{z}} [ z^-, -\infty] . 
\end{align}
The argument $\beta s$ of the polarized dipole amplitudes in Eqs.~\eqref{G_adj_defs} denotes a fraction $\beta$ of the projectile--target center of mass energy squared $s$. Since our projectile is minus-moving, here $\beta = k^-/p_2^-$, such that $\beta s = 2 k^- p_1^+$ with $p_1^+$ the large plus momentum of the target. The double angle brackets denote the averaging in the shock wave, now with the polarized target, multiplied by a factor of energy, such that $\llangle \ldots \rrangle = \beta \, s \, \left\langle \ldots \right\rangle$ (see \cite{Kovchegov:2015pbl} for details) .

Moving on to the diagram B we derive
\begin{align}\label{BCF}
& \frac{d \sigma^{B \, (C^* + F^*) + \mbox{c.c.}} (\lambda)}{d^2 k_T \, dy} =  \lambda \, \frac{4 \, \as}{(2 \pi)^4}  \frac{k^-}{p_2^-} \, \frac{N_c}{N_c^2 - 1} \, \int d^2 x \, d^2 y \, d^2 b \, e^{- i {\un k} \cdot ({\un x} - {\un y})} \, \left\{  \frac{{\un x} - {\un b}}{|{\un x} - {\un b}|^2} \, \cdot \, \frac{{\un y} - {\un b}}{|{\un y} - {\un b}|^2}  \left\langle  \Tr \left[ U_{\un y}^\dagger \, U_{{\un x}}^{\textrm{G} [1]} \right] - \Tr \left[ U_{\un b}^\dagger \, U_{{\un x}}^{\textrm{G} [1]} \right]  \right\rangle  \right. \notag \\ 
& \hspace*{3cm} \left. + \, i \int d^2 x' \, \frac{{\un x}'- {\un b}}{|{\un x}' - {\un b}|^2}  \, \times \, \frac{{\un y} - {\un b}}{|{\un y} - {\un b}|^2}   \left\langle  \Tr \left[ U_{\un y}^\dagger \, U_{{\un x}, {\un x}'}^{\textrm{G} [2]} \right] - \Tr \left[ U_{\un b}^\dagger \, U_{{\un x}, {\un x}'}^{\textrm{G} [2]} \right]  \right\rangle \right\} + \mbox{c.c.} ,
\end{align}
doubling the contribution in \eq{AAeik}.  

Similarly, from the diagram D we obtain
\begin{align}\label{DAeik}
& \frac{d \sigma^{D \, A^*_\textrm{eik} + \mbox{c.c.}} (\lambda)}{d^2 k_T \, dy} = \frac{\lambda}{N_c^2 - 1} \, \frac{4 \, \as}{(2 \pi)^4}  \int d^2 x \, d^2 y \, d^2 b \, e^{- i {\un k} \cdot ({\un x} - {\un y})} \, \frac{{\un x} - {\un b}}{|{\un x} - {\un b}|^2} \, \cdot \, \frac{{\un y} - {\un b}}{|{\un y} - {\un b}|^2} \left\langle \Tr \left[ U_{\un b}^\dagger \, U_{{\un b}}^{\textrm{G} [1]} \, T^a \, U_{\un x}^\dagger \,  \left( U_{\un y} -  U_{\un b} \right)  \, T^a \right]  \right\rangle + \mbox{c.c.} \notag \\ 
& = - \frac{\lambda}{N_c^2 - 1} \, \frac{4 \, \as}{(2 \pi)^4}  \int d^2 x \, d^2 y \, d^2 b \, e^{- i {\un k} \cdot ({\un x} - {\un y})} \, \frac{{\un x} - {\un b}}{|{\un x} - {\un b}|^2} \, \cdot \, \frac{{\un y} - {\un b}}{|{\un y} - {\un b}|^2} \left\langle \Tr \left[ U_{\un b}^\dagger \, U_{{\un b}}^{\textrm{G} [1]} \, T^a \, U_{\un x}^\dagger \, U_{\un b} \, T^a \right]  \right\rangle + \mbox{c.c.} \notag \\ 
& = -  \lambda \, \frac{\as}{4 \pi^4} \frac{1}{s} \, N_c \,   \int d^2 x \, d^2 y \, d^2 b \, e^{- i {\un k} \cdot ({\un x} - {\un y})} \, \frac{{\un x} - {\un b}}{|{\un x} - {\un b}|^2} \, \cdot \, \frac{{\un y} - {\un b}}{|{\un y} - {\un b}|^2} \, G^\textrm{adj}_{{\un b}, {\un x}} (2 k^- p_1^+) ,
\end{align}
where one can readily show that 
\begin{align}
\Tr \left[ U_{\un b}^\dagger \, U_{{\un b}}^{\textrm{G} [1]} \, T^a \, U_{\un x}^\dagger \,  U_{\un y}  \, T^a \right] + \mbox{c.c.} \ (\mbox{with} \ {\un x} \leftrightarrow {\un y}) = 0 
\end{align}
and that 
\begin{align}
\Tr \left[ U_{\un b}^\dagger \, U_{{\un b}}^{\textrm{G} [1]} \, T^a \, U_{\un x}^\dagger \, U_{\un b} \, T^a \right] = \frac{N_c}{2} \,  \Tr \left[ U_{\un x}^\dagger \, U_{{\un b}}^{\textrm{G} [1]} \right].
\end{align}
The latter relation may be derived by using 
\begin{align}\label{relation1}
 \Tr \left[ U_{\un b}^\dagger \, T^a \, U_{{\un b}}^{\textrm{G} [1]} \, T^{c} \right]  \, \left( U_{\un y}  \right)^{ac}  = \frac{N_c}{2} \,  \Tr \left[ U_{\un y}^\dagger \, U_{{\un b}}^{\textrm{G} [1]} \right], 
\end{align}
which we will also employ below.

For the diagram E we get the following contribution to the cross section: 
\begin{align}\label{EAeik}
& \frac{d \sigma^{E \, A^*_\textrm{eik} + \mbox{c.c.}} (\lambda)}{d^2 k_T \, dy} = - \frac{\lambda}{N_c^2 - 1} \, \frac{4 \, \as}{(2 \pi)^4}  \int d^2 x \, d^2 y \, d^2 b \, e^{- i {\un k} \cdot ({\un x} - {\un y})} \, \frac{{\un x} - {\un b}}{|{\un x} - {\un b}|^2} \, \cdot \, \frac{{\un y} - {\un b}}{|{\un y} - {\un b}|^2} \left\langle \Tr \left[ U_{\un b}^\dagger \, T^a \, U_{{\un b}}^{\textrm{G} [1]} \, T^{c} \right]  \, \left( U_{\un y} -  U_{\un b} \right)^{ac} \right\rangle + \mbox{c.c.} \notag \\
& = - \frac{\lambda}{N_c^2 - 1} \, \frac{4 \, \as}{(2 \pi)^4}  \int d^2 x \, d^2 y \, d^2 b \, e^{- i {\un k} \cdot ({\un x} - {\un y})} \, \frac{{\un x} - {\un b}}{|{\un x} - {\un b}|^2} \, \cdot \, \frac{{\un y} - {\un b}}{|{\un y} - {\un b}|^2} \left\langle \Tr \left[ U_{\un b}^\dagger \, T^a \, U_{{\un b}}^{\textrm{G} [1]} \, T^{c} \right]  \, \left( U_{\un y}  \right)^{ac} \right\rangle + \mbox{c.c.} \notag  \\
& = -  \lambda \, \frac{\as}{4 \pi^4} \frac{1}{s} \, N_c \,   \int d^2 x \, d^2 y \, d^2 b \, e^{- i {\un k} \cdot ({\un x} - {\un y})} \, \frac{{\un x} - {\un b}}{|{\un x} - {\un b}|^2} \, \cdot \, \frac{{\un y} - {\un b}}{|{\un y} - {\un b}|^2} \, G^\textrm{adj}_{{\un b}, {\un y}} (2 k^- p_1^+) ,
\end{align}
where we have used
\begin{align}\label{zero}
\left( T^a \, U_{{\un b}}^{\textrm{G} [1]} \right)^{cd} \, \left( T^a \, U_{\un b} \right)^{cd} = N_c \, \Tr \left[ U_{\un b}^\dagger \, U_{{\un b}}^{\textrm{G} [1]} \right] = 0 
\end{align}
along with \eq{relation1}.

A more careful treatment of the $(D+E) \, A^*_\textrm{eik} +$~c.c. contributions, presented in Appendix~\ref{A}, generates the following term, in addition to Eqs.~\eqref{DAeik} and \eqref{EAeik}:
\begin{align}\label{UVterm}
     -  \lambda \, \frac{\as}{4 \pi^4} \frac{1}{s} \, N_c \,   \int d^2 x \, d^2 y \, d^2 b \, e^{- i {\un k} \cdot ({\un x} - {\un y})} \, \frac{{\un x} - {\un b}}{|{\un x} - {\un b}|^2} \, \cdot \, \frac{{\un y} - {\un b}}{|{\un y} - {\un b}|^2} \, \left[ - 2 \, G^\textrm{adj}_{{\un b}, {\un b}'} (2 k^- p_1^+) \right],
\end{align}
where $\un b$ and ${\un b}'$ are the positions of the incoming gluon to the left and to the right of the cut with $|{\un b}' - {\un b}|^2 = 1/(2 k^+ p_2^-)$. The notation is illustrated in the last diagram of \fig{fig:cross_section}. The term in \eq{UVterm} is calculated in the linearized case, appropriate for the DLA used in helicity evolution of \cite{Kovchegov:2015pbl, Kovchegov:2018znm, Cougoulic:2019aja, Cougoulic:2022gbk} which resums powers of $\as \, \ln^2 (1/x)$. Two different positions of the incoming gluon on the two sides of the cut appear due to the lifetime ordering of emissions \cite{Cougoulic:2019aja}: for the $x^-$-lifetime of the incoming gluon with momentum $p_2$ to be much longer than the lifetime of the produced gluon with momentum $k$ (see \fig{fig:cross_section}), one requires that 
\begin{align}
\frac{2 p_2^-}{{\un p}_2^2} \gg \frac{2 k^- }{{\un k}^2},    
\end{align}
which gives ${\un p}_2^2 \ll {\un k}^2/\beta$, resulting in the position mismatch of $|{\un b}' - {\un b}|^2 = \beta/{\un k}^2 = 1/(2 k^+ p_2^-)$ for the incoming gluon.\footnote{One has to distinguish this impact parameter offset resulting from the lifetime ordering condition from the offset in $\un b$ and ${\un b}'$ due to the produced gluon having different positions $\un x$ and $\un y$ on the two sides of the cut (which is, in turn, due to the fixed momentum $\un k$ of the produced gluon in our production calculation). The latter offset is $|{\un b}' - {\un b}|^2 = \beta^2 \, |\un x - \un y|^2 \approx \beta^2/{\un k}^2$ (see, e.g., \cite{Kovchegov:2012ga}) and is much smaller that the $|{\un b}' - {\un b}|^2 = \beta/{\un k}^2$ offset due to the lifetime ordering, since $\beta \ll 1$: therefore, we will only keep the offset due to the lifetime ordering, neglecting the other much smaller offset.} 

This is, indeed, a very small position mismatch, suppressed by a power of energy: one may wonder why we need to keep it. The answer to this question is in the fact that the approach to zero of $\left\langle \Tr \left[ U_{{\un b}'}^\dagger \, U_{{\un b}}^{\textrm{G} [1]} \right] \right\rangle$ as ${\un b}' \to {\un b}$ is different from what one observes in unpolarized scattering, where the dipole amplitude goes to zero as a positive power of its transverse size \cite{Kovchegov:2012mbw}: for instance, at Born level one has \cite{Kovchegov:2016zex}
\begin{align}
    \left\langle \Tr \left[ U_{{\un b}'}^\dagger \, U_{{\un b}}^{\textrm{G} [1]} \right] \right\rangle_0 (\beta s) \propto \ln ( \beta s |{\un b}' - {\un b}|^2 ),
\end{align} 
such that zero amplitude is achieved at small but finite dipole size $|{\un b}' - {\un b}|^2 = 1/\beta s$, i.e., this zero is achieved at the dipole size squared given by the ultraviolet (UV) cutoff of the gluon--target scattering --- the inverse of the center of mass energy squared ($\beta s$). At the same time, the transverse positions mismatch of the incoming gluons is $|{\un b}' - {\un b}|^2 = 1/(2 k^+ p_2^-) = 1/\alpha s$ with $\alpha = k^+/p_1^+$, that is, given by the inverse center of mass energy squared $\alpha s$ of the gluon--projectile system. We observe that we have two different (physical) UV cutoffs in the calculation: $\alpha s$ for the gluon--projectile system and $\beta s$ for the gluon--target system. Due to the existence of these two UV scales, we see that the zero we obtained in \eq{zero} becomes, at Born level,  \begin{align}\label{ba}
    \left\langle \Tr \left[ U_{{\un b}'}^\dagger \, U_{{\un b}}^{\textrm{G} [1]} \right] \right\rangle_0 (\beta s) \Bigg|_{|{\un b}' - {\un b}|^2 = \frac{1}{\alpha s}} \propto \ln \frac{\beta}{\alpha} .
\end{align}   
This is, in general, not zero. It becomes zero only when $\alpha = \beta$. Unlike the unpolarized scattering at high energy, where all the transverse distances of the order of the inverse powers of energy can be safely put to zero in a calculation, here we need to keep track of the logarithms of such very short distances, since they contribute to the evolution \cite{Kirschner:1983di, Bartels:1995iu, Bartels:1996wc, Itakura:2003jp, Kovchegov:2015pbl, Kovchegov:2018znm, Cougoulic:2022gbk, Kovchegov:2022kyy}, and, as follows from our calculation, to the production cross section.

Assembling Eqs.~\eqref{AAeik}, \eqref{BCF}, \eqref{DAeik}, \eqref{EAeik}, and \eqref{UVterm} we obtain the helicity-dependent inclusive gluon production cross section in the pure glue sector,
\begin{tcolorbox}[ams align]\label{cross_sect_1}
& \frac{d \sigma (\lambda)}{d^2 k_T \, dy} = \lambda \, \frac{\as}{\pi^4}  \frac{1}{s} \, N_c \, \int d^2 x \, d^2 y \, d^2 b \, e^{- i {\un k} \cdot ({\un x} - {\un y})} \, \left\{  \frac{{\un x} - {\un b}}{|{\un x} - {\un b}|^2} \cdot \frac{{\un y} - {\un b}}{|{\un y} - {\un b}|^2} \bigg[ G^\textrm{adj}_{{\un x}, {\un y}} (2 k^- p_1^+) - G^\textrm{adj}_{{\un x}, {\un b}} (2 k^- p_1^+)  \right.  \notag \\ 
&  \left.  - \frac{1}{4} \left( G^\textrm{adj}_{{\un b}, {\un y}} (2 k^- p_1^+) +  G^\textrm{adj}_{{\un b}, {\un x}} (2 k^- p_1^+) - 2 \, G^\textrm{adj}_{{\un b}, {\un b}'} (2 k^- p_1^+) \right)  \bigg]
 - 2 i \, k^i \, \frac{{\un x} - {\un b}}{|{\un x} - {\un b}|^2}  \, \times \, \frac{{\un y} - {\un b}}{|{\un y} - {\un b}|^2}  \, G^{i \, \textrm{adj}}_{{\un x}, {\un b}} (2 k^- p_1^+)   \right\}  .
\end{tcolorbox}


\subsection{Target--projectile symmetric form for the gluon production cross section}
\label{sec:TP-symm}

The expression \eqref{cross_sect_1} gives us the helicity-dependent part of the inclusive gluon production cross section in the gluon--target scattering, with the pure-glue dynamics. By the nature of its derivation, the cross section \eqref{cross_sect_1} appears to treat the projectile (gluon) and the target (could be a proton or another gluon) differently: it is not explicitly target--projectile symmetric. At the same time, since both the target and the projectile are assumed to be longitudinally polarized protons, or, for the purpose of our gluon-sector derivation, gluons, and since we are producing a gluon in the central rapidity region, not close in rapidity to either projectile or to target, one would expect the gluon production cross section to be target--projectile symmetric (if one also interchanges the rapidity intervals between the produced gluon and the target/projectile). Therefore, one may wonder whether \eq{cross_sect_1} can be written in a target--projectile symmetric form.  Indeed, the interaction with the target in \eq{cross_sect_1} is described by the polarized dipole amplitudes, which may include multiple additional gluon emissions due to small-$x$ evolution, while the projectile is simply a gluon emitting the produced gluon, without any evolution corrections at the moment: hence, \eq{cross_sect_1} is target--projectile asymmetric. However, if the polarized dipole amplitudes in \eq{cross_sect_1} are taken at the lowest (Born) order, then the expression should become target--projectile symmetric. Conversely, inspired by the inclusive gluon production cross section in the unpolarized case \cite{Kovchegov:2001sc}, and by the target--projectile symmetry of the physical problem, one may be able to find a form of \eq{cross_sect_1} where the interaction with the projectile can be also encoded in the polarized dipole amplitudes, which, when taken at Born level, give \eq{cross_sect_1}. Here we will do just that: we will rewrite \eq{cross_sect_1} in the target--projectile symmetric form, which will facilitate inclusion of evolution corrections in the rapidity interval between the produced gluon and the projectile.

Defining
\begin{align}
{\un {\tilde x}} = {\un x} - {\un b}, \ \ \ {\un {\tilde y}} = {\un y} - {\un b},
\end{align}
we rewrite \eq{cross_sect_1} as
\begin{align}\label{cross_sect_3}
& \frac{d \sigma (\lambda)}{d^2 k_T \, dy} = \lambda \, \frac{\as}{\pi^4}  \frac{1}{s} \, N_c \, \int d^2 {\tilde x} \, d^2 {\tilde y} \, d^2 b \, e^{- i {\un k} \cdot ({\un {\tilde x}} - {\un {\tilde y}})} \, \left\{  \frac{{\un {\tilde x}}}{|{\un {\tilde x}}|^2} \cdot \frac{{\un {\tilde y}} }{|{\un {\tilde y}} |^2} \bigg[ G^\textrm{adj}_{{\un {\tilde x}} + {\un b}, {\un {\tilde y}} + {\un b}} (2 k^- p_1^+) - G^\textrm{adj}_{{\un {\tilde x}} + {\un b}, {\un b}} (2 k^- p_1^+)  \right.  \notag \\ 
&  \left.  - \frac{1}{4} \left( G^\textrm{adj}_{{\un b}, {\un {\tilde y}} + {\un b}} (2 k^- p_1^+) +  G^\textrm{adj}_{{\un b}, {\un {\tilde x}} + {\un b}} (2 k^- p_1^+) - 2 \, G^\textrm{adj}_{{\un b}, {\un b}'} (2 k^- p_1^+) \right)  \bigg]
 - 2 i \, k^i \,  \frac{{\un {\tilde x}}}{|{\un {\tilde x}}|^2} \, \times \, \frac{{\un {\tilde y}} }{|{\un {\tilde y}} |^2} \, G^{i \, \textrm{adj}}_{{\un {\tilde x}} + {\un b}, {\un b}} (2 k^- p_1^+)   \right\} . 
\end{align}
Integrating over $\un b$ we obtain
\begin{align}\label{cross_sect_4}
& \frac{d \sigma (\lambda)}{d^2 k_T \, dy} = \lambda \, \frac{\as}{\pi^4}  \frac{1}{s} \, N_c \, \int d^2 {\tilde x} \, d^2 {\tilde y} \, e^{- i {\un k} \cdot ({\un {\tilde x}} - {\un {\tilde y}})} \,  \frac{{\un {\tilde x}}}{|{\un {\tilde x}}|^2} \cdot \frac{{\un {\tilde y}} }{|{\un {\tilde y}} |^2} \bigg[ G^\textrm{adj} (|{\un {\tilde x}} -{\un {\tilde y}}|^2 ,  2 k^- p_1^+) - G^\textrm{adj} ( {\tilde x}_\perp^2 , 2 k^- p_1^+)  \notag \\ 
& \hspace*{1cm}   - \frac{1}{4} \left( G^\textrm{adj} ({\tilde y}_\perp^2,  2 k^- p_1^+) +  G^\textrm{adj} ({\tilde x}_\perp^2 , 2 k^- p_1^+) - 2 \, G^\textrm{adj} (1/(2 k^+ p_2^-)  , 2 k^- p_1^+) \right) - 2 \, G_2^\textrm{adj} ({\tilde x}_\perp^2 , 2 k^- p_1^+)  \bigg] .
\end{align}
Here $x_\perp = |{\un x}|$. We have employed the following definition and a decomposition \cite{Kovchegov:2016zex, Kovchegov:2017lsr, Cougoulic:2022gbk}:
\begin{subequations}
    \begin{align}
        & \int d^2 b \, G^\textrm{adj}_{{\un b}, {\un b} - {\un x}} (\beta s) = G^\textrm{adj} (x_\perp^2 , \beta s), \\
        & \int d^2 b \, G^{i \, \textrm{adj}}_{{\un b}, {\un b} - {\un x}} (\beta s) = x^i \, G_1^\textrm{adj} (x_\perp^2 , \beta s) + \epsilon^{ij} \, x^j \, G_2^\textrm{adj} (x_\perp^2 , \beta s), \label{Gi_decomp}
    \end{align}
\end{subequations}
along with $|{\un b}' - {\un b}|^2 = 1/(2 k^+ p_2^-)$.

Note that all the polarized dipole amplitudes, like $G^\textrm{adj} (x_\perp^2 , \beta s)$, are defined only in the physical region where $x_\perp^2 > 1/\beta s$ \cite{Kovchegov:2015pbl, Kovchegov:2018znm, Cougoulic:2019aja, Cougoulic:2022gbk}. Outside this region they are zero. With this in mind we define a ``subtracted" dipole amplitude 
\begin{align}\label{subtr_def}
    G^\textrm{adj}_\textrm{subtr} ({ x}_\perp^2 , 2 k^- p_1^+) \equiv G^\textrm{adj} ({x}_\perp^2 , 2 k^- p_1^+) -  G^\textrm{adj} (1/(2 k^+ p_2^-)  , 2 k^- p_1^+) 
\end{align}
which is zero at ${x}_\perp^2 = 1/(2 k^+ p_2^-) = 1/\alpha s$.

Employing the definition \eqref{subtr_def} and performing some of the two-dimensional integral we can reduce \eq{cross_sect_4} to
\begin{align}\label{cross_sect_66}
& \frac{d \sigma (\lambda)}{d^2 k_T \, dy} = \lambda \, \frac{2 \, \as}{\pi^3}  \frac{1}{s} \, N_c \, \int d^2 x \, e^{- i {\un k} \cdot {\un x}} \,   \bigg[ \ln \left( \frac{1}{x_\perp \, \Lambda} \right)  \, G^\textrm{adj}_\textrm{subtr} (x_\perp^2 ,  2 k^- p_1^+) \notag \\ 
& \hspace*{3cm} - i \, \frac{{\un x}}{|{\un x}|^2} \cdot \frac{{\un k} }{|{\un k} |^2} \left(  \frac{3}{2} \, G^\textrm{adj}_\textrm{subtr} ( x_\perp^2 , 2 k^- p_1^+)   + 2 \, G_2^\textrm{adj} (x_\perp^2 , 2 k^- p_1^+)  \right) \bigg] ,
\end{align}
with $\Lambda$ the infrared (IR) cutoff.

Our goal is to rewrite the cross section in the target--projectile symmetric form following \cite{Kovchegov:2001sc, Kovchegov:2012mbw}. We will use the following pure-glue initial conditions \cite{Kovchegov:2016zex, Kovchegov:2017lsr}: 
\begin{align}\label{GG000}
G^{\textrm{adj} \, (0)} ( x_\perp^2 , \alpha s) = 2 \, \as^2 \, \pi \, \frac{N_c}{C_F} \, \ln (\alpha s \, x_\perp^2), \ \ \ G_2^{\textrm{adj} \, (0)} ( x_\perp^2 , \alpha s) = 2 \, \as^2 \, \pi \, \frac{N_c}{C_F} \, \ln \left( \frac{1}{x_\perp \, \Lambda} \right).
\end{align} 
The sign of $G^{\textrm{adj} \, (0)}$ in \eq{GG000} is different from that found in \cite{Kovchegov:2016zex}: this is due to the anti-Brodsky-Lepage spinors \cite{Lepage:1980fj} used in \cite{Cougoulic:2022gbk} and here (see \eq{anti-BLspinors} below) having polarization index $\sigma$ different by a minus sign compared to that used in \cite{Kovchegov:2016zex, Kovchegov:2017lsr}. We have also changed the color factors $C_F/(2 N_c) \to N_c / (2 C_F)$, as compared to \cite{Kovchegov:2016zex, Kovchegov:2017lsr}, to account for the gluon--gluon scattering we consider here (with $C_F$ the fundamental Casimir operator of SU($N_c$)). In addition, we had to scale up the amplitudes by a factor of 2 due to the difference between the gluon and quark spins (the $\eta$-factor in the notation of \cite{Cougoulic:2020tbc}).

We proceed by rewriting \eq{cross_sect_66} as 
\begin{align}\label{cross_sect_9}
& \frac{d \sigma (\lambda)}{d^2 k_T \, dy} = \lambda \, \frac{2 \, \as}{\pi^3}  \frac{1}{s} \, N_c \, \int d^2 x \, e^{- i {\un k} \cdot {\un x}} \,   \bigg\{ \gamma \, \left[ \ln \left( \frac{1}{x_\perp \, \Lambda} \right) - 2 i \, \frac{{\un x}}{|{\un x}|^2} \cdot \frac{{\un k} }{|{\un k} |^2} \right]  \, G^\textrm{adj}_\textrm{subtr} (x_\perp^2 ,  2 k^- p_1^+)  \notag \\ 
& + (1-\gamma) \, \ln \left( \frac{1}{x_\perp \, \Lambda} \right) \, G_\textrm{subtr}^\textrm{adj} (x_\perp^2 ,  2 k^- p_1^+) - i \, \delta \, \frac{{\un x}}{|{\un x}|^2} \cdot \frac{{\un k} }{|{\un k} |^2} \left(  \frac{3}{2} - 2 \, \gamma \right) \, G^\textrm{adj}_\textrm{subtr} ( x_\perp^2 , 2 k^- p_1^+) \notag \\ 
& - i \, (1-  \delta ) \, \frac{{\un x}}{|{\un x}|^2} \cdot \frac{{\un k} }{|{\un k} |^2} \left(  \frac{3}{2} - 2 \, \gamma \right) \, G^\textrm{adj}_\textrm{subtr} ( x_\perp^2 , 2 k^- p_1^+)   - 2 i \,\frac{{\un x}}{|{\un x}|^2} \cdot \frac{{\un k} }{|{\un k} |^2}  \, G_2^\textrm{adj} (x_\perp^2 , 2 k^- p_1^+)  \bigg\} 
\end{align}
with some unknown constants $\gamma$ and $\delta$. Since $G_\textrm{subtr}^\textrm{adj} ( x_\perp^2 =1/(2 k^+ p_2^-), 2 k^- p_1^+) =0$, we can modify the first term in the curly brackets of \eq{cross_sect_9} as \cite{Kovchegov:2001sc}
\begin{align}\label{sub1}
& \int d^2 x \, e^{- i {\un k} \cdot {\un x}} \, \gamma \, \left[ \ln \left( \frac{1}{x_\perp \, \Lambda} \right) - 2 i \, \frac{{\un x}}{|{\un x}|^2} \cdot \frac{{\un k} }{|{\un k} |^2} \right]  \, G_\textrm{subtr}^\textrm{adj} (x_\perp^2 ,  2 k^- p_1^+) \notag \\ 
& = - \int d^2 x \, e^{- i {\un k} \cdot {\un x}} \, \gamma \, \ln \left( \frac{1}{x_\perp \, \Lambda} \right)  \, \frac{1}{k_T^2} \, \nabla_\perp^2 G_\textrm{subtr}^\textrm{adj} (x_\perp^2 ,  2 k^- p_1^+). 
\end{align}
In addition, we can rewrite the fourth term in the curly brackets as 
\begin{align}\label{sub2}
& \int d^2 x \, e^{- i {\un k} \cdot {\un x}} \, (- i) \, (1-  \delta ) \, \frac{{\un x}}{|{\un x}|^2} \cdot \frac{{\un k} }{|{\un k} |^2} \left(  \frac{3}{2} - 2 \, \gamma \right) \, G_\textrm{subtr}^\textrm{adj} ( x_\perp^2 , 2 k^- p_1^+) \notag \\ 
& = \int d^2 x \, e^{- i {\un k} \cdot {\un x}} \, (1-  \delta ) \, \frac{x^i}{|{\un x}|^2} \cdot \frac{1}{k_T^2} \left(  \frac{3}{2} - 2 \, \gamma \right) \, \pd^i G_\textrm{subtr}^\textrm{adj} ( x_\perp^2 , 2 k^- p_1^+) 
\end{align}
by replacing $k^i \to - i \pd^i$ with the derivative operator acting on the exponential and integrating by parts. Here we are assuming that the $\ln (1/x_\perp \Lambda)$ and ${\un x}/|{\un x}|^2$ terms in \eq{cross_sect_9} are associated with the gluon--projectile system, such that their UV scale is $2 k^+ p_2^- = \alpha s$. Therefore, we have
\begin{align}
    \nabla_\perp^2 \ln \left( \frac{1}{x_\perp \, \Lambda} \right)  = \pd^i \, \frac{x^i}{|{\un x}|^2} = - 2 \pi \, \delta^2_{1/\alpha s} ({\un x}), 
\end{align}
where $\delta^2_{1/\alpha s} ({\un x})$ denotes a two-dimensional delta function with the coarse-graining scale of $1/\sqrt{\alpha s}$, such that
\begin{align}
    \int d^2 x \, \delta^2_{1/\alpha s} ({\un x}) \, f(x_\perp^2) = f \left( x_\perp^2 =1/(2 k^+ p_2^-) = 1/\alpha s \right).
\end{align}
After the substitutions \eqref{sub1} and \eqref{sub2}, \eq{cross_sect_9} becomes
\begin{align}\label{cross_sect_10}
& \frac{d \sigma (\lambda)}{d^2 k_T \, dy} = \lambda \, \frac{2 \, \as}{\pi^3}  \frac{1}{s} \, N_c \, \int d^2 x \, e^{- i {\un k} \cdot {\un x}} \,   \bigg\{ - \gamma \, \ln \left( \frac{1}{x_\perp \, \Lambda} \right)  \, \frac{1}{k_T^2} \, \nabla_\perp^2 G^\textrm{adj}_\textrm{subtr} (x_\perp^2 ,  2 k^- p_1^+)  \notag \\ 
& + (1-\gamma) \, \ln \left( \frac{1}{x_\perp \, \Lambda} \right) \, G^\textrm{adj}_\textrm{subtr} (x_\perp^2 ,  2 k^- p_1^+) - i \, \delta \, \frac{{\un x}}{|{\un x}|^2} \cdot \frac{{\un k} }{|{\un k} |^2} \left(  \frac{3}{2} - 2 \, \gamma \right) \, G^\textrm{adj}_\textrm{subtr} ( x_\perp^2 , 2 k^- p_1^+) \notag \\ 
& + (1-  \delta ) \, \frac{x^i}{|{\un x}|^2} \cdot \frac{1}{k_T^2} \left(  \frac{3}{2} - 2 \, \gamma \right) \, \pd^i G^\textrm{adj}_\textrm{subtr} ( x_\perp^2 , 2 k^- p_1^+)  - 2 i \,\frac{{\un x}}{|{\un x}|^2} \cdot \frac{{\un k} }{|{\un k} |^2}  \, G_2^\textrm{adj} (x_\perp^2 , 2 k^- p_1^+)  \bigg\} .
\end{align}

Employing \eq{GG000}, we make the following replacements in \eq{cross_sect_10}:
\begin{subequations}\label{substitutions}
\begin{align}
& \ln \left( \frac{1}{x_\perp \, \Lambda} \right) \to \frac{C_F}{2 \, \as^2 \, \pi \, N_c} \, G_2^{\textrm{adj} \, (0)} ( x_\perp^2 , 2 k^+ p_2^-), \\
& \delta \, \frac{x^i}{|{\un x}|^2} = \delta \, \pd^i  \ln \left( \frac{1}{x_\perp \, \Lambda} \right) \to \frac{C_F}{2 \, \as^2 \, \pi \, N_c} \, \pd^i \left[ \alpha_1 \, G^{\textrm{adj} \, (0)} ( x_\perp^2 , 2 k^+ p_2^-) + \beta_1 \, G_2^{\textrm{adj} \, (0)} ( x_\perp^2 , 2 k^+ p_2^-)\right], \\
& (1-\delta) \, \frac{x^i}{|{\un x}|^2} \to \frac{C_F}{2 \, \as^2 \, \pi \, N_c} \, \pd^i \left[ \alpha_2 \, G^{\textrm{adj} \, (0)} ( x_\perp^2 , 2 k^+ p_2^-) + \beta_2 \, G_2^{\textrm{adj} \, (0)} ( x_\perp^2 , 2 k^+ p_2^-)\right], \\
& \frac{x^i}{|{\un x}|^2} \to \frac{C_F}{2 \, \as^2 \, \pi \, N_c} \, \pd^i \left[ \alpha_3 \, G^{\textrm{adj} \, (0)} ( x_\perp^2 , 2 k^+ p_2^-) + \beta_3 \, G_2^{\textrm{adj} \, (0)} ( x_\perp^2 , 2 k^+ p_2^-)\right],
\end{align}
\end{subequations}
introducing new unknown parameters $\alpha_1, \beta_2, \ldots , \beta_3$ while eliminating $\delta$. Note that the substitutions \eqref{substitutions} are valid only when these parameters satisfy
\begin{subequations}
\begin{align}
& \beta_1 + \beta_2 - 2 \, \alpha_1 - 2 \, \alpha_2 = 1 ,  \label{cond12} \\
& \beta_3 - 2 \, \alpha_3 = 1. \label{cond3}
\end{align}
\end{subequations}

Performing the substitutions \eqref{substitutions} in \eq{cross_sect_10}, replacing the remaining $k^i \to - i \pd^i$ with the derivative acting on the exponential, and integrating by parts, yields
\begin{align}\label{cross_sect_11}
& \frac{d \sigma (\lambda)}{d^2 k_T \, dy} = \lambda \, \frac{C_F}{\as \, \pi^4}  \frac{1}{s \, k_T^2} \,  \int d^2 x \, e^{- i {\un k} \cdot {\un x}} \,   \bigg\{ - \gamma  \, G^\textrm{adj}_{2P} \, \nabla_\perp^2 G^\textrm{adj}_T  + (1-\gamma) \, G^\textrm{adj}_{2P} \, G^\textrm{adj}_T \notag \\ 
& + \left(  \frac{3}{2} - 2 \, \gamma \right) \, \pd^i \left[ \left( \alpha_1 \,  \pd^i  G^\textrm{adj}_P + \beta_1 \,  \pd^i  G^\textrm{adj}_{2P} \right) \, G^\textrm{adj}_T \right] + \left(  \frac{3}{2} - 2 \, \gamma \right) \, \left( \alpha_2 \,  \pd^i  G^\textrm{adj}_P + \beta_2 \,  \pd^i  G^\textrm{adj}_{2P} \right) \, \pd^i G^\textrm{adj}_T  \notag \\
& + 2 \, \pd^i \left[ \left( \alpha_3 \,  \pd^i  G^\textrm{adj}_P + \beta_3 \,  \pd^i  G^\textrm{adj}_{2P} \right) \, G^\textrm{adj}_{2T} \right]  \bigg\} .
\end{align}
In arriving at \eq{cross_sect_11}, we have replaced
\begin{align}\label{sub2}
G^\textrm{adj}_\textrm{subtr} \to G_T^\textrm{adj}, \ \ \ G_2^\textrm{adj}  \to G_{2 T}^\textrm{adj} , \ \ \ G^{\textrm{adj} \, (0)}  \to G_P^\textrm{adj}, \ \ \  G_2^{\textrm{adj} \, (0)} \to G_{2 P}^\textrm{adj} ,
\end{align}
trying to describe the target and the projectile in terms of the same-type quantities. (The fact that $G_T^\textrm{adj}$ denotes a subtracted dipole amplitude while $G_P^\textrm{adj}$ denotes the un-subtracted one will be addressed shortly.) We have also suppressed the arguments of the dipole scattering amplitudes for brevity. 

Analyzing this expression \eqref{cross_sect_11}, we notice that the $(1-\gamma) \, G^\textrm{adj}_{2P} \, G^\textrm{adj}_T$ has no target--projectile (T~$\leftrightarrow$~P) ``dual", i.e, there is no $G^\textrm{adj}_{2T} \, G^\textrm{adj}_P$ term in the expression. Since we are looking for a T~$\leftrightarrow$~P symmetric expression for the cross section, we conclude that this term must vanish, that is,
\begin{align}
\gamma = 1. 
\end{align}
Equation \eqref{cross_sect_11} then becomes
\begin{align}\label{cross_sect_12}
& \frac{d \sigma (\lambda)}{d^2 k_T \, dy} = \lambda \, \frac{C_F}{\as \, \pi^4}  \frac{1}{s \, k_T^2} \,  \int d^2 x \, e^{- i {\un k} \cdot {\un x}} \,   \bigg\{ - G^\textrm{adj}_{2P} \, \nabla_\perp^2 G^\textrm{adj}_T  - \frac{\alpha_1}{2}  \, \left( \nabla_\perp^2  G^\textrm{adj}_P \right) \, G^\textrm{adj}_T - \frac{\beta_1}{2}  \,  \left( \nabla_\perp^2 G^\textrm{adj}_{2P} \right) \, G^\textrm{adj}_T  \notag \\ 
& - \frac{\alpha_1 + \alpha_2}{2} \, \pd^i  G^\textrm{adj}_P \, \pd^i G^\textrm{adj}_T - \frac{\beta_1+ \beta_2}{2} \,  \pd^i  G^\textrm{adj}_{2P} \, \pd^i G^\textrm{adj}_T + 2 \, \alpha_3 \,  \left( \nabla_\perp^2  G^\textrm{adj}_P \right) \, G^\textrm{adj}_{2T}  + 2 \, \beta_3 \, \left( \nabla_\perp^2  G^\textrm{adj}_{2P} \right) \, G^\textrm{adj}_{2T}  \notag \\ 
& + 2 \, \alpha_3 \,  \pd^i  G^\textrm{adj}_P \, \pd^i   G^\textrm{adj}_{2T} + 2 \, \beta_3 \,  \pd^i  G^\textrm{adj}_{2P} \, \pd^i   G^\textrm{adj}_{2T}  \bigg\} .
\end{align}
Similar to the above, the $- \frac{\alpha_1}{2}  \, \left( \nabla_\perp^2  G^\textrm{adj}_P \right) \, G^\textrm{adj}_T$, $- \frac{\beta_1}{2}  \,  \left( \nabla_\perp^2 G^\textrm{adj}_{2P} \right) \, G^\textrm{adj}_T$ and $2 \, \beta_3 \,  \pd^i  G^\textrm{adj}_{2P} \, \pd^i   G^\textrm{adj}_{2T}$ terms have no T~$\leftrightarrow$~P ``duals". Hence, 
\begin{align}
\alpha_1 = \beta_1 = \beta_3 = 0.
\end{align}
The T~$\leftrightarrow$~P imposed on the remaining terms containing $\nabla_\perp^2$ yields
\begin{align}
\alpha_3 = - \frac{1}{2} .
\end{align}
Note that $\alpha_3 = -1/2, \beta_3 =0$ satisfy \eq{cond3}. 

Further, equating the coefficients of the $\pd^i  G^\textrm{adj}_{2P} \, \pd^i G^\textrm{adj}_T$ and $\pd^i  G^\textrm{adj}_P \, \pd^i   G^\textrm{adj}_{2T}$ terms in \eq{cross_sect_12} in order to achieve the T~$\leftrightarrow$~P symmetry gives
\begin{align}
 - \frac{\beta_1+ \beta_2}{2} = 2 \, \alpha_3  \ \ \ \Rightarrow  \ \ \ \beta_2 = 2. 
\end{align}
Finally, employing the condition \eqref{cond12}, we arrive at 
\begin{align}
\alpha_2 = \frac{1}{2}. 
\end{align}

With all the coefficient fixed, we rewrite \eq{cross_sect_12} as
\begin{tcolorbox}[ams align]\label{cross_sect_13}
& \frac{d \sigma}{d^2 k_T \, dy} = \frac{C_F}{\as \, \pi^4}  \frac{1}{s \, k_T^2} \,  \int d^2 x \, e^{- i {\un k} \cdot {\un x}} \,   \bigg\{  G^\textrm{adj}_{2P} \, \nabla_\perp^2 G^\textrm{adj}_T  + \left( \nabla_\perp^2  G^\textrm{adj}_P \right) \, G^\textrm{adj}_{2T} + \frac{1}{4} \, \pd^i  G^\textrm{adj}_P \, \pd^i G^\textrm{adj}_T  \notag \\ 
& \hspace*{5.9cm} + \pd^i  G^\textrm{adj}_{2P} \, \pd^i G^\textrm{adj}_T  + \pd^i  G^\textrm{adj}_P \, \pd^i   G^\textrm{adj}_{2T}   \bigg\} .
\end{tcolorbox}
Here we have absorbed $- \lambda$ into $G_P^\textrm{adj}$ and $G_{2 P}^\textrm{adj}$, since this is the projectile polarization which is now included in those amplitudes. The minus sign accounts for the fact that the projectile gluon is left-moving, while we have been using the polarization vector basis for a right-moving gluon. Target polarization enters through $G_T^\textrm{adj}$ and $G_{2 T}^\textrm{adj}$, as before. 

Note that the amplitude $G^\textrm{adj}_T$ enters \eq{cross_sect_13} only through its derivatives, i.e., through $\nabla_\perp^2 G^\textrm{adj}_T$ and $\pd^i G^\textrm{adj}_T$. As one can readily see from \eq{subtr_def}, the difference between the subtracted and un-subtracted dipole amplitudes vanishes after any transverse coordinate differentiation. Therefore, $G^\textrm{adj}_T$  in \eq{cross_sect_13} can be thought of as the ``true" un-subtracted polarized dipole amplitude on the target proton. 

We observe that \eq{cross_sect_13} is completely symmetric with respect to the projectile--target interchange. For completeness, let us rewrite it restoring the arguments of the dipole amplitudes. Employing the rapidity variable 
\begin{align}
y = \frac{1}{2} \ln \frac{k^-}{k^+}
\end{align}
such that $y=0$ is mid-rapidity, we get $2 p_1^+ k^- = \sqrt{2} \, p_1^+ \, k_T \, e^y$ and $2 p_2^- k^+ = \sqrt{2} \, p_2^- \, k_T \, e^{-y}$. In the center-of-mass frame we have $2 p_1^+ k^- = \sqrt{s} \, k_T \, e^y$ and $2 p_2^- k^+ = \sqrt{s} \, k_T \, e^{-y}$.

Including the arguments of the dipole amplitudes, we rewrite \eq{cross_sect_13} as
\begin{tcolorbox}[ams align]\label{kT_fact_final}
& \frac{d \sigma }{d^2 k_T \, dy}  = \frac{C_F}{\as \, \pi^4}  \frac{1}{s \, k_T^2} \,   \int d^2 x \, e^{- i {\un k} \cdot {\un x}} \notag \\ 
& \times \,  
\left( 
\begin{matrix}
G_P^\textrm{adj} & G_{2 P}^\textrm{adj} 
\end{matrix}
\right) (x_\perp^2, \sqrt{2} \, p_2^- \, k_T \, e^{-y}) \,
\left(
\begin{matrix}
\frac{1}{4} {\cev \nabla}_\perp \cdot {\vec \nabla}_\perp   & {\cev \nabla}_\perp^2 + {\cev \nabla}_\perp \cdot {\vec \nabla}_\perp \\
{\vec \nabla}_\perp^2 + {\cev \nabla}_\perp \cdot {\vec \nabla}_\perp & 0
\end{matrix}
\right) \, 
\left( 
\begin{matrix}
G_T^\textrm{adj} \\ 
G_{2 T}^\textrm{adj} 
\end{matrix}
\right) (x_\perp^2, \sqrt{2} \, p_1^+ \, k_T \, e^y) .
\end{tcolorbox}

This cross section is the main result of this work. Let us reiterate here that the dipole amplitudes are defined in the regions $x_\perp^2 > 1/(\sqrt{2} \, p_2^- \, k_T \, e^{-y})$ for the projectile and $x_\perp^2 > 1/(\sqrt{2} \, p_1^+ \, k_T \, e^y)$ for the target. This implies that the integration in \eq{kT_fact_final} is bounded by
\begin{align}\label{x_perp_min}
    x_\perp^2 > \max \left\{ \frac{1}{\sqrt{2} \, p_2^- \, k_T \, e^{-y}} ,  \frac{1}{\sqrt{2} \, p_1^+ \, k_T \, e^y} \right\}
\end{align}
from below.

Note that at large $N_c$ we have the following relation between the adjoint and fundamental ($G, G_2$) polarized dipole amplitudes \cite{Kovchegov:2018znm, Cougoulic:2022gbk},
\begin{align}\label{largeNc}
G^\textrm{adj} = 4 \, G, \ \ \  G_2^\textrm{adj} = 2 \, G_2 . 
\end{align}

We can rewrite \eq{kT_fact_final} in terms of the gluon transverse-momentum dependent  distributions (TMDs). The fundamental-dipole gluon helicity TMD at small $x$ and at large $N_c$ can be written at DLA as (see, e.g., Eq.~(41) of \cite{Cougoulic:2022gbk}; here, in DLA, we have dropped the derivative with respect to the transverse size squared in that formula) 
\begin{align}\label{glue_hel_TMD57}
g_{1L}^{G \, dip} (x, k_T^2) = \frac{N_c}{\as \, 4 \pi^4} \, \int d^2 x \, e^{- i \un{k} \cdot \un{x} } \, G^\textrm{adj}_2 \left(  x_\perp^2,  \frac{Q^2}{x} \right).
\end{align}
Thus, the polarized dipole amplitude $G^\textrm{adj}_2$ is related to $g_{1L}^{G \, dip} (x, k_T^2)$. (Here $Q^2$ is the renormalization scale.)

In Appendix~\ref{sec:small-x_TMDs} we show that the polarized dipole amplitude $G^\textrm{adj}$ is, in turn, related to the twist-three helicity-flip dipole gluon TMD $\Delta H^\perp_{3L} (x, k_T^2)$ \cite{Mulders:2000sh} at small $x$ and at large $N_c$ by\footnote{Note that our definition of $\Delta H^\perp_{3L}$ is equal to $-2 x$ times the definition of this TMD in \cite{Mulders:2000sh}.} 
\begin{align}\label{helicity_flip_TMD}
\Delta H^{\perp \, dip}_{3L} (x, k_T^2) = \frac{N_c}{\as \, 16 \pi^4} \, \int d^2 x \, e^{- i \un{k} \cdot \un{x} } \, G^\textrm{adj} \left(  x_\perp^2, \frac{Q^2}{x} \right). 
\end{align}

Using Eqs.~\eqref{glue_hel_TMD57} and \eqref{helicity_flip_TMD} we can rewrite \eq{kT_fact_final} at large-$N_c$ and at small $x$ as
\begin{align}\label{kT_fact_TMDs}
    & \frac{d \sigma }{d^2 k_T \, dy}  = - \frac{32 \pi^4 \, \as}{N_c}  \frac{1}{s \, k_T^2} \,   \int \frac{d^2 q}{(2 \pi)^2}  \notag \\ 
& \times \,  
\bigg( 
\begin{matrix}
\Delta H^{\perp \, dip \, P}_{3L} & g_{1L}^{G \, dip \, P} 
\end{matrix}
\bigg) \left( q_T^2, \frac{k_T}{\sqrt{2} \, p_2^-} \, e^{y} \right) \,
\left(
\begin{matrix}
{\un q} \cdot ({\un k}  - {\un q})   & \,  {\un q} \cdot {\un k} \\~~\\
{\un k} \cdot ({\un k} - {\un q})  & \, 0
\end{matrix}
\right) \, 
\left( 
\begin{matrix}
\Delta H^{\perp \, dip \, T}_{3L} \\~~\\ 
g_{1L}^{G \, dip \, T} 
\end{matrix}
\right) \left( ({\un k} - {\un q})^2, \frac{k_T}{\sqrt{2} \, p_1^+ } \, e^{-y} \right) ,
\end{align}
with the renormalization scale chosen to be equal to $k_T^2$. The P and T superscripts on the TMDs denote the TMDs of the projectile and of the target, respectively. \eq{kT_fact_TMDs} is the polarized proton--proton scattering analogue of the unpolarized $k_T$-factorization formula for gluon production derived in \cite{Kovchegov:2001sc, Braun:2000bh}. Just as the result of \cite{Kovchegov:2001sc, Braun:2000bh}, our $k_T$-factorization expression \eqref{kT_fact_TMDs} is valid in the quasi-classical GGM/MV approximation and, as we will show below, for the leading-order small-$x$ evolution of the dipole amplitudes (that is, in DLA for helicity evolution). We do not expect \eq{kT_fact_TMDs} to apply beyond the small-$x$ regime.


\subsection{Lowest-order calculations}
\label{sec:fixed}

Let us cross-check our main result \eqref{kT_fact_final} by performing the lowest-order gluon production calculation in the polarized parton--parton scattering. For simplicity, we will work in the large-$N_c$ approximation. Even in this approximation, finding the lowest non-trivial contribution to the polarization-dependent part of the $GG \to GGG$ process can be a significant calculation. Our strategy is to do the calculation of the polarization-dependent part of the $qq \to qqG$ process instead, and then use the observations obtained in \cite{Cougoulic:2020tbc} to deduce the (large-$N_c$ part of) of the polarization-dependent  $GG \to GGG$ cross section.

\begin{figure}[ht]
    \centering
\includegraphics[width= 0.8 \textwidth]{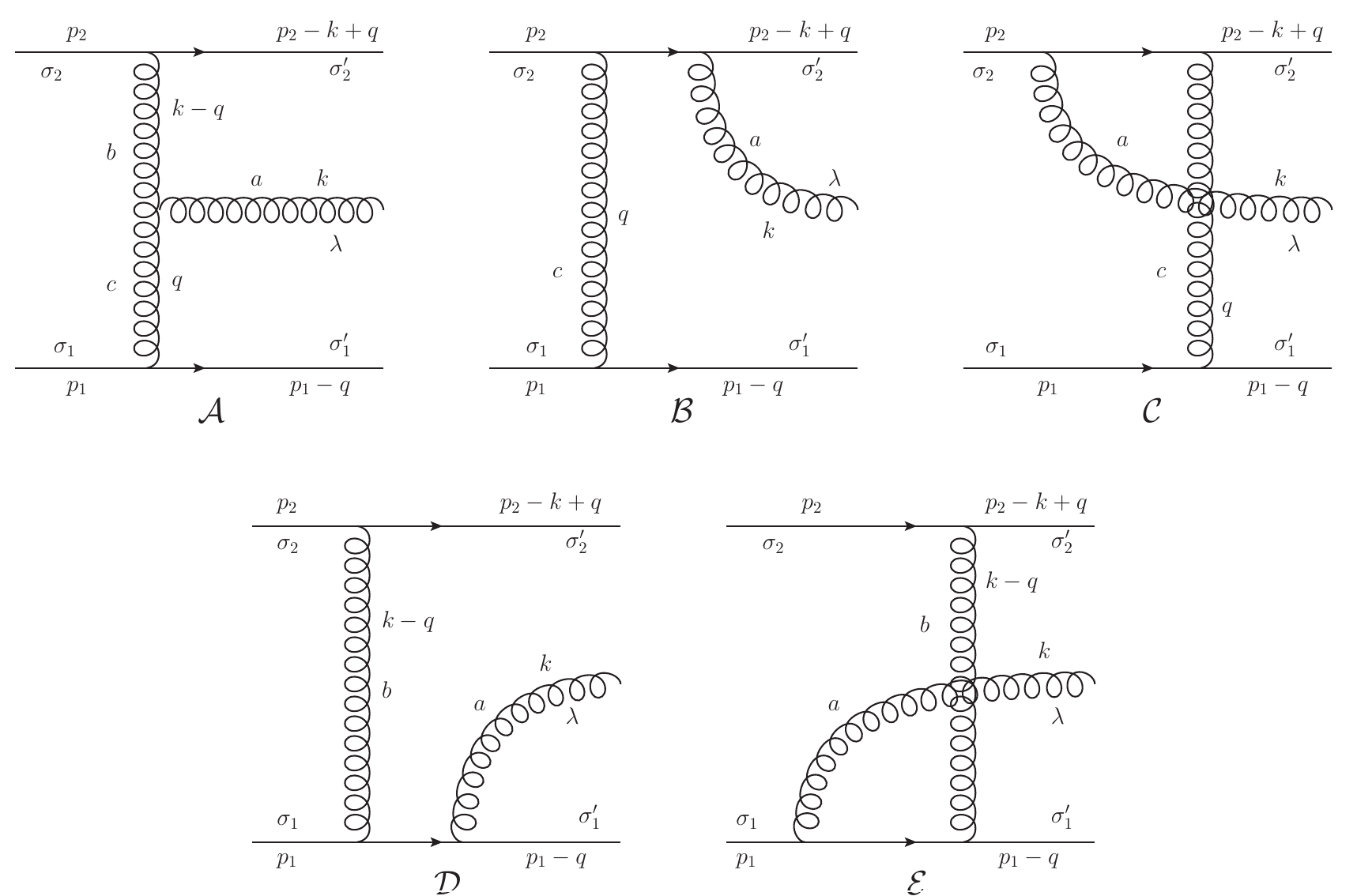}
    \caption{Diagrams contributing to the leading-$N_c$ gluon production cross section in the polarized quark--quark scattering. }
    \label{fig:gluon_production_LO}
\end{figure}

The diagrams contributing to gluon production amplitude in the $qq \to qqG$ process at large $N_c$ are shown in \fig{fig:gluon_production_LO}. The diagrams we are neglecting here may not be immediately seen as $N_c$-suppressed, but they lead to $N_c$-suppressed contributions in the polarization-dependent part of the cross section at the sub-eikonal level of our calculation. 

As before, our incoming ``projectile" quark has a large minus momentum component $p_2^-$, while the ``target" quark has a large plus momentum $p_1^+$.  We will work in the multi-peripheral regime for particle production, defined by 
\begin{align}
(p_2 + q - k)^- \gg k^- \gg (p_1 - q)^- , \ \ \ (p_2 + q - k)^+ \ll k^+ \ll (p_1 - q)^+ .
\end{align}
It requires that $\alpha \ll 1$ and $\beta \ll 1$, along with
\begin{align}\label{MPP}
q_T^2 \ll \beta \, s ,  \ \ \ ({\un k} - {\un q})^2 \ll \alpha \, s,
\end{align}
where $\alpha = k^+/p_1^+$, $\beta = k^-/p_2^-$ and $s = 2 p_1^+ p_2^-$. Note that $k_T^2 = \alpha \beta s$, which is much smaller than $\alpha s$ and $\beta s$, such that the upper bound on $q_T$ becomes
\begin{align}\label{qT_max}
    q_T^2 \ll \min \{ \alpha, \beta \} \, s . 
\end{align}

For the plus-moving (target) quark line we use the Brodsky-Lepage spinors \cite{Lepage:1980fj}. For the minus-moving (projectile) quark line we use the $(\pm)$-interchanged Brodsky-Lepage spinors, which we will refer to as the anti-BL spinors following \cite{Kovchegov:2018znm, Kovchegov:2021iyc}. These are
\begin{align}\label{anti-BLspinors}
u_\sigma (p) = \frac{1}{\sqrt{\sqrt{2} \, p^-}} \, [\sqrt{2} \, p^- + m \, \gamma^0 +  \gamma^0 \, {\un \gamma} \cdot {\un p} ] \,  \rho (\sigma), \ \ \ v_\sigma (p) = \frac{1}{\sqrt{\sqrt{2} \, p^-}} \, [\sqrt{2} \, p^- - m \, \gamma^0 +  \gamma^0 \, {\un \gamma} \cdot {\un p} ] \,  \rho (-\sigma),
\end{align}
with $p^\mu = \left( \frac{{\un p}^2+ m^2}{2 p^-}, p^-, {\un p} \right)$ and
\begin{align}
  \rho (+1) \, = \, \frac{1}{\sqrt{2}} \, \left(
  \begin{array}{c}
      1 \\ 0 \\ -1 \\ 0
  \end{array}
\right), \ \ \ \rho (-1) \, = \, \frac{1}{\sqrt{2}} \, \left(
  \begin{array}{c}
        0 \\ 1 \\ 0 \\ 1
  \end{array}
\right) .
\end{align}

Calculating the diagrams in \fig{fig:gluon_production_LO} in $A^-=0$ gauge while including only the sub-eikonal helicity-dependent terms, we get the following contributions to the scattering amplitude: 
\begin{subequations}\label{ABCDE_Feyn}
\begin{align}
& i \, M_{\cal A} = g^3 f^{abc} (t^b)_2 (t^c)_1 \, \delta_{\sigma_1 \sigma'_1} \, \delta_{\sigma_2 \sigma'_2} \, \frac{1}{{\un q}^2 \, ({\un k} - {\un q})^2} \Bigg\{ - 4 \, s \, {\un \epsilon}^*_\lambda \cdot ({\un k} - {\un q}) + 2 i \frac{p_1^+}{k^+} \,  \sigma_2 \, \left[ 2 \, {\un k} \times {\un q} \, {\un \epsilon}^*_\lambda \cdot {\un k} - (k_T^2 - q_T^2) \, {\un \epsilon}^*_\lambda \times {\un q} \right] \notag \\
& \hspace*{2.5cm} + 4 i p_2^- k^+  \sigma_1 \, {\un \epsilon}^*_\lambda \times ({\un k} - {\un q}) + 2 \, \sigma_1 \, \sigma_2 \, \left[ {\un \epsilon}^*_\lambda \cdot  {\un q} \, {\un q} \cdot ({\un k} - {\un q}) +  {\un \epsilon}^*_\lambda \times {\un k} \, {\un k} \times {\un q}  \right] \Bigg\}, \\
& i \, M_{\cal B} = i g^3 (t^a t^c)_2 (t^c)_1 \, \delta_{\sigma_1 \sigma'_1} \, \delta_{\sigma_2 \sigma'_2} \, \frac{1}{q_T^2 \, 2 p_1^+ \, k^-} \, \Bigg\{ 4 \frac{p_1^+}{k^+} \, s \, {\un \epsilon}^*_\lambda \cdot {\un k}  - 2 i s \, {\un \epsilon}^*_\lambda \times {\un k} \, \sigma_1  - 2 p_1^+ \, \sigma_1 \, \sigma_2 \, \left[ {\un \epsilon}^*_\lambda \cdot {\un k} \, \frac{q_T^2}{k^+} - {\un \epsilon}^*_\lambda \cdot {\un q} \, k^- \right] \Bigg\} , \\
& i \, M_{\cal C} = - i g^3 (t^c t^a)_2 (t^c)_1 \, \delta_{\sigma_1 \sigma'_1} \, \delta_{\sigma_2 \sigma'_2} \, \frac{1}{q_T^2 \, 2 p_1^+ \, k^-} \, \Bigg\{ 4 \frac{p_1^+}{k^+} \, s \, {\un \epsilon}^*_\lambda \cdot {\un k}  - 2 i s \, {\un \epsilon}^*_\lambda \times {\un k} \, \sigma_1 - 2 p_1^+ \, \sigma_1 \, \sigma_2 \, \left[ {\un \epsilon}^*_\lambda \cdot {\un k} \, \frac{q_T^2}{k^+} - {\un \epsilon}^*_\lambda \cdot {\un q} \, k^- \right] \Bigg\} , \\
& i \, M_{\cal D} = i g^3 (t^b)_2 (t^a t^b)_1 \, \delta_{\sigma_1 \sigma'_1} \,  \delta_{\sigma_2 \sigma'_2} \, \, \frac{1}{2 p_2^- k^+ \, ({\un q} - {\un k})^2} \Bigg\{ 2 i s \, \sigma_2 \, {\un \epsilon}^*_\lambda \times ({\un q} - {\un k}) 
+  \sigma_1 \, \sigma_2 \, 2 \, p_2^- \, k^+ \, {\un \epsilon}^*_\lambda \cdot ({\un k} - {\un q}) \Bigg\}, \\
& i \, M_{\cal E} = - i g^3 (t^b)_2 (t^b t^a)_1 \, \delta_{\sigma_1 \sigma'_1} \, \delta_{\sigma_2 \sigma'_2} \, \frac{1}{2 p_2^- k^+ \, ({\un q} - {\un k})^2} \Bigg\{ 2 i s \, \sigma_2 \, {\un \epsilon}^*_\lambda \times ({\un q} - {\un k})  +  \sigma_1 \, \sigma_2 \, 2 \, p_2^- \, k^+ \, {\un \epsilon}^*_\lambda \cdot ({\un k} - {\un q}) \Bigg\}.
\end{align}
\end{subequations}
All momenta, colors and polarizations are as labeled in \fig{fig:gluon_production_LO}. We also keep only the leading term in each channel (with the channels defined by their $\sigma$-structure). The subscripts 1 and 2 on the (products of) color matrices indicate the color spaces of quarks 1 and 2. 

Adding up the contributions in Eqs.~\eqref{ABCDE_Feyn} we get
\begin{align}\label{netM}
& i \, M = i \, (M_{\cal A} + M_{\cal B} + M_{\cal C} + M_{\cal D} + M_{\cal E}) = g^3 f^{abc} (t^b)_2 (t^c)_1 \, \delta_{\sigma_1 \sigma'_1} \, \delta_{\sigma_2 \sigma'_2} \, \Bigg\{ - \frac{4 s}{q_T^2} \, {\un \epsilon}^*_\lambda \cdot \left( \frac{{\un k } - {\un q}}{({\un k } - {\un q})^2} - \frac{\un k}{{\un k}^2} \right) \notag \\
& + 2 i \sigma_2 \, \frac{p_1^+}{k^+} \, \frac{ 2 \, {\un k} \times {\un q} \, {\un \epsilon}^*_\lambda \cdot {\un k} - k_T^2  \, {\un \epsilon}^*_\lambda \times {\un q} + q_T^2  \, {\un \epsilon}^*_\lambda \times {\un k} }{{\un q}^2 \, ({\un k } - {\un q})^2} + 2 i \sigma_1 \, \frac{p_2^-}{k^-} \, \frac{k_T^2 \, {\un \epsilon}^*_\lambda \times ({\un k} - {\un q}) - ({\un k} - {\un q})^2 \, {\un \epsilon}^*_\lambda \times {\un k} }{{\un q}^2 \, ({\un k } - {\un q})^2} \notag \\
& + \sigma_1 \, \sigma_2 \, \left[ 2 \, \frac{{\un \epsilon}^*_\lambda \cdot  {\un q} \, {\un q} \cdot ({\un k} - {\un q}) +  {\un \epsilon}^*_\lambda \times {\un k} \, {\un k} \times {\un q} }{{\un q}^2 \, ({\un k } - {\un q})^2} - {\un \epsilon}^*_\lambda \cdot \left( 2 \frac{\un k}{{\un k}^2} - \frac{\un q}{{\un q}^2} + \frac{{\un k} - {\un q}}{({\un k } - {\un q})^2}\right) \right] \Bigg\}.
\end{align}
The corresponding $\sigma_1 \, \sigma_2$-dependent part of the amplitude squared (and averaged over the incoming quarks' colors) is
\begin{align}\label{total}
\langle |M|^2 \rangle = -4 g^6 C_F \, \sigma_1 \, \sigma_2 \, s \, \frac{{\un k}^2 + {\un q}^2 + ({\un k} - {\un q})^2}{{\un k}^2 \, {\un q}^2 \, ({\un k} - {\un q})^2} .
\end{align}
This agrees, up to an overall sign, with Eq.~(B4) in \cite{Kovchegov:2016zex}, where the same process was calculated in Feynman gauge. The sign difference is due to the fact that the anti-BL spinors \eqref{anti-BLspinors} have a different sign convention for the quark polarization $\sigma$ compared to the extraction of the polarization-dependent part of the cross section using the projection with the $\sigma_2 \gamma^5 /2$ operator employed in \cite{Kovchegov:2016zex}.

The amplitude squared in \eq{total} is for the polarization-dependent part of the $qq \to qqG$ process, while we are interested in the $GG \to GGG$ process. To construct the latter, we first concentrate on the contribution of diagrams ${\cal B}$ and $\cal C$ from \fig{fig:gluon_production_LO}. If we replace the quark lines by gluons in those diagrams, the $qq \to qq$ sub-process would become $GG \to GG$, which, at the sub-eikonal level, can proceed through the $s$- and $u$-channel exchange, along with a 4-gluon vertex contribution, in addition to the $t$-channel exchange shown in  \fig{fig:gluon_production_LO}, as demonstrated in \cite{Cougoulic:2020tbc}. The effect of the $s$- and $u$-channel exchanges and the 4-gluon vertex contribution is to reduce the $t$-channel exchange contribution in half \cite{Cougoulic:2020tbc}. Thus, before scaling \eq{total} up by the factors appropriate to the conversion from incoming quarks to gluons, we need to reduce the contribution of diagrams ${\cal B}$ and $\cal C$ by 2.

The $\sigma_1 \, \sigma_2$-dependent part of the diagrams ${\cal B}$ and $\cal C$ from \fig{fig:gluon_production_LO}, interfering with the eikonal (first in the curly brackets) term in \eq{netM}, gives the following contribution to $\langle |M|^2 \rangle$, 
\begin{align}\label{BC}
4 g^6 C_F \, \sigma_1 \, \sigma_2 \, s \, \frac{1}{{\un q}^2} \, \left( \frac{{\un k } - {\un q}}{({\un k } - {\un q})^2} - \frac{\un k}{{\un k}^2} \right) \cdot \left( 2 \, \frac{\un k}{{\un k}^2} -  \frac{\un q}{{\un q}^2} \right) . 
\end{align}
The last ${\un q}/{\un q}^2$ term in the second parentheses can be neglected: indeed,
\begin{align}
\int d^2 q_\perp \frac{1}{{\un q}^2} \, \frac{\un k}{{\un k}^2} \cdot \frac{\un q}{{\un q}^2} = 0,
\end{align}
while
\begin{align}
\int d^2 q_\perp \frac{1}{{\un q}^2} \, \frac{{\un k } - {\un q}}{({\un k } - {\un q})^2} \cdot \frac{\un q}{{\un q}^2} = - \frac{\pi}{k_T^2} 
\end{align}
and is non-logarithmic, suppressed by a power of the logarithm compared to the rest of the expression (integrated over $\un q$). Such contribution probably corresponds to emission from inside the shock wave in the shock wave formalism. 

The remainder of \eq{BC} gives
\begin{align}\label{BC2}
- 4 g^6 C_F \, \sigma_1 \, \sigma_2 \, s \, \frac{- {\un k}^2 + {\un q}^2 + ({\un k} - {\un q})^2}{{\un k}^2 \, {\un q}^2 \, ({\un k} - {\un q})^2} . 
\end{align} 
To go from $qq \to qqG$ to $GG \to GGG$, we need to subtract \eq{BC2} from \eq{total}, and add $1/2$ of \eq{BC2} to the resulting expression: this amounts to subtracting $1/2$ of \eq{BC2} from \eq{total}. In addition, we need to multiply everything by 4 to account for the color factor difference between $qq \to qqG$ and $GG \to GGG$ (at large $N_c$) and by 4 to account for the spin difference between gluons and quarks (see \cite{Cougoulic:2020tbc}). We get (at large $N_c$)
\begin{align}
\langle |M|^2 \rangle_{GG \to GGG} = -32 g^6 C_F \, \sigma_1 \, \sigma_2 \, s \, \frac{3 {\un k}^2 + {\un q}^2 + ({\un k} - {\un q})^2}{{\un k}^2 \, {\un q}^2 \, ({\un k} - {\un q})^2} 
\end{align}
with the corresponding cross section
\begin{align}\label{GG->GGG_LO}
\frac{d \sigma_{LO}^{GG \to GGG}}{d^2 k_T \, dy} = - \frac{4 \, \as^3}{\pi^2} \frac{N_c}{s} \int d^2 q \, \frac{3 {\un k}^2 + {\un q}^2 + ({\un k} - {\un q})^2}{{\un k}^2 \, {\un q}^2 \, ({\un k} - {\un q})^2}.
\end{align}
The $\un q$ integral is bounded from above by the limits stated in \eq{qT_max}. We will use the IR cutoff $\Lambda$ to regulate the singularities at ${\un q} = 0$ and ${\un q} = {\un k}$. Performing the $\un q$ integration in \eq{GG->GGG_LO} yields
\begin{align}\label{GG->GGG_LO2}
\frac{d \sigma_{LO}^{GG \to GGG}}{d^2 k_T \, dy} = - \frac{8 \, \as^3}{\pi} \frac{N_c}{s \, k_T^2} \, \left\{ 3 \, \ln \frac{k_T^2}{\Lambda^2} + \ln \left( \frac{\min \{ \alpha, \beta \} \, s}{\Lambda^2} \right) \right\}.
\end{align}

This result needs to be compared to our main result \eqref{kT_fact_final} evaluated at the lowest non-trivial order. Indeed, substituting the Born-level polarized dipole amplitudes from \eq{GG000} into \eq{kT_fact_final}, both for the projectile and for the target (while replacing $\alpha \to \beta$ for the latter), and integrating over $\un x$ while keeping the lower bound \eqref{x_perp_min} in mind, such that 
\begin{align}
    \nabla_\perp^2 \ln (\alpha s x_\perp^2) = \nabla_\perp^2 \ln (\beta s x_\perp^2) = 4 \pi \delta^2_{\max \left\{ 1/\alpha s, 1/\beta s \right\}} ({\un x}),
\end{align}
we readily arrive at \eq{GG->GGG_LO2} (up to the minus sign mentioned before due to the anti-BL spinor convention from \cite{Kovchegov:2018znm, Kovchegov:2021iyc}). This accomplishes the cross check of \eq{kT_fact_final} at the lowest non-trivial order. 

Let us add that the inclusive gluon production cross section in unpolarized gluon--gluon collisions at the same lowest order in the coupling (order-$\as^3$) is (see \cite{Kovchegov:2012mbw} and references therein)
\begin{align}
    \frac{d \sigma_{LO, \, \textrm{unpolarized}}^{GG \to GGG}}{d^2 k_T \, dy} = \frac{4 \, \as^3 \, N_c^2}{\pi \, C_F} \frac{1}{k_T^4} \, \ln \frac{k_T^2}{\Lambda^2} .
\end{align}
Using this result, along with \eq{GG->GGG_LO2} in \eq{ALL}, we conclude that, at this lowest order, the double-spin asymmetry scales as
\begin{align}
    A_{LL} \sim \frac{k_T^2}{s},
\end{align}
in qualitative agreement with the experimental data reported at RHIC \cite{PHENIX:2014gbf,PHENIX:2015fxo,STAR:2014wox,STAR:2021mqa}. 


\section{Including small-$x$ evolution}
\label{sec:evolution}

The aim of this Section is to show that our main result \eqref{kT_fact_final} for the gluon production cross section in polarized proton--proton collisions in the pure glue sector is valid when the DLA small-$x$ helicity evolution \cite{Kovchegov:2015pbl, Kovchegov:2016zex, Kovchegov:2018znm, Cougoulic:2022gbk} is included in the rapidity intervals between the produced gluon and the projectile and between the produced gluon and the target. We will show that the evolution would just enter the polarized dipole amplitudes in \eq{kT_fact_final}, both for the dipole scattering on the projectile and on the target.  

The evolution is easier to discuss in terms of rapidity variables. Assuming that both the projectile and the target are characterized by the same transverse momentum scale $\Lambda$, our IR cutoff, we define the rapidity of the projectile as 
\begin{align}
    Y_P = \frac{1}{2} \ln \frac{p_2^-}{p_2^+} = \ln \frac{\sqrt{2} \, p_2^-}{\Lambda}
\end{align}
and the rapidity of the target as
\begin{align}
    Y_T = \frac{1}{2} \ln \frac{p_1^-}{p_1^+} = \ln \frac{\Lambda}{\sqrt{2} \, p_1^+}.
\end{align}
Using these two variables, we rewrite \eq{kT_fact_final} as
\begin{align}\label{kT_fact_final_rapidity}
& \frac{d \sigma }{d^2 k_T \, dy}  = \frac{C_F}{\as \, \pi^4}  \frac{1}{s \, k_T^2} \,   \int d^2 x \, e^{- i {\un k} \cdot {\un x}} \notag \\ 
& \times \,  
\left( 
\begin{matrix}
G_P^\textrm{adj} & G_{2 P}^\textrm{adj} 
\end{matrix}
\right) (x_\perp^2, \Lambda \, k_T \, e^{Y_P -y}) \,
\left(
\begin{matrix}
\frac{1}{4} {\cev \nabla}_\perp \cdot {\vec \nabla}_\perp   & {\cev \nabla}_\perp^2 + {\cev \nabla}_\perp \cdot {\vec \nabla}_\perp \\
{\vec \nabla}_\perp^2 + {\cev \nabla}_\perp \cdot {\vec \nabla}_\perp & 0
\end{matrix}
\right) \, 
\left( 
\begin{matrix}
G_T^\textrm{adj} \\ 
G_{2 T}^\textrm{adj} 
\end{matrix}
\right) (x_\perp^2, \Lambda \, k_T \, e^{y- Y_T}) .
\end{align}
Hence, the evolution in the target dipole amplitudes will correspond to the evolution in the $y- Y_T$ rapidity interval, while the evolution in the projectile dipole amplitudes would be given by the evolution in the $Y_P -y$ rapidity interval.


\subsection{Target side}

Including the DLA evolution \cite{Kovchegov:2015pbl, Kovchegov:2016zex, Kovchegov:2018znm, Cougoulic:2022gbk} in the $y- Y_T$ rapidity interval is quite straightforward. For the central rapidity gluon production in unpolarized proton--proton or proton--nucleus collisions, the inclusion of small-$x$ evolution between the gluon and the target was done in \cite{Kovchegov:2001sc}. For that unpolarized case, this is now a standard procedure in the field, detailed in \cite{Kovchegov:2012mbw}: one can show that to include small-$x$ evolution in the $y- Y_T$ rapidity interval, one has to simply evolve the unpolarized dipole amplitude $N$ on the target with the non-linear small-$x$ evolution equations \cite{  Balitsky:1995ub,Balitsky:1998ya,Kovchegov:1999yj,Kovchegov:1999ua,Jalilian-Marian:1997dw,Jalilian-Marian:1997gr,Weigert:2000gi,Iancu:2001ad,Iancu:2000hn,Ferreiro:2001qy}, leaving the expression for the production cross section the same. This conclusion is perhaps easiest to see in the operator language: writing the dipole amplitude as a correlator of light-cone Wilson line operators makes the expression equally valid with and without the small-$x$ evolution. 

The same logic applies to the helicity evolution at hand. While the target-projectile symmetric expression \eqref{kT_fact_final} treats the target and projectile polarized dipole amplitudes on the same footing by the virtue of substitutions \eqref{substitutions} and \eqref{sub2}, the expression we derived above in the operator language is \eq{cross_sect_1}. In \eq{cross_sect_1} the gluon production cross section is expressed in terms of the target polarized dipole amplitudes, defined via correlators of polarized and regular Wilson line operators in Eqs.~\eqref{G_adj_defs}. These operator matrix elements can be evaluated in the quasi--classical approximation of the helicity-dependent MV (hMV) model \cite{Cougoulic:2020tbc}, or by including the small-$x$ helicity evolution \cite{Kovchegov:2015pbl, Kovchegov:2016zex, Kovchegov:2018znm, Cougoulic:2022gbk} into them. In this sense, the effects of the DLA helicity evolution are already included into \eq{kT_fact_final} (or, equivalently, into \eq{kT_fact_final_rapidity}) if we evaluate $G_T^\textrm{adj}$ and $G_{2 T}^\textrm{adj}$ in those expressions by solving the equations derived in \cite{Cougoulic:2022gbk}. Since the helicity evolution equations \cite{Kovchegov:2015pbl, Kovchegov:2016zex, Kovchegov:2018znm, Cougoulic:2022gbk} close only in the large-$N_c$ and large-$N_c \& N_f$ limits, and as the large-$N_c$ limit is the one appropriate for our gluons-only calculation here, the relations in \eq{largeNc} are needed to connect the adjoint dipole amplitudes $G_T^\textrm{adj}$ and $G_{2 T}^\textrm{adj}$ to the fundamental ones, $G_T$ and $G_{2 T}$, entering the large-$N_c$ helicity evolution equations in \cite{Cougoulic:2022gbk}. For finite $N_c$, the polarized dipole amplitudes $G_T^\textrm{adj}$ and $G_{2 T}^\textrm{adj}$ can be evaluated using the helicity-dependent extension \cite{Cougoulic:2019aja} of the Jalilian-Marian--Iancu--McLerran--Weigert--Leonidov--Kovner (JIMWLK) evolution equation \cite{Jalilian-Marian:1997dw,Jalilian-Marian:1997gr,Weigert:2000gi,Iancu:2001ad,Iancu:2000hn,Ferreiro:2001qy}.  

To summarize this Subsection, we see that the small-$x$ evolution effects in the $y- Y_T$ rapidity interval are already included in \eq{kT_fact_final} if the target polarized dipole amplitudes are evaluated using the DLA helicity evolution from \cite{Kovchegov:2015pbl, Kovchegov:2016zex, Kovchegov:2018znm, Cougoulic:2022gbk}.


\subsection{Projectile side}

Including the small-$x$ evolution on the projectile side, in the $Y_P -y$ rapidity interval, is less straightforward than on the target side. Here we will follow the strategy applied in \cite{Kovchegov:2001sc}, with more details provided in Chapter~8 of \cite{Kovchegov:2012mbw}. We will also work in the large-$N_c$ limit: while this may be a little less general than the any-$N_c$ applicability of \eq{kT_fact_final}, currently the helicity evolution equations are solved in the large-$N_c$ limit \cite{Cougoulic:2022gbk, Borden:2023ugd} (and at large-$N_c \& N_f$ \cite{Adamiak:2023yhz}), with the solution of helicity-JIMWLK (hJIMWLK) evolution \cite{Cougoulic:2019aja} still lacking (and, with the hJIMWLK evolution \cite{Cougoulic:2019aja} itself in need of revisiting in light of the corrections found in \cite{Cougoulic:2022gbk}). 

\begin{figure}[th]
    \centering
\includegraphics[width= 0.9 \textwidth]{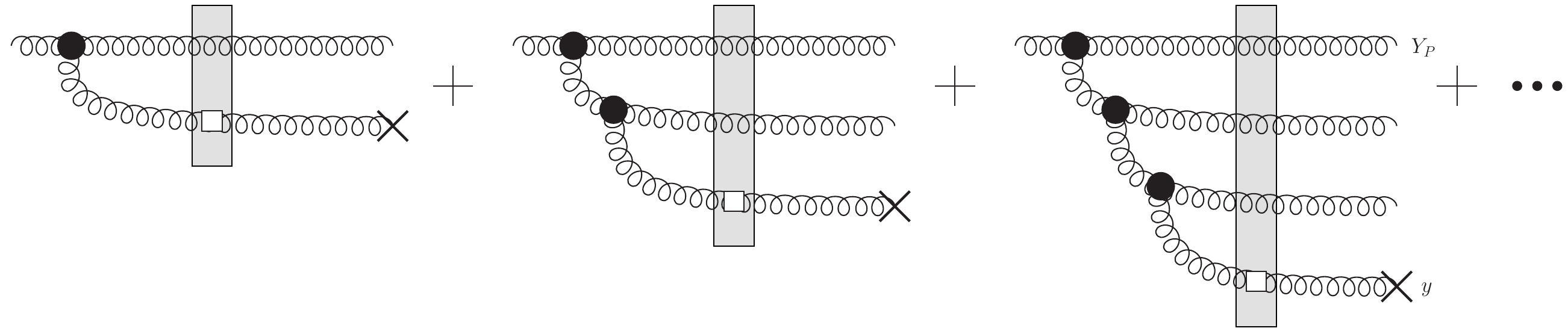}
    \caption{Diagrams illustrating gluon emissions in the rapidity interval $Y_P -y$ between the projectile gluon at rapidity $Y_P$ and the produced gluon at rapidity $y$. The produced gluon is marked by a cross.}
    \label{fig:Pevolution}
\end{figure}

Following \cite{Kovchegov:2001sc},
we first argue that, in the shock wave picture, the gluon emissions in the $Y_P -y$ rapidity interval take place {\sl before} the interaction with the shock wave. The argument completely parallels that in \cite{Kovchegov:2001sc} and will not be repeated here: we just remark that the hared gluon emissions (that is, emissions in the $Y_P -y$ rapidity interval) happening after the interaction with the shock wave cancel after squaring the amplitude (see \cite{Kovchegov:2012mbw} for more details). The result is illustrated in \fig{fig:Pevolution}, where the emission of gluons harder (in the longitudinal momentum fraction $\beta$) than the produced gluon are shown for a gluon projectile. Per the above-mentioned argument, the surviving emissions all take place to the left of the shock wave (before the shock wave interaction).

In the large-$N_c$ limit, 
these early emissions of harder in $\beta$ gluons result in a dipole cascade, with the produced gluon being emitted from one of the dipoles formed this way. Unlike the eikonal unpolarized evolution, where there is only one type of dipoles \cite{Mueller:1994rr,Mueller:1994jq,Mueller:1995gb,Balitsky:1995ub,Balitsky:1998ya,Kovchegov:1999yj,Kovchegov:1999ua}, in the sub-eikonal helicity evolution case in the gluon sector we have two types of dipole amplitudes, as given by Eqs.~\eqref{G_adj_defs}. Therefore, the produced gluon can be emitted by either of the two possible dipoles, constructed out of the ${\cal F}^{12}$ and $\mathscr{D}^i - \cev{\mathscr{D}}^i$ operators, respectively. These emissions are a generalization of the above calculation for a single projectile gluon and are considered in detail in Appendix~\ref{sec:dipole} below. In Appendix~\ref{sec:dipole} we demonstrate that when the produced gluon is emitted by a color-dipole made using either the ${\cal F}^{12}$ or $\mathscr{D}^i - \cev{\mathscr{D}}^i$ operators, the gluon production cross section still factorizes into the part describing the gluon emission by the projectile dipole, and the interaction of the system with the target. Importantly, the interaction with the target is described by the same dipole amplitudes $G_T^\textrm{adj} (\beta s)$ and $G_T^{i \, \textrm{adj}} (\beta s)$ from Eqs.~\eqref{G_adj_defs}, just like in the gluon projectile case of \eq{cross_sect_1}. We can then conclude that the interaction with the target for the single gluon inclusive production cross section with multiple higher-$\beta$ gluon emissions (generating a dipole cascade to the left of the shock wave) is also described by  $G_T^\textrm{adj} (\beta s)$ and $G_T^{i \, \textrm{adj}} (\beta s)$ dipole amplitudes. Owing to \eq{Gi_decomp}, this means that the interaction with the target is expressible in terms of $G_T^\textrm{adj} (\beta s)$ and $G_{2 T}^{\textrm{adj}} (\beta s)$. 

We now employ the following argument: based on the above observation, we can write the inclusive gluon production cross section with the the full higher-$\beta$ gluon cascade at large $N_c$ as 
\begin{align}\label{conv}
    \frac{d \sigma }{d^2 k_T \, dy}  =  C_1 (\alpha \, s) \otimes G_T^\textrm{adj} (\beta s) + C_2 (\alpha \, s) \otimes G_{2 T}^{\textrm{adj}} (\beta s),
\end{align}
with some unknown functions $C_1$ and $C_2$ resulting from re-summing the higher-$\beta$ cascade in the DLA. The convolution ($\otimes$) is over the dipole sizes which are not shown explicitly. If we now ``turn off" the evolution between the produced gluon and the target by replacing $G_T^\textrm{adj} (\beta s) \to G_T^{\textrm{adj} \, (0)} (\beta s)$ and $G_{2 T}^{\textrm{adj}} (\beta s) \to G_{2 T}^{\textrm{adj} \, (0)} (\beta s)$ in \eq{conv}, we would reduce the cross section to that for the scattering of an un-evolved target on the projectile, that is, to \eq{cross_sect_1} with the target and projectile interchanged. Following the above calculations, we can then rewrite this cross section as in \eq{kT_fact_final}, but with $G_T^{\textrm{adj} \, (0)}$ and $G_{2 T}^{\textrm{adj} \, (0)}$ instead of $G_T^{\textrm{adj}}$ and $G_{2 T}^{\textrm{adj}}$. At this point, we readily observe that we can simply reverse the earlier replacement of the dipole amplitudes, $G_T^{\textrm{adj} \, (0)} (\beta s) \to G_T^\textrm{adj} (\beta s)$ and $G_{2 T}^{\textrm{adj} \, (0)} (\beta s) \to G_{2 T}^{\textrm{adj}} (\beta s)$, reinstating the evolution on the target side, and arriving at \eq{kT_fact_final} with the DLA small-$x$ helicity evolution included both on the target and on the projectile side. 

We thus conclude that our main result \eqref{kT_fact_final} for the projectile and target polarization-dependent inclusive gluon production cross section is still valid when the small-$x$ DLA helicity evolution \cite{Kovchegov:2015pbl, Kovchegov:2016zex, Kovchegov:2018znm, Cougoulic:2022gbk} is included both on the target and on the projectile sides. This makes our result \eqref{kT_fact_final} more general, and, potentially, applicable to the hadron or jet production phenomenology in the central rapidity region of the polarized proton--proton collisions at RHIC \cite{STAR:2014wox, PHENIX:2015fxo, Aschenauer:2013woa, Aschenauer:2015eha}.


\section{Conclusions and outlook}
\label{sec:conclusions}

In this paper we have derived the inclusive gluon production cross section for the scattering of a longitudinally polarized projectile on a longitudinally polarized target. The cross section was calculated in the pure gluon sector, with the inclusion of quarks left for the future work. Our main result, written in a $k_T$-factorized form, is shown in \eq{kT_fact_final}. It employs the impact-parameter integrated polarized dipole amplitudes $G_T^{\textrm{adj}} (2 k^- p_1^+)$ and $G_{2 T}^{\textrm{adj}} (2 k^- p_1^+)$ for a gluon dipole scattering on the target and the dipole amplitudes $G_P^{\textrm{adj}} (2 k^+ p_2^-)$ and $G_{2 P}^{\textrm{adj}} (2 k^+ p_2^-)$ for the scattering on the projectile. These dipole amplitudes can be found by solving the DLA evolution equations from \cite{Kovchegov:2015pbl, Kovchegov:2016zex, Kovchegov:2018znm, Cougoulic:2022gbk}. We have, therefore, constructed a theoretical prediction for the small-$x$ gluon production in the central rapidity region of longitudinally polarized nucleon--nucleon collisions. Our result parallels that of \cite{Kovchegov:2001sc, Braun:2000bh} for the small-$x$ gluon production at mid-rapidity in unpolarized scattering.    

While our calculation was done for gluon production only, it may still be applicable to phenomenology of hadron and jet production at mid-rapidity in the polarized proton--proton collisions at RHIC \cite{STAR:2014wox, PHENIX:2015fxo, Aschenauer:2013woa, Aschenauer:2015eha}, where gluon contribution is known to dominate over the quark one. The data analysis can be done following the approach pioneered in \cite{Adamiak:2021ppq} for polarized DIS at small $x$, which has recently been extended to polarized SIDIS in \cite{Adamiak:2023okq}. Our cross section \eqref{kT_fact_final}, when convoluted with the appropriate fragmentation functions, would give the numerator of the double spin asymmetry $A_{LL}$ for hadron production measured in the polarized proton--proton collisions at RHIC.

In the future, our result needs to be expanded to include the contribution of quarks, both to inclusive gluon production and to inclusive quark production at central rapidity in polarized proton--proton collisions. This can be done along the lines of \cite{Chirilli:2021lif, Kovchegov:2015pbl, Kovchegov:2018znm, Cougoulic:2022gbk}. Such calculation would lead to a complete expression for the inclusive small-$x$ parton production cross section in longitudinally polarized nucleon--nucleon scattering. It would allow to fully complete the small-$x$ polarized data analysis program started in \cite{Adamiak:2021ppq, Adamiak:2023okq} and to make as precise a prediction as possible (based on the small-$x$ formalism) for the longitudinal spin observables to be measured at the EIC.


\section*{Acknowledgments}
The authors would like to thank Daniel Adamiak for his contributions at the early stages of this project. We are also grateful to Nicholas Baldonado, Daniel Pitonyak, and Matthew Sievert for helpful and encouraging discussions. 

This material is based upon work supported by
the U.S. Department of Energy, Office of Science, Office of Nuclear
Physics under Award Number DE-SC0004286 and within the framework of the Saturated Glue (SURGE) Topical Theory Collaboration. \\


\appendix


\section{Ultraviolet contribution to $(D+E) \, A^*_\textrm{eik} +$~c.c. interference terms}
\label{A}

The mismatch between the transverse positions of the incoming gluon to the left and to the right of the final state cut only affects the $(D+E) \, A^*_\textrm{eik} +$~c.c. contributions to the production cross section calculated in Sec.~\ref{sec:cross_section} in the main text, as follows from an analysis of the diagrams shown in \fig{fig:cross_section}. Indeed, in \fig{fig:cross_section}, only diagrams D and E come in with the sub-eikonal interaction, denoted by the white square, taking place when the gluon at position $\un b$ crosses the shock wave, potentially leading to the issue discussed around \eq{ba} above. 

\begin{figure}[h]
    \centering
\includegraphics[width= 0.99 \textwidth]{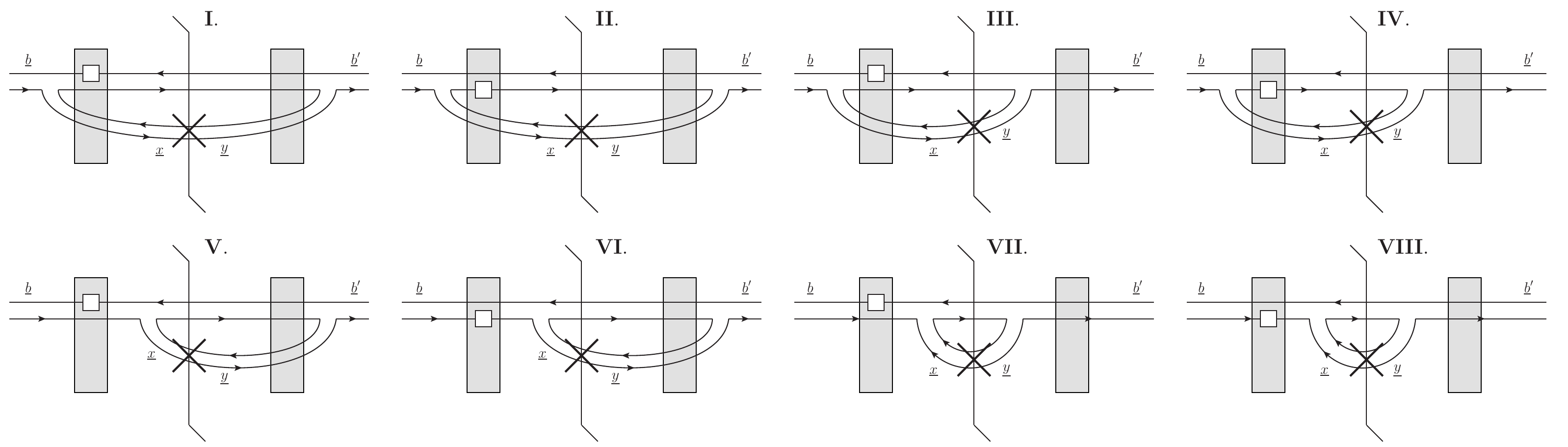}
    \caption{Large-$N_c$ diagrams contributing to $(D+E) \, A^*_\textrm{eik}$ interference terms. The shaded rectangle represents the shock wave, the white square denotes the sub-eikonal interactions, while the cross denotes the produced gluon. All double lines represent gluons at large-$N_c$ \cite{tHooft:1973alw}. The complex conjugate terms are not shown explicitly.}
    \label{fig:UV_divergence}
\end{figure}

Here we analyse the $(D+E) \, A^*_\textrm{eik}$ diagrams a little more carefully. Our discussion employs the large-$N_c$ limit, which is exact for the pure-glue diagrams of \fig{fig:cross_section}, as can be readily checked. The $(D+E) \, A^*_\textrm{eik}$ diagrams from \fig{fig:cross_section} are depicted again in \fig{fig:UV_divergence}, utilizing the large-$N_c$ limit (with the gluons shown as double lines). The top row of diagrams in \fig{fig:UV_divergence} corresponds to the $D \, A^*_\textrm{eik}$ interference, while the bottom row corresponds to $E \, A^*_\textrm{eik}$ contributions. 

Working in the large-$N_c$ limit, we immediately observe that diagrams IV and VI (and their complex conjugates) in \fig{fig:UV_divergence} do not contain a color-dipole or quadrupole involving both lines at $\un b$ and ${\un b}'$. These contributions are included in equations \eqref{DAeik} and \eqref{EAeik} and have no potential new terms arising from the ${\un b}' \to {\un b}$ limit. Therefore, we do not need to consider those further. 

Moreover, combining diagrams I, III, V, and VII, along with their complex conjugates, we notice that their contribution is proportional to the unpolarized gluon production cross section (see \cite{Kovchegov:1998bi, Kovchegov:2012mbw}), such that
\begin{align}\label{S4}
    \textrm{I} + \textrm{III} + \textrm{V} + \textrm{VII} \propto \left\langle \tr \left[ V_{\un b}^{\textrm{G} [1] \, \dagger} \, V_{{\un b}'} \right]  \right\rangle \, \left[ S_{{\un x}, {\un y}} - (S_{{\un x}, {\un b}})^2 - (S_{{\un b}, {\un y}})^2 + 1 \right]. 
\end{align}
Here $S_{{\un x}, {\un y}}$ is the unpolarized dipole $S$-matrix \cite{Balitsky:1995ub, Kovchegov:1999yj} and we put ${\un b}' = {\un b}$ in the second square brackets, since this limit is regular for the unpolarized dipole $S$-matrices. With the DLA accuracy of our calculation, all $S=1$ in \eq{S4}, giving
\begin{align}\label{40}
    \textrm{I} + \textrm{III} + \textrm{V} + \textrm{VII} =0. 
\end{align}
We can thus discard diagrams I, III, V, and VII, along with their complex conjugates. It appears that those diagrams would contribute if one includes single-logarithmic corrections (that is, powers of $\as \, \ln (1/x)$) into the gluon production calculation. However, while some of these single-logarithmic corrections to small-$x$ helicity evolution were found in \cite{Kovchegov:2021lvz}, it appears that for a complete calculation of small-$x$ helicity evolution at the single logarithmic order one would need to go beyond the shock-wave picture we are employing here. 

We are left with the diagrams II and VIII from \fig{fig:UV_divergence}. While the diagram II contains a polarized color-quadrupole amplitude made out of the ${\un x}, {\un y}, {\un b}, {\un b}'$ lines, the part of diagram II giving new terms in the ${\un b}' \to {\un b}$ limit is obtained when the ${\un x}$ and ${\un y}$ lines do not interact and the quadrupole amplitude reduces to the ${\un b}, {\un b}'$ polarized dipole. This part of diagram II is identical to diagram VII. Therefore, in the DLA, we conclude that the ${\un b}' \to {\un b}$ limit of the diagrams in \fig{fig:UV_divergence}, not accounted for in Eqs.~\eqref{DAeik} and \eqref{EAeik}, is given by the sum of diagrams VII and VIII (and their complex conjugates). This is a part of the $E \, A^*_\textrm{eik}$ interference, which follows from \eq{EAeik} and is given by  
\begin{align}\label{EAeik2}
& - \frac{\lambda}{N_c^2 - 1} \, \frac{4 \, \as}{(2 \pi)^4}  \int d^2 x \, d^2 y \, d^2 b \, e^{- i {\un k} \cdot ({\un x} - {\un y})} \, \frac{{\un x} - {\un b}}{|{\un x} - {\un b}|^2} \, \cdot \, \frac{{\un y} - {\un b}}{|{\un y} - {\un b}|^2} \left\langle \Tr \left[ U_{{\un b}'}^\dagger \, T^a \, U_{{\un b}}^{\textrm{G} [1]} \, T^{c} \right]  \, \left( -  U_{{\un b}'} \right)^{ac} \right\rangle + \mbox{c.c.} \notag \\
& = \frac{\lambda}{N_c^2 - 1} \, \frac{4 \, \as}{(2 \pi)^4}  \int d^2 x \, d^2 y \, d^2 b \, e^{- i {\un k} \cdot ({\un x} - {\un y})} \, \frac{{\un x} - {\un b}}{|{\un x} - {\un b}|^2} \, \cdot \, \frac{{\un y} - {\un b}}{|{\un y} - {\un b}|^2} \left\langle \Tr \left[ T^a \, U_{{\un b}}^{\textrm{G} [1]} \, U_{{\un b}'}^\dagger \, T^{a} \right] \right\rangle + \mbox{c.c.} \notag  \\
& = -  \lambda \, \frac{\as}{4 \pi^4} \frac{1}{s} \, N_c \,   \int d^2 x \, d^2 y \, d^2 b \, e^{- i {\un k} \cdot ({\un x} - {\un y})} \, \frac{{\un x} - {\un b}}{|{\un x} - {\un b}|^2} \, \cdot \, \frac{{\un y} - {\un b}}{|{\un y} - {\un b}|^2} \, \left[ - 2 \, G^\textrm{adj}_{{\un b}, {\un b}'} (2 k^- p_1^+) \right] .
\end{align}
This is exactly the result in \eq{UVterm} of the main text.


\section{Small-$x$ limit of twist-3 gluon helicity-flip TMD}
\label{sec:small-x_TMDs}

The general gauge-invariant two-gluon correlation functions are defined as \cite{Mulders:2000sh}
\begin{equation}\label{Gamma_corr}
\Gamma^{\mu\nu; \rho\sigma}(k, P, S)= \int \frac{d^4\xi}{(2\pi)^4} e^{ik\cdot\xi} \, \bra{P, S} \mathrm{Tr}\left[F^{\mu\nu}(0) \, \mathcal{U}^{[+]}(0, \xi) \, F^{\rho\sigma}(\xi) \, \mathcal{U}^{[-]}(\xi, 0)\right] \ket{P, S}  .
\end{equation}
We choose the future- ($\mathcal{U}^{[+]}$) and past-pointing ($\mathcal{U}^{[-]}$) fundamental Wilson line staples relevant to the gluon production process under consideration. This way, the TMDs one can construct out of the correlator \eqref{Gamma_corr} will be the  dipole TMDs. Here $\ket{P, S}$ is the proton state with momentum $P$ and spin $S$.

Among all the possible values of $\mu, \nu, \rho, \sigma$, the leading twist-2 gluon correlation function
\begin{equation}\label{app:twist-2}
 \Gamma^{+i; +j}(k, P, S)= \int \frac{d^4\xi}{(2\pi)^4} e^{ik\cdot\xi} \bra{P, S} \mathrm{Tr}\left[F^{+i}(0) \, \mathcal{U}^{[+]}(0, \xi) \, F^{+j}(\xi) \, \mathcal{U}^{[-]}(\xi, 0)\right] \ket{P, S}
\end{equation}
is related to the gluon helicity TMD for a longitudinally polarized proton and if projected onto $\epsilon^{ij}$. For longitudinally polarized proton states, the small $x$ limit of \eq{app:twist-2} (projected onto $\epsilon^{ij}$) was derived in \cite{Cougoulic:2022gbk}. 

In this Appendix, we derive the small $x$ limit of a twist-3 gluon helicity-flip TMD, which results from the correlator
\begin{equation}\label{Gamma_L}
\Gamma^{ij; l+} (k; P, S_L) = \int \frac{d^4\xi}{(2\pi)^4}e^{ik\cdot\xi} \bra{P, S_L} \mathrm{Tr}\left[F^{ij}(0) \, \mathcal{U}^{[+]}(0, \xi) \, F^{l+}(\xi) \, \mathcal{U}^{[-]}(\xi, 0)\right] \ket{P, S_L} 
\end{equation}
with the longitudinally polarized proton state having spin $S_L$. Further, following \cite{Mulders:2000sh}, we define
\begin{align}\label{MGamma}
    M \, \Gamma_L^{ij, l}(x, {\un k}) = \int dk^- \, \Gamma^{ij; l+}(k; P, S_L) 
\end{align}
with $k^+ = x P^+$. Here $M$ is the mass of the proton. 

For a longitudinally polarized proton state, the twist-3 gluon helicity-flip TMD $\Delta H_{3L}^{\perp}(x, k_T^2)$ is defined by
\begin{align}\label{Hdef}
    M \, \Gamma_L^{ij, l}(x, {\un k}) = - \frac{i}{4} S_L \, \epsilon^{ij} \, k^l \, \Delta H_{3L}^{\perp}(x, k_T^2). 
\end{align}
Note that we do not include the additional factor $x$ on the right of \eq{Hdef} compared to that in \cite{Mulders:2000sh} and we use the normalization in \cite{Cougoulic:2022gbk} which differs from that in \cite{Mulders:2000sh} by a factor of $(-1/2)$ on the right of \eqref{Hdef}: overall, our \eq{Hdef} differs from Eq.~(22) in \cite{Mulders:2000sh} by an additional factor of $-1/(2 x)$ on its right-hand side. Removing the factor of $x$ gives the resulting TMD $\Delta H_{3L}^{\perp}$ the right ``eikonality": as will become apparent shortly, this is a sub-eikonal quantity, and should scale as $(1/x)^0$ at small $x$ before the evolution corrections are included.  

Inverting \eq{Hdef} we get
\begin{align}
\Delta H_{3L}^{\perp}(x, k_T^2)  = 2i\epsilon^{ij}\frac{k^l}{{\un k}^2}\frac{1}{2}\sum_{S_L} S_L M \, \Gamma_L^{ij;l}(x, {\un k}).
\end{align}

Using the definition \eqref{Gamma_L}, one can show that 
\begin{align}\label{GG*}
    \Gamma^{l+; ij}(k; P, S_L) = \Gamma^{ij; l+ \, *}(k; P, S_L).
\end{align}
Since the definition \eqref{Hdef} implies that the TMD is real, and, hence, $\Gamma_L^{ij, l}$ is imaginary, then so is $\Gamma^{ij; l+}$. Therefore, \eq{GG*} implies that
\begin{align}\label{GG-}
    \Gamma^{l+; ij}(k; P, S_L) = - \Gamma^{ij; l+}(k; P, S_L)
\end{align}
allowing us to rewrite \eq{MGamma} as 
\begin{equation}\label{G-G}
M \, \Gamma_L^{ij, l}(x, {\un k}) = \int dk^-\frac{1}{2} \left[ \Gamma^{ij; l+}(k; P, S_L) - \Gamma^{l+; ij}(k; P, S_L)\right]. 
\end{equation}

Combining Eqs.~\eqref{Hdef} and \eqref{G-G} we obtain 
\begin{equation}\label{H1}
\begin{split}
&\Delta H_{3L}^{\perp}(x, k_T^2)  
= \frac{2i}{V^-(2\pi)^3} \epsilon^{ij}\frac{k^l}{{\un k}^2} \sum_{S_L} \frac{1}{2} S_L \int d\xi^- d^2 \xi  \, d \zeta^- d^2 \zeta \,  e^{ik^+(\xi^--\zeta^-)}e^{ -i{\un k} \cdot({\un \xi}-{\un \zeta})}\\
&\qquad \times \frac{1}{2} \Bigg[ \bra{P, S_L} \mathrm{Tr}\left[F^{ij}(\zeta) \mathcal{U}^{[+]}(\zeta, \xi)F^{l+}(\xi) \mathcal{U}^{[-]}(\xi, \zeta)\right] \ket{P, S_L} |_{\xi^+=\zeta^+=0}\\
&\qquad \qquad - \bra{P, S_L} \mathrm{Tr}\left[F^{l+}(\zeta) \mathcal{U}^{[+]}(\zeta, \xi)F^{ij}(\xi) \mathcal{U}^{[-]}(\xi, \zeta)\right] \ket{P, S_L} |_{\xi^+=\zeta^+=0} \Bigg] .\\
\end{split}
\end{equation}
Here $V^- = \int dx^- d^2 x$.

To obtain the small-$x$ limit of  $\Delta H_{3L}^{\perp}(x, k_T^2) $, we first note that in any gauge where $A_\perp \to 0$ as $x^- \to \pm \infty$, we can write the gauge links as 
\begin{align}\label{UUdef}
   \mathcal{U}^{[+]}(\zeta, \xi) = V_{\un \zeta} [\zeta^-, \infty] \, V_{\un \xi} [\infty, \xi^-], \ \ \  \mathcal{U}^{[-]}(\xi, \zeta) = V_{\un \xi} [\xi^-, -\infty] \, V_{\un \zeta} [-\infty, \zeta^-].
\end{align}
We then can approximate (cf. Eq.~(22) in \cite{Cougoulic:2022gbk})
\begin{equation}\label{Weik}
\begin{split}
&\frac{k^l}{{\un k}^2}\int d^2  \xi \,  e^{-i{\un k}\cdot{\un \xi}}\int_{-\infty}^{\infty}d\xi^-  \, e^{ixP^+ \xi^-} \,  V_{{\un \xi}} [\infty, \xi^-] \, F^{l+}(\xi^-, {\un \xi}) \, V_{{\un \xi}} [\xi^-, -\infty]\\ 
=&\frac{k^l}{{\un k}^2}\int d^2 \xi e^{-i{\un k}\cdot{\un \xi}}\int_{-\infty}^{\infty}d\xi^- \, e^{ixP^+ \xi^-} \,  V_{{\un \xi}}[\infty, \xi^-] \, \left[ \partial^l A^+(\xi^-, {\un \xi}) + i x P^+ A^l (\xi^-, {\un \xi}) \right] \, V_{{\un \xi}}[\xi^-, -\infty] \\ 
\approx &\frac{1}{ig}\frac{k^l}{{\un k}^2}\int d^2 \xi \, e^{-i{\un k}\cdot{\un \xi}}\, \partial^l V_{{\un \xi}} 
= -\frac{1}{g}  \int d^2 \xi \, e^{-i{\un k}\cdot{\un \xi}}\, V_{{\un \xi}} .
\end{split}
\end{equation}
In the last line of the above derivation we have neglected terms suppressed by powers of $x \ll 1$.

On the other hand, one gets, again by expanding in $x$,
\begin{equation}\label{VG11}
\begin{split}
&\epsilon^{ij} \int d\zeta^- \, e^{- ixP^+ \zeta^-} \, V_{{\un \zeta}} [ -\infty, \zeta^-] \, F^{ij}(\zeta^-, {\un \zeta}) \, V_{{\un \zeta}}[\zeta^-, \infty]\\
\approx & \ 2\int d\zeta^- V_{{\un \zeta}}[-\infty, \zeta^-] \, F^{12}(\zeta^-, {\un \zeta}) \, V_{{\un \zeta}} [\zeta^-, \infty] \\
=&\frac{4k^-}{-ig} V^{\rm{G}[1]\dagger}_{{\un \zeta}},
\end{split}
\end{equation}
where (cf. \eq{UG1}) \cite{Cougoulic:2022gbk}
\begin{align}
    V_{\un x}^{\textrm{G} [1]}  = \frac{i \, g \, P^+}{s} \int\limits_{-\infty}^{\infty} d{x}^- V_{\un{x}} [ \infty, x^-] \, F^{12} (x^-, {\un x}) \, \, V_{\un{x}} [ x^-, -\infty]  \label{VG1} .
\end{align}

Employing Eqs.~\eqref{UUdef}, \eqref{Weik}, and \eqref{VG11} in \eq{H1}, and remembering that 
\begin{align}
\Big\langle \ldots \Big\rangle \equiv \half\sum_{S_L} S_L \, \frac{1}{2 P^+ V^-} \, \bra{P, S_L} \ldots \ket{P, S_L}
\end{align}
and $\llangle \ldots \rrangle = s \, \left\langle \ldots \right\rangle$, we obtain the final expression for the small $x$ limit of $\Delta H^{\perp}_{3L}(x, k_T^2)$, 
\begin{equation}\label{H3}
\begin{split}
\Delta H_{3L}^{\perp}(x, k_T^2) 
=&\frac{8}{g^2(2\pi)^3} \int d^2 \xi d^2 \zeta \, e^{-i{\un k}\cdot({\un \xi}-{\un \zeta})} \, \frac{1}{2}\llangle \tr \left[V_{{\un \xi}} \, V^{\rm{G}[1]\dagger}_{{\un \zeta}} + V^{\dagger}_{{\un \zeta}} \, V^{G[1]}_{{\un \xi}}\right]\rrangle \\
=&\frac{N_c}{\alpha_s 4\pi^4} \int d^2 x \, e^{-i{\un k}\cdot{\un x}}\, G \left(x^2_{\perp}, \frac{Q^2}{x} \right),  \\
\end{split}
\end{equation}
where 
\begin{align}
    G (x^2_{\perp}, s) = \int d^2 b \, G_{{\un b} + {\un x}, {\un b}} (s)
\end{align}
with
\begin{align}\label{Ggdef}
 G_{{\un x}, {\un y}} (s) \equiv \frac{1}{2 \, N_c} \, \mbox{Re} \, \llangle \mbox{T} \, \tr \left[ V_{\un x} \,  V_{\un y}^{\textrm{G} [1] \,\dagger} \right] + \mbox{T} \,  \tr \left[ V_{\un y}^{\textrm{G} [1]} \, V_{\un x}^\dagger \right]   \rrangle (s) .
\end{align}
Since, at large $N_c$, $G^\textrm{adj} = 4 G$, \eq{H3} is equivalent to \eq{helicity_flip_TMD} in the main text.  

To conclude, the small-$x$ limit of the twist-3 gluon helicity-flip TMD is given by the dipole amplitude containing the chromo-magnetically polarized Wilson line $V^{G[1]}$.


\section{Gluon production in an adjoint dipole scattering on a polarized target}
\label{sec:dipole}

In this Appendix we show that gluon production cross section for the scattering on a longitudinally polarized target is expressible in terms of the $G_T^{\textrm{adj}}$ and $G_{2 T}^{\textrm{adj}}$ polarized dipole amplitudes for the incoming projectile being either an ${\cal F}^{12}$- or $\left( \mathscr{D}^i - \cev{\mathscr{D}}^i \right)$-type dipole.  


\subsection{${\cal F}^{12}$-type dipole projectile}

\begin{figure}[h]
    \centering
\includegraphics[width= \textwidth]{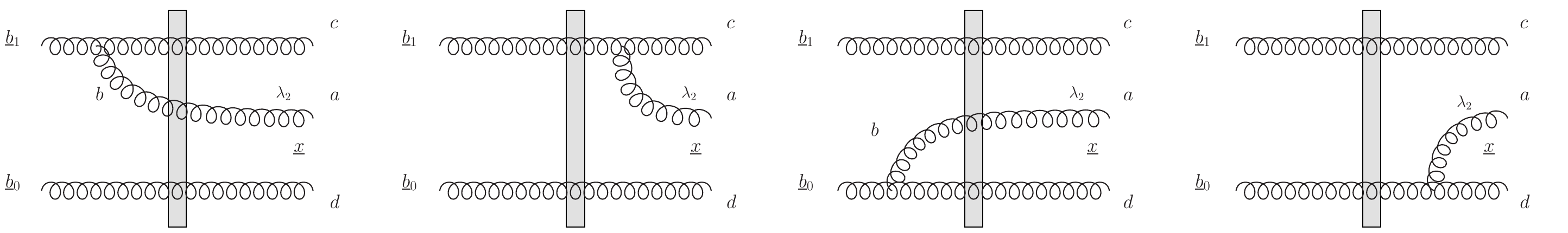}
    \caption{Diagrams contributing to eikonal inclusive gluon production for the gluon dipole projectile.}
    \label{fig:eikonal_dipole}
\end{figure}

We start with the eikonal gluon production for the dipole projectile, as shown in \fig{fig:eikonal_dipole}. 
The corresponding amplitude is 
\begin{align}
A_\textrm{eik}^\textrm{LO} ({\un x}, {\un b}) = \frac{g}{\pi} \, \delta_{\lambda, \lambda'} \,  \left( U_{{\un b}_1} \, T^b \, U_{{\un b}_0}^\dagger \right)^{cd}  \, \left\{ \frac{{\un \epsilon}_{\lambda_2}^* \cdot ({\un x} - {\un b}_1)}{|{\un x} - {\un b}_1|^2} \, \left[ \left( U_{\un x} \right)^{ab} - \left( U_{{\un b}_1} \right)^{ab} \right] - \frac{{\un \epsilon}_{\lambda_2}^* \cdot ({\un x} - {\un b}_0)}{|{\un x} - {\un b}_0|^2} \, \left[ \left( U_{\un x} \right)^{ab} - \left( U_{{\un b}_0} \right)^{ab} \right] \right\}.
\end{align}

\begin{figure}[bh]
    \centering
\includegraphics[width= \textwidth]{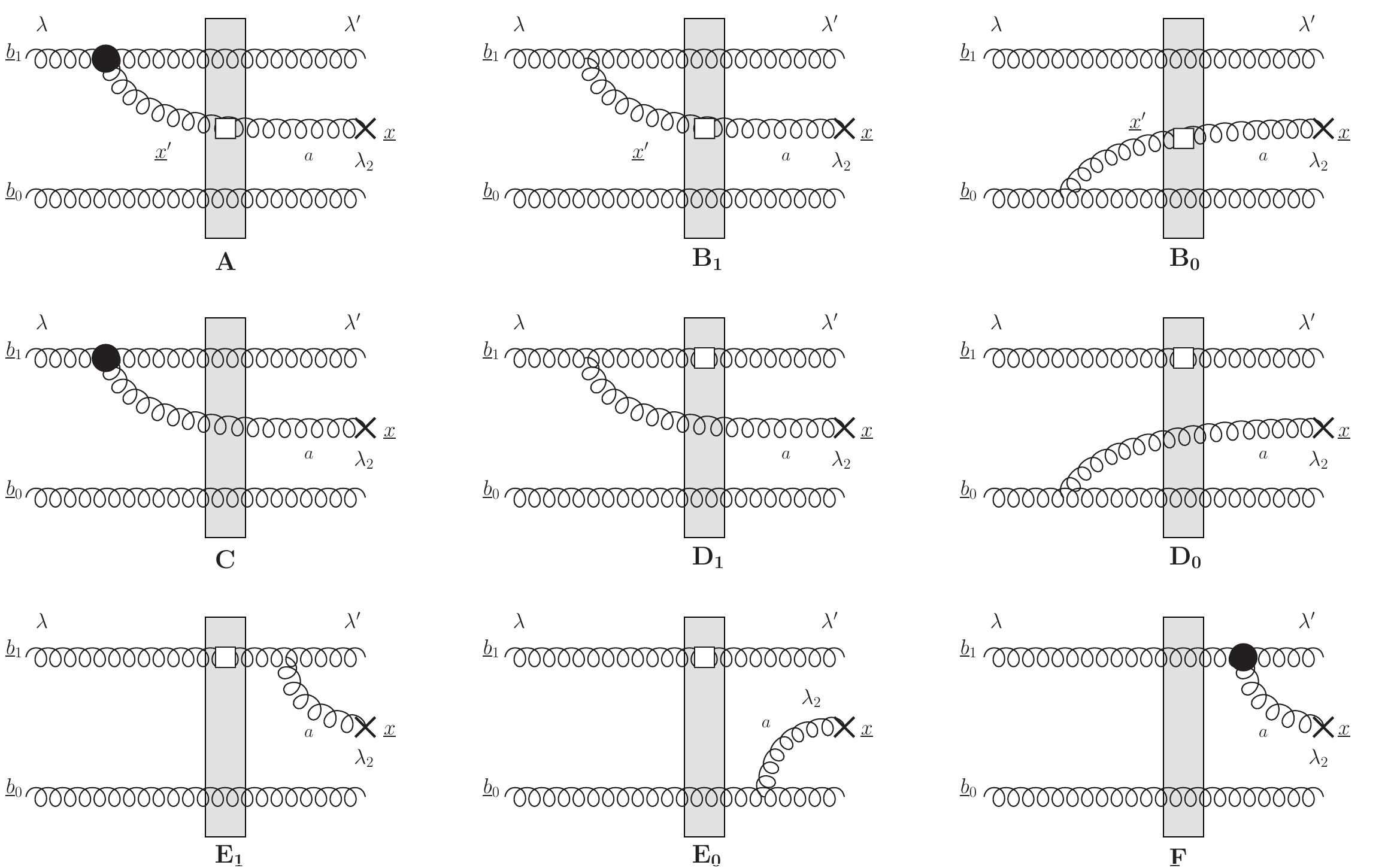}
    \caption{Diagrams contributing to inclusive sub-eikonal gluon production for the gluon dipole projectile.}
    \label{fig:F12_dipole}
\end{figure}

Next, let us consider inclusive gluon production amplitudes for a projectile being a gluon dipole. If we choose the polarized gluon line to be at ${\un b}_1$ in the transverse plane, we would need to generalize diagrams B, D and E to include emission from the eikonal gluon line at ${\un b}_0$, as shown in \fig{fig:F12_dipole}. As before, the black circle denotes the insertion of the ${\cal F}^{12}$ operator from \eq{UG1}, while the white boxes in the diagrams $D_1, D_0, E_1$ and $E_0$ denote $U^{\textrm{G} [1]}$. The white boxes in $B_1$ and $B_0$ denote the entire sub-eikonal polarized Wilson line (with the gluon background fields only).

Calculating the diagrams in \fig{fig:F12_dipole} in a similar way to the calculation in the main text, we can write
\begin{align}\label{AAeik_dip}
& \frac{d \sigma^{A \, A^*_\textrm{eik} + \mbox{c.c.}} (\lambda)}{d^2 k_T \, dy} = \lambda \, \frac{4 \, \as}{(2 \pi)^4}  \frac{k^-}{p_2^-} \, \frac{N_c}{N_c^2 - 1} \, \int d^2 x \, d^2 y \, d^2 b_1 \, e^{- i {\un k} \cdot ({\un x} - {\un y})} \, \sum_{i=0}^1 (-1)^{i+1}  \left\{  \frac{{\un x} - {\un b}_1}{|{\un x} - {\un b}_1|^2} \, \cdot \, \frac{{\un y} - {\un b}_i}{|{\un y} - {\un b}_i|^2} \right. \notag \\ 
& \left. \times  \left\langle  \Tr \left[ U_{\un y}^\dagger \, U_{{\un x}}^{\textrm{G} [1]} \right] - \Tr \left[ U_{{\un b}_i}^\dagger \, U_{{\un x}}^{\textrm{G} [1]} \right]  \right\rangle  
+ \, i \int d^2 x' \, \frac{{\un x}'- {\un b}_1}{|{\un x}' - {\un b}_1|^2}  \, \times \, \frac{{\un y} - {\un b}_i}{|{\un y} - {\un b}_i|^2}   \left\langle  \Tr \left[ U_{\un y}^\dagger \, U_{{\un x}, {\un x}'}^{\textrm{G} [2]} \right] - \Tr \left[ U_{{\un b}_i}^\dagger \, U_{{\un x}, {\un x}'}^{\textrm{G} [2]} \right]  \right\rangle \right\} + \mbox{c.c.} \notag \\
& =  \lambda \, \frac{\as}{2 \pi^4}  \frac{1}{s} \, N_c \, \int d^2 x \, d^2 y \, d^2 b_1 \, e^{- i {\un k} \cdot ({\un x} - {\un y})} \, \sum_{i=0}^1 (-1)^{i+1} \left\{  \frac{{\un x} - {\un b}_1}{|{\un x} - {\un b}_1|^2} \, \cdot \, \frac{{\un y} - {\un b}_i}{|{\un y} - {\un b}_i|^2} \left[ G^\textrm{adj}_{{\un x}, {\un y}} (2 k^- p_1^+) - G^\textrm{adj}_{{\un x}, {\un b}_i} (2 k^- p_1^+) \right] \right. \notag \\ 
& \hspace*{3cm}\left. - 2 i \, k^j \, \frac{{\un x} - {\un b}_1}{|{\un x} - {\un b}_1|^2}  \, \times \, \frac{{\un y} - {\un b}_i}{|{\un y} - {\un b}_i|^2}  \, G^{j \, \textrm{adj}}_{{\un x}, {\un b}_i} (2 k^- p_1^+)   \right\}  .
\end{align}
Note that when integrating $d^2 b_1$ we keep ${\un b}_{10}$ fixed (and, hence, ${\un b}_0$ varies). Calculating the diagrams $B_1$ and $B_0$ from \fig{fig:F12_dipole} we find their contribution to the cross section to be 
\begin{align}\label{BCF_dip}
& \frac{d \sigma^{(B_1 + B_0) \, (C^* + F^*) + \mbox{c.c.}} (\lambda)}{d^2 k_T \, dy} =  \lambda \, \frac{4 \, \as}{(2 \pi)^4}  \frac{k^-}{p_2^-} \, \frac{N_c}{N_c^2 - 1} \, \int d^2 x \, d^2 y \, d^2 b_1 \, e^{- i {\un k} \cdot ({\un x} - {\un y})} \, \sum_{i=0}^1 (-1)^{i+1}  \, \left\{  \frac{{\un x} - {\un b}_i}{|{\un x} - {\un b}_i|^2} \, \cdot \, \frac{{\un y} - {\un b}_1}{|{\un y} - {\un b}_1|^2}  \right. \notag \\ 
& \times \left.  \left\langle  \Tr \left[ U_{\un y}^\dagger \, U_{{\un x}}^{\textrm{G} [1]} \right] - \Tr \left[ U_{{\un b}_1}^\dagger \, U_{{\un x}}^{\textrm{G} [1]} \right]  \right\rangle + \, i \int d^2 x' \, \frac{{\un x}'- {\un b}_i}{|{\un x}' - {\un b}_i|^2}  \, \times \, \frac{{\un y} - {\un b}_1}{|{\un y} - {\un b}_1|^2}   \left\langle  \Tr \left[ U_{\un y}^\dagger \, U_{{\un x}, {\un x}'}^{\textrm{G} [2]} \right] - \Tr \left[ U_{{\un b}_1}^\dagger \, U_{{\un x}, {\un x}'}^{\textrm{G} [2]} \right]  \right\rangle \right\} + \mbox{c.c.} \notag \\
& =  \lambda \, \frac{\as}{2 \pi^4}  \frac{1}{s} \, N_c \, \int d^2 x \, d^2 y \, d^2 b_1 \, e^{- i {\un k} \cdot ({\un x} - {\un y})} \, \sum_{i=0}^1 (-1)^{i+1} \left\{  \frac{{\un x} - {\un b}_i}{|{\un x} - {\un b}_i|^2} \, \cdot \, \frac{{\un y} - {\un b}_1}{|{\un y} - {\un b}_1|^2} \left[ G^\textrm{adj}_{{\un x}, {\un y}} (2 k^- p_1^+) - G^\textrm{adj}_{{\un x}, {\un b}_1} (2 k^- p_1^+) \right] \right. \notag \\ 
& \hspace*{3cm}\left. - 2 i \, k^j \, \frac{{\un x} - {\un b}_i}{|{\un x} - {\un b}_i|^2}  \, \times \, \frac{{\un y} - {\un b}_1}{|{\un y} - {\un b}_1|^2}  \, G^{j \, \textrm{adj}}_{{\un x}, {\un b}_1} (2 k^- p_1^+)   \right\}  .
\end{align}

Diagrams $D_1$ and $D_0$ give
\begin{align}\label{DAeik_dip}
& \frac{d \sigma^{(D_1 + D_0) \, A^*_\textrm{eik} + \mbox{c.c.}} (\lambda)}{d^2 k_T \, dy} = \frac{\lambda}{N_c^2 - 1} \, \frac{4 \, \as}{(2 \pi)^4}  \int d^2 x \, d^2 y \, d^2 b_1 \, e^{- i {\un k} \cdot ({\un x} - {\un y})} \, \sum_{i,j = 0}^1 \, (-1)^{i+j} \, \frac{{\un x} - {\un b}_i}{|{\un x} - {\un b}_i|^2} \, \cdot \, \frac{{\un y} - {\un b}_j}{|{\un y} - {\un b}_j|^2} \notag \\ 
& \times \left\langle \Tr \left[ U_{{\un b}_1}^\dagger \, U_{{\un b}_1}^{\textrm{G} [1]} \, T^a \, U_{\un x}^\dagger \,  \left( U_{\un y} -  U_{{\un b}_j} \right)  \, T^a \right]  \right\rangle + \mbox{c.c.} \notag \\ 
& = - \frac{\lambda}{N_c^2 - 1} \, \frac{4 \, \as}{(2 \pi)^4}  \int d^2 x \, d^2 y \, d^2 b_1 \, e^{- i {\un k} \cdot ({\un x} - {\un y})} \, \sum_{i,j = 0}^1 \, (-1)^{i+j} \, \frac{{\un x} - {\un b}_i}{|{\un x} - {\un b}_i|^2} \, \cdot \, \frac{{\un y} - {\un b}_j}{|{\un y} - {\un b}_j|^2} \notag \\ 
& \times \left\langle \Tr \left[ U_{{\un b}_1}^\dagger \, U_{{\un b}_1}^{\textrm{G} [1]} \, T^a \, U_{\un x}^\dagger \, U_{{\un b}_j}   \, T^a \right]  \right\rangle + \mbox{c.c.} \notag \\ 
& = - \frac{\lambda}{N_c^2 - 1} \, \frac{4 \, \as}{(2 \pi)^4}  \int d^2 x \, d^2 y \, d^2 b_1 \, e^{- i {\un k} \cdot ({\un x} - {\un y})} \, \left( \frac{{\un x} - {\un b}_1}{|{\un x} - {\un b}_1|^2} - \frac{{\un x} - {\un b}_0}{|{\un x} - {\un b}_0|^2} \right) \, \cdot \, \frac{{\un y} - {\un b}_1}{|{\un y} - {\un b}_1|^2} \notag \\ 
& \times \left\langle \Tr \left[ U_{{\un b}_1}^\dagger \, U_{{\un b}_1}^{\textrm{G} [1]} \, T^a \, U_{\un x}^\dagger \, U_{{\un b}_1}   \, T^a \right]  \right\rangle + \mbox{c.c.} \notag \\ 
& = -  \lambda \, \frac{\as}{4 \pi^4} \frac{1}{s} \, N_c \,   \int d^2 x \, d^2 y \, d^2 b_1 \, e^{- i {\un k} \cdot ({\un x} - {\un y})} \, \left( \frac{{\un x} - {\un b}_1}{|{\un x} - {\un b}_1|^2} - \frac{{\un x} - {\un b}_0}{|{\un x} - {\un b}_0|^2} \right) \, \cdot \, \frac{{\un y} - {\un b}_1}{|{\un y} - {\un b}_1|^2} \, G^\textrm{adj}_{{\un b}_1, {\un x}} (2 k^- p_1^+) ,
\end{align}
where we have employed cancellations of some of the contributions with parts of $(E_1 + E_0) \, A^*_\textrm{eik} + \mbox{c.c.}$, which, in turn, is
\begin{align}\label{EAeik_dip}
& \frac{d \sigma^{(E_1 + E_0) \, A^*_\textrm{eik} + \mbox{c.c.}} (\lambda)}{d^2 k_T \, dy} = - \frac{\lambda}{N_c^2 - 1} \, \frac{4 \, \as}{(2 \pi)^4}  \int d^2 x \, d^2 y \, d^2 b_1 \, e^{- i {\un k} \cdot ({\un x} - {\un y})} \, \left\langle \left[ (T^a \, U_{{\un b}_1}^{\textrm{G} [1]} \, U_{{\un b}_0}^\dagger )^{cd} \,  \frac{{\un x} - {\un b}_1}{|{\un x} - {\un b}_1|^2} - (U_{{\un b}_1}^{\textrm{G} [1]} \, U_{{\un b}_0}^\dagger \, T^a  )^{cd} \right. \right. \notag \\
& \left. \left. \times \,  \frac{{\un x} - {\un b}_0}{|{\un x} - {\un b}_0|^2} \right] \cdot \left[  \left[ \left( U_{\un y} \right)^{ac} - \left( U_{{\un b}_1} \right)^{ac} \right] \, \frac{{\un y} - {\un b}_1}{|{\un y} - {\un b}_1|^2} -  \left[ \left( U_{\un y} \right)^{ac} - \left( U_{{\un b}_0} \right)^{ac} \right] \, \frac{{\un y} - {\un b}_0}{|{\un y} - {\un b}_0|^2} \right] \, (U_{{\un b}_1} \, T^{*c} \, U_{{\un b}_0}^\dagger )^{cd} \right\rangle + \mbox{c.c.}  \notag \\ 
& = - \frac{\lambda \, N_c}{N_c^2 - 1} \, \frac{2 \, \as}{(2 \pi)^4}  \int d^2 x \, d^2 y \, d^2 b_1 \, e^{- i {\un k} \cdot ({\un x} - {\un y})} \, \left\{ \frac{{\un x} - {\un b}_1}{|{\un x} - {\un b}_1|^2} \cdot \left( \frac{{\un y} - {\un b}_1}{|{\un y} - {\un b}_1|^2} - \frac{{\un y} - {\un b}_0}{|{\un y} - {\un b}_0|^2} \right) \, \left\langle \Tr \left[ U^\dagger_{\un y} \, U_{{\un b}_1}^{\textrm{G} [1]} \right] \right\rangle \right. \notag \\ 
& \left. + \left( \frac{{\un x} - {\un b}_1}{|{\un x} - {\un b}_1|^2} \cdot \frac{{\un y} - {\un b}_0}{|{\un y} - {\un b}_0|^2} + \frac{{\un x} - {\un b}_0}{|{\un x} - {\un b}_0|^2} \cdot \frac{{\un y} - {\un b}_1}{|{\un y} - {\un b}_1|^2} \right) \, \left\langle \Tr \left[ U^\dagger_{{\un b}_0} \, U_{{\un b}_1}^{\textrm{G} [1]} \right] \right\rangle \right\} + \mbox{c.c.} \notag \\
& = -  \lambda \, \frac{\as}{4 \pi^4} \frac{1}{s} \, N_c \,   \int d^2 x \, d^2 y \, d^2 b_1 \, e^{- i {\un k} \cdot ({\un x} - {\un y})} \, \left\{ \frac{{\un x} - {\un b}_1}{|{\un x} - {\un b}_1|^2} \cdot \left( \frac{{\un y} - {\un b}_1}{|{\un y} - {\un b}_1|^2} - \frac{{\un y} - {\un b}_0}{|{\un y} - {\un b}_0|^2} \right) \, G^\textrm{adj}_{{\un b}_1, {\un y}} (2 k^- p_1^+) \right. \notag \\
& \left. + \left( \frac{{\un x} - {\un b}_1}{|{\un x} - {\un b}_1|^2} \cdot \frac{{\un y} - {\un b}_0}{|{\un y} - {\un b}_0|^2} + \frac{{\un x} - {\un b}_0}{|{\un x} - {\un b}_0|^2} \cdot \frac{{\un y} - {\un b}_1}{|{\un y} - {\un b}_1|^2} \right) \, G^\textrm{adj}_{{\un b}_1, {\un b}_0} (2 k^- p_1^+) \right\} .
\end{align}

Assembling all the above contributions together and adding \eq{EAeik2} with ${\un b} \to {\un b}_1$ to account for the UV-divergent terms calculated in Appendix~\ref{A} (such that $|{\un b}'_1 - {\un b}_1|^2 = 1/(2 k^+ p_2^-)$) yields
\begin{align}\label{cross_sect_dip1}
& \frac{d \sigma (\lambda)}{d^2 k_T \, dy} = \lambda \, \frac{\as}{4 \pi^4}  \frac{1}{s} \, N_c \, \int d^2 x \, d^2 y \, d^2 b_1 \, e^{- i {\un k} \cdot ({\un x} - {\un y})}  \notag \\ 
& \times \, \left\{  \frac{{\un x} - {\un b}_1}{|{\un x} - {\un b}_1|^2} \cdot \frac{{\un y} - {\un b}_1}{|{\un y} - {\un b}_1|^2} \bigg[ 4\,  G^\textrm{adj}_{{\un x}, {\un y}} (2 k^- p_1^+) - 4 \, G^\textrm{adj}_{{\un x}, {\un b}_1} (2 k^- p_1^+)  - G^\textrm{adj}_{{\un b}_1, {\un y}} (2 k^- p_1^+) - G^\textrm{adj}_{{\un b}_1, {\un x}} (2 k^- p_1^+) + 2 \, G^\textrm{adj}_{{\un b}_1, {\un b}'_1} (2 k^- p_1^+) \bigg] \right. \notag \\ 
& - \frac{{\un x} - {\un b}_1}{|{\un x} - {\un b}_1|^2} \cdot \frac{{\un y} - {\un b}_0}{|{\un y} - {\un b}_0|^2} \bigg[ 2 \, G^\textrm{adj}_{{\un x}, {\un y}} (2 k^- p_1^+) - 2 \, G^\textrm{adj}_{{\un x}, {\un b}_0} (2 k^- p_1^+) - G^\textrm{adj}_{{\un b}_1, {\un y}} (2 k^- p_1^+) + G^\textrm{adj}_{{\un b}_1, {\un b}_0} (2 k^- p_1^+)  \bigg] \notag \\
& - \frac{{\un x} - {\un b}_0}{|{\un x} - {\un b}_0|^2} \cdot \frac{{\un y} - {\un b}_1}{|{\un y} - {\un b}_1|^2} \bigg[ 2 \, G^\textrm{adj}_{{\un x}, {\un y}} (2 k^- p_1^+) - 2 \, G^\textrm{adj}_{{\un x}, {\un b}_1} (2 k^- p_1^+) - G^\textrm{adj}_{{\un b}_1, {\un x}} (2 k^- p_1^+) + G^\textrm{adj}_{{\un b}_1, {\un b}_0} (2 k^- p_1^+)  \bigg] \notag \\
&  - 8 i \, k^j \, \frac{{\un x} - {\un b}_1}{|{\un x} - {\un b}_1|^2}  \, \times \, \frac{{\un y} - {\un b}_1}{|{\un y} - {\un b}_1|^2}  \, G^{j \, \textrm{adj}}_{{\un x}, {\un b}_1} (2 k^- p_1^+) + 4 i \, k^j \, \frac{{\un x} - {\un b}_1}{|{\un x} - {\un b}_1|^2}  \, \times \, \frac{{\un y} - {\un b}_0}{|{\un y} - {\un b}_0|^2}  \, G^{j \, \textrm{adj}}_{{\un x}, {\un b}_0} (2 k^- p_1^+)  \notag \\
& \left. + 4 i \, k^j \, \frac{{\un x} - {\un b}_0}{|{\un x} - {\un b}_0|^2}  \, \times \, \frac{{\un y} - {\un b}_1}{|{\un y} - {\un b}_1|^2}  \, G^{j \, \textrm{adj}}_{{\un x}, {\un b}_1} (2 k^- p_1^+)  \right\}  .
\end{align}
We see that the inclusive gluon production cross section \eqref{cross_sect_dip1} is expressed in terms of the polarized dipole amplitudes $G_T^{\textrm{adj}}$ and $G_{T}^{i \, \textrm{adj}}$. Since the amplitudes $G_{T}^{i \, \textrm{adj}}$ enter \eq{cross_sect_dip1} multiplied by the cross-product, which contains $\epsilon^{ij}$, we need another Levy-Civita symbol for the term to survive. The decomposition \eq{Gi_decomp} indicates that only the dipole amplitude $G_{2 T}^{\textrm{adj}}$ comes in with the needed Levy-Civita symbol. We thus conclude that the the inclusive gluon production cross section \eqref{cross_sect_dip1} for the ${\cal F}^{21}$-type gluon dipole projectile is expressed in terms of $G_T^{\textrm{adj}}$ and $G_{2 T}^{\textrm{adj}}$. 


\subsection{$\left( \mathscr{D}^i - \cev{\mathscr{D}}^i \right)$-type dipole projectile}

Next we consider inclusive gluon production for the projectile dipole with a type-2 operator. That is, we consider the same diagrams in \fig{fig:F12_dipole}, except that now the black circles of the diagrams A, C and F come in with the $\left( \mathscr{D}^i - \cev{\mathscr{D}}^i \right)$ operator (and with the prefactor) from \eq{Ui}. The white boxes in diagrams $D_1, D_0, E_1$ and $E_0$ denote $U^{i \, \textrm{G} [2]}$, while the same boxes in diagrams $B_1$ and $B_0$ still denote the entire sub-eikonal polarized Wilson line. This means that all diagrams, when squared, give contributions which carry a transverse index $i$, which we will use to distinguish them from the diagrams in the previous Subsection of this Appendix.

Our goal here is to show that even for the projectile being a dipole with a type-2 sub-eikonal operator, the corresponding cross section would still depend only on $G_T^{\textrm{adj}}$ and $G_{2 T}^{\textrm{adj}}$. We demonstrate this by an explicit calculation. Similar to the above we write\footnote{Indeed it is not quite right to call the object in \eq{AiAeik_dip} a cross section, since the projectile here is not a physical object, but a dipole made with the \eqref{Ui} polarized Wilson line, which carries a transverse index $i$. We will proceed with this notation, however, with the understanding that this cross section has to be convoluted with some $i$-dependent function describing how the polarized Wilson line \eqref{Ui} arises in the projectile through small-$x$ evolution.} 
\begin{align}\label{AiAeik_dip}
& \frac{d \sigma^{A^i \, A^*_\textrm{eik} + \mbox{c.c.}} }{d^2 k_T \, dy} = \frac{\as}{(2 \pi)^4}  \frac{k^-}{p_2^-} \, \frac{N_c}{N_c^2 - 1} \, \int d^2 x \, d^2 y \, d^2 b_1 \, e^{- i {\un k} \cdot ({\un x} - {\un y})} \, \sum_{m=0}^1 \, (-1)^{m+1} \,  \\ 
& \left\{ \epsilon^{kj} \, \left( \delta^{ik} - \frac{2 (x-b_1)^i (x-b_1)^k}{|{\un x} - {\un b}_1|^2} \right)  \, \frac{({y} - {b}_m)^j}{|{\un y} - {\un b}_m|^2}    \left\langle  \Tr \left[ U_{\un y}^\dagger \, U_{{\un x}}^{\textrm{G} [1]} \right] - \Tr \left[ U_{{\un b}_m}^\dagger \, U_{{\un x}}^{\textrm{G} [1]} \right]  \right\rangle  \right. \notag \\ 
&  \left. - i \int d^2 x' \,  \left( \delta^{ij} - \frac{2 (x'-b_1)^i (x'-b_1)^j}{|{\un x}' - {\un b}_1|^2} \right)  \, \frac{(y-b_m)^j}{|{\un y} - {\un b}_m|^2}   \left\langle  \Tr \left[ U_{\un y}^\dagger \, U_{{\un x}, {\un x}'}^{\textrm{G} [2]} \right] - \Tr \left[ U_{{\un b}_m}^\dagger \, U_{{\un x}, {\un x}'}^{\textrm{G} [2]} \right]  \right\rangle \right\}  + \mbox{c.c.}  . \notag 
\end{align}
The $m=1$ term vanishes for the following reason: $U_{{\un x}}^{\textrm{G} [1]}$ is real, while $U_{{\un x}, {\un x}'}^{\textrm{G} [2]}$ is imaginary. Hence, the first correlator in \eqref{AiAeik_dip} is real, and ``$i$" times the second one is real too. The object in the curly brackets of \eqref{AiAeik_dip} is, therefore, also real, and has one Lorentz index $i$. Integration over the impact parameters yields, for $m=1$,
\begin{align}\label{impacts}
\int d^2 b_1 \, d^2 \left( \frac{x+y}{2} \right) \, \{ \ldots \}^i_{m=1} = (x-y)^i \, f_1 [({\un x} - {\un y})^2] + \epsilon^{im} \, (x-y)^m \, f_2 [({\un x} - {\un y})^2]
\end{align}
with some real functions $f_1$ and $f_2$. Fourier-transforming \eq{impacts} yields a purely imaginary expression, which is cancelled in \eq{AiAeik_dip} when we add the complex conjugate term.

We are left with the $m=0$ term in \eq{AiAeik_dip} giving
\begin{align}\label{AiAeik_dip2}
& \frac{d \sigma^{A^i \, A^*_\textrm{eik} + \mbox{c.c.}} }{d^2 k_T \, dy} = - \frac{\as}{(2 \pi)^4}  \frac{k^-}{p_2^-} \, \frac{N_c}{N_c^2 - 1} \, \int d^2 x \, d^2 y \, d^2 b_1 \, e^{- i {\un k} \cdot ({\un x} - {\un y})} \,  \left\{ \epsilon^{kj} \, \left( \delta^{ik} - \frac{2 (x-b_1)^i (x-b_1)^k}{|{\un x} - {\un b}_1|^2} \right)  \, \frac{({y} - {b}_0)^j}{|{\un y} - {\un b}_0|^2}  \right. \\ 
&  \left\langle  \Tr \left[ U_{\un y}^\dagger \, U_{{\un x}}^{\textrm{G} [1]} \right] - \Tr \left[ U_{{\un b}_0}^\dagger \, U_{{\un x}}^{\textrm{G} [1]} \right]  \right\rangle \left. - i \int d^2 x' \,  \left( \delta^{ij} - \frac{2 (x'-b_1)^i (x'-b_1)^j}{|{\un x}' - {\un b}_1|^2} \right)  \, \frac{(y-b_0)^j}{|{\un y} - {\un b}_0|^2}  \right.  \notag \\
& \times \, \left. \left\langle  \Tr \left[ U_{\un y}^\dagger \, U_{{\un x}, {\un x}'}^{\textrm{G} [2]} \right] - \Tr \left[ U_{{\un b}_0}^\dagger \, U_{{\un x}, {\un x}'}^{\textrm{G} [2]} \right]  \right\rangle \right\} + \mbox{c.c.}  . \notag 
\end{align}

Next we look at the interference of B with $C^i +F^i$, 
\begin{align}\label{BCiFi_dip}
& \frac{d \sigma^{B \, (C^{i*} + F^{i*}) + \mbox{c.c.}} }{d^2 k_T \, dy} =  \frac{\as}{(2 \pi)^4}  \frac{k^-}{p_2^-} \, \frac{N_c}{N_c^2 - 1} \, \int d^2 x \, d^2 y \, d^2 b_1 \, e^{- i {\un k} \cdot ({\un x} - {\un y})} \, \sum_{m=0}^1 \, (-1)^{m+1}  \\ 
& \times  \, \left\{ \epsilon^{kj} \, \left( \delta^{ik} - \frac{2 (y-b_1)^i (y-b_1)^k}{|{\un y} - {\un b}_1|^2} \right)  \, \frac{({x} - {b}_m)^j}{|{\un x} - {\un b}_m|^2}  \left\langle  \Tr \left[ U_{\un y}^\dagger \, U_{{\un x}}^{\textrm{G} [1]} \right] - \Tr \left[ U_{{\un b}_1}^\dagger \, U_{{\un x}}^{\textrm{G} [1]} \right]  \right\rangle  \right. \notag \\ 
& \left. + i \int d^2 x' \,  \left( \delta^{ij} - \frac{2 (y-b_1)^i (y-b_1)^j}{|{\un y} - {\un b}_1|^2} \right)  \, \frac{(x'-b_m)^j}{|{\un x}' - {\un b}_m|^2}   \left\langle  \Tr \left[ U_{\un y}^\dagger \, U_{{\un x}, {\un x}'}^{\textrm{G} [2]} \right] - \Tr \left[ U_{{\un b}_1}^\dagger \, U_{{\un x}, {\un x}'}^{\textrm{G} [2]} \right]  \right\rangle \right\} + \mbox{c.c.}  \notag \\
& =  - \frac{\as}{(2 \pi)^4}  \frac{k^-}{p_2^-} \, \frac{N_c}{N_c^2 - 1} \, \int d^2 x \, d^2 y \, d^2 b_1 \, e^{- i {\un k} \cdot ({\un x} - {\un y})} \notag   \\ 
& \times  \, \left\{ \epsilon^{kj} \, \left( \delta^{ik} - \frac{2 (y-b_1)^i (y-b_1)^k}{|{\un y} - {\un b}_1|^2} \right)  \, \frac{({x} - {b}_0)^j}{|{\un x} - {\un b}_0|^2}  \left\langle  \Tr \left[ U_{\un y}^\dagger \, U_{{\un x}}^{\textrm{G} [1]} \right] - \Tr \left[ U_{{\un b}_1}^\dagger \, U_{{\un x}}^{\textrm{G} [1]} \right]  \right\rangle  \right. \notag \\ 
& \left. + i \int d^2 x' \,  \left( \delta^{ij} - \frac{2 (y-b_1)^i (y-b_1)^j}{|{\un y} - {\un b}_1|^2} \right)  \, \frac{(x'-b_0)^j}{|{\un x}' - {\un b}_0|^2}   \left\langle  \Tr \left[ U_{\un y}^\dagger \, U_{{\un x}, {\un x}'}^{\textrm{G} [2]} \right] - \Tr \left[ U_{{\un b}_1}^\dagger \, U_{{\un x}, {\un x}'}^{\textrm{G} [2]} \right]  \right\rangle \right\} + \mbox{c.c.} .  \notag 
\end{align}

Let us take a closer look at the $U_{{\un x}, {\un x}'}^{\textrm{G} [2]}$-containing terms in the above expressions. In doing so, let us assume that the cross sections we are calculating will be convoluted with
\begin{align}\label{ebf}
\int d^2 b_{10} \, \epsilon^{ii'} \, b^{i'}_{10} \, f (b_{10}^2),
\end{align}
where $f(b_{10}^2)$ is some real-valued function. Indeed, after integrating over the impact parameters (transverse positions) of the (evolved) projectile, we would be left with some function of ${\un b}_{10}$ with one transverse index $i$, necessitating the form in \eq{ebf}. Since the Levi-Civita symbol always appears in helicity calculations, we have included it into \eq{ebf}.

We are interested in the following object
\begin{align}
& \int d^2 b_{10} \, \epsilon^{ii'} \, b^{i'}_{10} \, f (b_{10}^2) \, \left[ \frac{d \sigma^{A^i \, A^*_\textrm{eik} + \mbox{c.c.}} }{d^2 k_T \, dy} + \frac{d \sigma^{B \, (C^{i*} + F^{i*}) + \mbox{c.c.}} }{d^2 k_T \, dy} \right]_{U_{{\un x}, {\un x}'}^{\textrm{G} [2]} - \textrm{terms}} =  i \, \frac{\as}{(2 \pi)^4}  \frac{k^-}{p_2^-} \, \frac{N_c}{N_c^2 - 1} \, \int d^2 b_{10} \, \epsilon^{ii'} \, b^{i'}_{10} \, f (b_{10}^2) \notag \\ 
& \times \, 
 \int d^2 x \, d^2 y \, d^2 b_1 \, e^{- i {\un k} \cdot ({\un x} - {\un y})}  \, \Bigg\{  \int d^2 x' \,  \left( \delta^{ij} - \frac{2 (x'-b_1)^i (x'-b_1)^j}{|{\un x}' - {\un b}_1|^2} \right)  \, \frac{(y-b_0)^j}{|{\un y} - {\un b}_0|^2}   \left\langle  \Tr \left[ \left( U_{\un y}^\dagger - U_{{\un b}_0}^\dagger \right) \, U_{{\un x}, {\un x}'}^{\textrm{G} [2]} \right]  \right\rangle \notag \\
 & - \int d^2 x' \,  \left( \delta^{ij} - \frac{2 (y-b_1)^i (y-b_1)^j}{|{\un y} - {\un b}_1|^2} \right)  \, \frac{(x'-b_0)^j}{|{\un x}' - {\un b}_0|^2}   \left\langle  \Tr \left[ \left( U_{\un y}^\dagger - U_{{\un b}_1}^\dagger \right)  \, U_{{\un x}, {\un x}'}^{\textrm{G} [2]} \right]  \right\rangle \Bigg\} + \mbox{c.c.} .
\end{align}
We will next follow the calculations in \cite{Cougoulic:2022gbk} and in Appendix~A of \cite{Kovchegov:2023yzd}. Writing
\begin{subequations}
\begin{align}
& \delta^{ij} - \frac{2 (x'-b_1)^i (x'-b_1)^j}{|{\un x}' - {\un b}_1|^2} = - 4 \pi \, \int \frac{d^2 q}{(2 \pi)^2} \, e^{i {\un q} \cdot ({\un x}' - {\un b}_1)} \frac{1}{q_\perp^2} \, \left[ \delta^{ij} - \frac{2 q^i q^j}{q_\perp^2} \right], \\ 
& \frac{(y-b_0)^j}{|{\un y} - {\un b}_0|^2}  = - 2 \pi i \, \int \frac{d^2 l}{(2 \pi)^2} \, e^{i {\un l} \cdot ({\un y} - {\un b}_0)} \frac{l^j}{l_\perp^2} 
\end{align}
\end{subequations}
and using the definition \eqref{UG2} of $U_{{\un x}, {\un x}'}^{\textrm{G} [2]}$ we get
\begin{align}\label{AB_UG2}
& \int d^2 b_{10} \, \epsilon^{ii'} \, b^{i'}_{10} \, f (b_{10}^2) \, \left[ \frac{d \sigma^{A^i \, A^*_\textrm{eik} + \mbox{c.c.}} }{d^2 k_T \, dy} + \frac{d \sigma^{B \, (C^{i*} + F^{i*}) + \mbox{c.c.}} }{d^2 k_T \, dy} \right]_{U_{{\un x}, {\un x}'}^{\textrm{G} [2]} - \textrm{terms}} = i \, \frac{\as}{(2 \pi)^4}  \frac{k^-}{p_2^-} \, \frac{N_c}{N_c^2 - 1} \, \int d^2 b_{10} \, \epsilon^{ii'} \, b^{i'}_{10} \, f (b_{10}^2) \notag \\ 
& \times \, \frac{2 (2 \pi)^2}{2 k^-} \, \int\limits_{-\infty}^\infty d z^- \, \int d^2 x \, d^2 y \, d^2 b_1 \, e^{- i {\un k} \cdot ({\un x} - {\un y})}  \, \Bigg\{ \int \frac{d^2 q}{(2 \pi)^2} \, e^{i {\un q} \cdot ({\un x} - {\un b}_1)} \frac{1}{q_\perp^2} \, \left[ \delta^{ij} - \frac{2 q^i q^j}{q_\perp^2} \right] \, \int \frac{d^2 l}{(2 \pi)^2} \, e^{i {\un l} \cdot ({\un y} - {\un b}_0)} \frac{l^j}{l_\perp^2} \notag \\
& \times \, \left[ \left\langle  \Tr \left[ \left( U_{\un y}^\dagger - U_{{\un b}_0}^\dagger \right) \, \left( (\mathscr{D}^p + i \, k^p) U_{\un x} [\infty, z^-] \right) \, \left( (\mathscr{D}^p - i \, q^p) U_{\un x} [z^-, - \infty] \right)  \right]  \right\rangle \notag \right. \\ 
& \left. + \left\langle  \Tr \left[ \left( U_{\un y} - U_{{\un b}_0} \right) \, \left( (\mathscr{D}^p - i \, q^p) U_{\un x} [-\infty, z^-] \right) \, \left( (\mathscr{D}^p + i \, k^p) U_{\un x} [z^-, \infty] \right)  \right]  \right\rangle  \right] \notag \\
& - \int \frac{d^2 q}{(2 \pi)^2} \, e^{i {\un q} \cdot ({\un y} - {\un b}_1)} \frac{1}{q_\perp^2} \, \left[ \delta^{ij} - \frac{2 q^i q^j}{q_\perp^2} \right] \, \int \frac{d^2 l}{(2 \pi)^2} \, e^{i {\un l} \cdot ({\un x} - {\un b}_0)} \frac{l^j}{l_\perp^2} \notag \\
& \times \, \left[ \left\langle  \Tr \left[ \left( U_{\un y}^\dagger - U_{{\un b}_1}^\dagger \right) \, \left( (\mathscr{D}^p + i \, k^p) U_{\un x} [\infty, z^-] \right) \, \left( (\mathscr{D}^p - i \, l^p) U_{\un x} [z^-, - \infty] \right)  \right]  \right\rangle \notag \right. \\ 
& \left. + \left\langle  \Tr \left[ \left( U_{\un y} - U_{{\un b}_1} \right) \, \left( (\mathscr{D}^i - i \, l^i) U_{\un x} [-\infty, z^-] \right) \, \left( (\mathscr{D}^p + i \, k^p) U_{\un x} [z^-, \infty] \right)  \right]  \right\rangle  \right] \Bigg\} ,
\end{align}
where we explicitly show the c.c. terms, having employed the fact that the cross section, along with each term in the above expression, are symmetric under ${\un k} \to - {\un k}$: this simplified the complex conjugation.

Now, let is employ the passive PT transformation. Under this transformation we have
\begin{align}\label{UPT}
U_{\ul x} \ \xrightarrow[]{\textrm{PT}} \ U_{- \ul x}^\dagger , \ \ \ U_{\un{x}} [\infty,  z^-]  \ \xrightarrow[]{\textrm{PT}} \ U_{-\un{x}} [  - \infty , -z^-] , \ \ \ U_{\un{x}} [ z^- ,  -\infty ] \ \xrightarrow[]{\textrm{PT}} \ U_{-\un{x}} [ -z^-, \infty ], \ \ \ \int\limits_{-\infty}^\infty d z^-  \ \xrightarrow[]{\textrm{PT}} \ \int\limits_{\infty}^{-\infty} d z^- .
\end{align}
Note that the sign of the integration variables can always be reverted, compensating for the sign change in the integrand due to PT transformation: for instance, in the arguments of Wilson lines in \eq{UPT} we have ${\un x} \ \xrightarrow[]{\textrm{PT}} \  - {\un x}$, which can be eliminated by redefining the integration variable ${\un x} \to - {\un x}$, generating no overall minus sign. Therefore, one needs to pay attention to the sign change in the integration limits only. 

Under such a passive PT transformation, \eq{AB_UG2} becomes
\begin{align}\label{AB_UG2_2}
& \int d^2 b_{10} \, \epsilon^{ii'} \, b^{i'}_{10} \, f (b_{10}^2) \, \left[ \frac{d \sigma^{A^i \, A^*_\textrm{eik} + \mbox{c.c.}} }{d^2 k_T \, dy} + \frac{d \sigma^{B \, (C^{i*} + F^{i*}) + \mbox{c.c.}} }{d^2 k_T \, dy} \right]_{U_{{\un x}, {\un x}'}^{\textrm{G} [2]} - \textrm{terms}} \!\!\!\!\! \xrightarrow[]{\textrm{PT}} \, - i \, \frac{\as}{(2 \pi)^4}  \frac{k^-}{p_2^-} \, \frac{N_c}{N_c^2 - 1} \, \int d^2 b_{10} \, \epsilon^{ii'} \, b^{i'}_{10} \, f (b_{10}^2) \notag \\ 
& \times \, \frac{2 (2 \pi)^2}{2 k^-} \, \int\limits_{-\infty}^\infty d z^- \, \int d^2 x \, d^2 y \, d^2 b_1 \, e^{- i {\un k} \cdot ({\un x} - {\un y})}  \, \Bigg\{ \int \frac{d^2 q}{(2 \pi)^2} \, e^{i {\un q} \cdot ({\un x} - {\un b}_1)} \frac{1}{q_\perp^2} \, \left[ \delta^{ij} - \frac{2 q^i q^j}{q_\perp^2} \right] \, \int \frac{d^2 l}{(2 \pi)^2} \, e^{i {\un l} \cdot ({\un y} - {\un b}_0)} \frac{l^j}{l_\perp^2} \notag \\
& \times \, \left[ \left\langle  \Tr \left[ \left( U_{\un y} - U_{{\un b}_0} \right) \, \left( (\mathscr{D}^p + i \, k^p) U_{\un x} [-\infty, z^-] \right) \, \left( (\mathscr{D}^p - i \, q^p) U_{\un x} [z^-, \infty] \right)  \right]  \right\rangle \notag \right. \\ 
& \left. + \left\langle  \Tr \left[ \left( U_{\un y}^\dagger - U_{{\un b}_0}^\dagger \right) \, \left( (\mathscr{D}^p - i \, q^p) U_{\un x} [\infty, z^-] \right) \, \left( (\mathscr{D}^p + i \, k^p) U_{\un x} [z^-, - \infty] \right)  \right]  \right\rangle  \right] \notag \\
& - \int \frac{d^2 q}{(2 \pi)^2} \, e^{i {\un q} \cdot ({\un y} - {\un b}_1)} \frac{1}{q_\perp^2} \, \left[ \delta^{ij} - \frac{2 q^i q^j}{q_\perp^2} \right] \, \int \frac{d^2 l}{(2 \pi)^2} \, e^{i {\un l} \cdot ({\un x} - {\un b}_0)} \frac{l^j}{l_\perp^2} \notag \\
& \times \, \left[ \left\langle  \Tr \left[ \left( U_{\un y} - U_{{\un b}_1} \right) \, \left( (\mathscr{D}^p + i \, k^p) U_{\un x} [-\infty, z^-] \right) \, \left( (\mathscr{D}^p - i \, l^p) U_{\un x} [z^-, \infty] \right)  \right]  \right\rangle \notag \right. \\ 
& \left. + \left\langle  \Tr \left[ \left( U_{\un y}^\dagger - U_{{\un b}_1}^\dagger \right) \, \left( (\mathscr{D}^p - i \, l^p) U_{\un x} [\infty, z^-] \right) \, \left( (\mathscr{D}^p + i \, k^p) U_{\un x} [z^-, - \infty] \right)  \right]  \right\rangle  \right] \Bigg\} .
\end{align}
Comparing Eqs.~\eqref{AB_UG2_2} and \eqref{AB_UG2}, we see that the terms containing $ \mathscr{D}^p \, \mathscr{D}^p$, $q^p \, k^p$, and $l^p \, k^p$ are PT-odd (note the overall sign change between these two equations). Since the cross section (and the double spin asymmetry $A_{LL}$) should be PT-even, such terms must be zero. Therefore, we drop them. Concentrating on the terms containing one power of momentum and one covariant derivative, we see that the first trace in \eq{AB_UG2} contains
\begin{align}
i k^p \, \mathscr{D}^p - i q^p \, \cev{\mathscr{D}}^p = \frac{i}{2} \, (k^p + q^p) \, \left( \mathscr{D}^p -  \cev{\mathscr{D}}^p \right) + \frac{i}{2} \, (k^p - q^p) \, \left( \mathscr{D}^p +  \cev{\mathscr{D}}^p \right).
\end{align}  
This term should be compared to that contained in the second trace of \eq{AB_UG2_2}, which, otherwise, has the same Wilson line structure as the first trace in \eq{AB_UG2},
\begin{align}
i k^p \, \cev{\mathscr{D}}^p - i q^p \, \mathscr{D}^p = - \frac{i}{2} \, (k^p + q^p) \, \left( \mathscr{D}^p -  \cev{\mathscr{D}}^p \right) + \frac{i}{2} \, (k^p - q^p) \, \left( \mathscr{D}^p +  \cev{\mathscr{D}}^p \right).
\end{align}
Combining this with the overall sign change between Eqs.~\eqref{AB_UG2_2} and \eqref{AB_UG2}, we conclude that the $\tfrac{i}{2} \, (k^p + q^p) \, \left( \mathscr{D}^p -  \cev{\mathscr{D}}^p \right)$ term is PT-even and should be kept, while the $\tfrac{i}{2} \, (k^p - q^p) \, \left( \mathscr{D}^p +  \cev{\mathscr{D}}^p \right)$ term is PT-odd and should be zero: we can neglect this term. 

Applying similar treatment to other terms in Eqs.~\eqref{AB_UG2_2} and \eqref{AB_UG2} containing one power of momentum and one covariant derivative, we arrive at
\begin{align}\label{AB_UG2_3}
& \int d^2 b_{10} \, \epsilon^{ii'} \, b^{i'}_{10} \, f (b_{10}^2) \, \left[ \frac{d \sigma^{A^i \, A^*_\textrm{eik} + \mbox{c.c.}} }{d^2 k_T \, dy} + \frac{d \sigma^{B \, (C^{i*} + F^{i*}) + \mbox{c.c.}} }{d^2 k_T \, dy} \right]_{U_{{\un x}, {\un x}'}^{\textrm{G} [2]} - \textrm{terms}} = i \, \frac{\as}{(2 \pi)^4}  \frac{k^-}{p_2^-} \, \frac{N_c}{N_c^2 - 1} \, \int d^2 b_{10} \, \epsilon^{ii'} \, b^{i'}_{10} \, f (b_{10}^2) \notag \\ 
& \times \, \frac{2 (2 \pi)^2}{2 k^-} \, \int\limits_{-\infty}^\infty d z^- \, \int d^2 x \, d^2 y \, d^2 b_1 \, e^{- i {\un k} \cdot ({\un x} - {\un y})}  \, \Bigg\{ \int \frac{d^2 q}{(2 \pi)^2} \, e^{i {\un q} \cdot ({\un x} - {\un b}_1)} \frac{1}{q_\perp^2} \, \left[ \delta^{ij} - \frac{2 q^i q^j}{q_\perp^2} \right] \, \int \frac{d^2 l}{(2 \pi)^2} \, e^{i {\un l} \cdot ({\un y} - {\un b}_0)} \frac{l^j}{l_\perp^2} \notag \\
& \times \, \left[ \frac{i}{2} \, (k^p + q^p) \, \left\langle  \Tr \left[ \left( U_{\un y}^\dagger - U_{{\un b}_0}^\dagger \right) \, U_{\un x} [\infty, z^-] \, \left( \mathscr{D}^p -  \cev{\mathscr{D}}^p \right) \, U_{\un x} [z^-, - \infty]  \right]  \right\rangle \notag \right. \\ 
& \left. - \frac{i}{2} \, (k^p + q^p) \,  \left\langle  \Tr \left[ \left( U_{\un y} - U_{{\un b}_0} \right) \, U_{\un x} [-\infty, z^-] \, \left( \mathscr{D}^p -  \cev{\mathscr{D}}^p \right) \, U_{\un x} [z^-, \infty]   \right]  \right\rangle  \right] \notag \\
& - \int \frac{d^2 q}{(2 \pi)^2} \, e^{i {\un q} \cdot ({\un y} - {\un b}_1)} \frac{1}{q_\perp^2} \, \left[ \delta^{ij} - \frac{2 q^i q^j}{q_\perp^2} \right] \, \int \frac{d^2 l}{(2 \pi)^2} \, e^{i {\un l} \cdot ({\un x} - {\un b}_0)} \frac{l^j}{l_\perp^2} \notag \\
& \times \, \left[ \frac{i}{2} \, (k^p + l^p) \,  \left\langle  \Tr \left[ \left( U_{\un y}^\dagger - U_{{\un b}_1}^\dagger \right) \, U_{\un x} [\infty, z^-] \, \left( \mathscr{D}^p -  \cev{\mathscr{D}}^p \right) \, U_{\un x} [z^-, - \infty] \right]  \right\rangle \notag \right. \\ 
& \left. - \frac{i}{2} \, (k^p + l^p) \, \left\langle  \Tr \left[ \left( U_{\un y} - U_{{\un b}_1} \right) \, U_{\un x} [-\infty, z^-] \, \left( \mathscr{D}^p -  \cev{\mathscr{D}}^p \right) \, U_{\un x} [z^-, \infty]  \right]  \right\rangle  \right] \Bigg\} ,
\end{align}
which we can rewrite using the definition of $U_{\un x}^{i \, \textrm{G} [2]}$ in \eq{Ui} as
\begin{align}\label{AB_UiG2}
& \int d^2 b_{10} \, \epsilon^{ii'} \, b^{i'}_{10} \, f (b_{10}^2) \, \left[ \frac{d \sigma^{A^i \, A^*_\textrm{eik} + \mbox{c.c.}} }{d^2 k_T \, dy} + \frac{d \sigma^{B \, (C^{i*} + F^{i*}) + \mbox{c.c.}} }{d^2 k_T \, dy} \right]_{U_{{\un x}, {\un x}'}^{\textrm{G} [2]} - \textrm{terms}} = - \frac{\as}{(2 \pi)^4}  \frac{k^-}{p_2^-} \, \frac{N_c}{N_c^2 - 1} \, \int d^2 b_{10} \, \epsilon^{ii'} \, b^{i'}_{10} \, f (b_{10}^2) \notag \\ 
& \times \, 2 (2 \pi)^2 \, \int d^2 x \, d^2 y \, d^2 b_1 \, e^{- i {\un k} \cdot ({\un x} - {\un y})}  \, \Bigg\{ \int \frac{d^2 q}{(2 \pi)^2} \, e^{i {\un q} \cdot ({\un x} - {\un b}_1)} \frac{1}{q_\perp^2} \, \left[ \delta^{ij} - \frac{2 q^i q^j}{q_\perp^2} \right] \, (k^p + q^p) \, \int \frac{d^2 l}{(2 \pi)^2} \, e^{i {\un l} \cdot ({\un y} - {\un b}_0)} \frac{l^j}{l_\perp^2}  \notag \\
& \times \, \left[  \left\langle  \Tr \left[ \left( U_{\un y}^\dagger - U_{{\un b}_0}^\dagger \right) \, U_{\un x}^{p \, \textrm{G} [2]}  \right]  \right\rangle +  \left\langle  \Tr \left[ \left( U_{\un y} - U_{{\un b}_0} \right) \, \left( U_{\un x}^{p \, \textrm{G} [2]} \right)^\dagger  \right]  \right\rangle  \right] \notag \\
& - \int \frac{d^2 q}{(2 \pi)^2} \, e^{i {\un q} \cdot ({\un y} - {\un b}_1)} \frac{1}{q_\perp^2} \, \left[ \delta^{ij} - \frac{2 q^i q^j}{q_\perp^2} \right] \, \int \frac{d^2 l}{(2 \pi)^2} \, e^{i {\un l} \cdot ({\un x} - {\un b}_0)} \frac{l^j}{l_\perp^2} \, (k^p + l^p) \notag \\
& \times \, \left[  \left\langle  \Tr \left[ \left( U_{\un y}^\dagger - U_{{\un b}_1}^\dagger \right) \, U_{\un x}^{p \, \textrm{G} [2]} \right]  \right\rangle +  \left\langle  \Tr \left[ \left( U_{\un y} - U_{{\un b}_1} \right) \, \left( U_{\un x}^{p \, \textrm{G} [2]} \right)^\dagger  \right]  \right\rangle  \right] \Bigg\} .
\end{align}

Note that in our expressions here we did not explicitly include the time-ordering and anti-time-ordering signs $\tord$ and $\atord$. Since each term in, say, \eq{AB_UG2} contains both the operators from the scattering amplitude and from the complex conjugate amplitude, some of the operators in each term should be time-ordered, while others should be anti-time-ordered. While strictly-speaking we should keep track of such (anti-)time-ordering, since both the PT transformation and the complex conjugation interchange the two orderings, $\tord \leftrightarrow \atord$, our above argument would be unaffected by using the more lengthy notation with $\tord$ and $\atord$.

We can rewrite \eq{AB_UiG2} as
\begin{align}\label{AB_UiG2_2}
& \int d^2 b_{10} \, \epsilon^{ii'} \, b^{i'}_{10} \, f (b_{10}^2) \, \left[ \frac{d \sigma^{A^i \, A^*_\textrm{eik} + \mbox{c.c.}} }{d^2 k_T \, dy} + \frac{d \sigma^{B \, (C^{i*} + F^{i*}) + \mbox{c.c.}} }{d^2 k_T \, dy} \right]_{U_{{\un x}, {\un x}'}^{\textrm{G} [2]} - \textrm{terms}} = - \frac{\as N_c}{(2 \pi)^4}  \frac{1}{s} \, \int d^2 b_{10} \, \epsilon^{ii'} \, b^{i'}_{10} \, f (b_{10}^2) \notag \\ 
& \times \, 4 (2 \pi)^2 \, \int d^2 x \, d^2 y \, d^2 b_1 \, e^{- i {\un k} \cdot ({\un x} - {\un y})}  \, \Bigg\{ \int \frac{d^2 q}{(2 \pi)^2} \, e^{i {\un q} \cdot ({\un x} - {\un b}_1)} \frac{1}{q_\perp^2} \, \left[ \delta^{ij} - \frac{2 q^i q^j}{q_\perp^2} \right] \, (k^p + q^p) \, \int \frac{d^2 l}{(2 \pi)^2} \, e^{i {\un l} \cdot ({\un y} - {\un b}_0)} \frac{l^j}{l_\perp^2}  \notag \\
& \times \, \left[ G^{p \, \textrm{adj}}_{{\un x}, {\un y}} (2 k^- p_1^+) - G^{p \, \textrm{adj}}_{{\un x}, {\un b}_0} (2 k^- p_1^+)  \right] \notag \\
& - \int \frac{d^2 q}{(2 \pi)^2} \, e^{i {\un q} \cdot ({\un y} - {\un b}_1)} \frac{1}{q_\perp^2} \, \left[ \delta^{ij} - \frac{2 q^i q^j}{q_\perp^2} \right] \, \int \frac{d^2 l}{(2 \pi)^2} \, e^{i {\un l} \cdot ({\un x} - {\un b}_0)} \frac{l^j}{l_\perp^2} \, (k^p + l^p) \, \left[  G^{p \, \textrm{adj}}_{{\un x}, {\un y}} (2 k^- p_1^+) - G^{p \, \textrm{adj}}_{{\un x}, {\un b}_1} (2 k^- p_1^+)   \right] \Bigg\} .
\end{align}
Performing the Fourier transformations we arrive at
\begin{align}\label{AB_UiG2_F}
& \int d^2 b_{10} \, \epsilon^{ii'} \, b^{i'}_{10} \, f (b_{10}^2) \, \left[ \frac{d \sigma^{A^i \, A^*_\textrm{eik} + \mbox{c.c.}} }{d^2 k_T \, dy} + \frac{d \sigma^{B \, (C^{i*} + F^{i*}) + \mbox{c.c.}} }{d^2 k_T \, dy} \right]_{U_{{\un x}, {\un x}'}^{\textrm{G} [2]} - \textrm{terms}} = - \frac{\as N_c}{(2 \pi)^4}  \frac{1}{s} \, \int d^2 b_{10} \, \epsilon^{ii'} \, b^{i'}_{10} \, f (b_{10}^2) \notag \\ 
& \times \, 2 \, \int d^2 x \, d^2 y \, d^2 b_1 \, e^{- i {\un k} \cdot ({\un x} - {\un y})}  \, \Bigg\{ \left[ (\pd_x^p - i \,  k^p )  \left( \delta^{ij} - \frac{2 (x -b_1)^i (x -b_1)^j}{|{\un x}  - {\un b}_1|^2} \right) \right] \, \frac{(y-b_0)^j}{|{\un y} - {\un b}_0|^2} \notag \\
& \times \, \left[ G^{p \, \textrm{adj}}_{{\un x}, {\un y}} (2 k^- p_1^+) - G^{p \, \textrm{adj}}_{{\un x}, {\un b}_0} (2 k^- p_1^+)  \right] \notag \\
& -  \left[ (\pd_x^p - i \,  k^p ) \, \frac{(x-b_0)^j}{|{\un x} - {\un b}_0|^2}  \right] \, \left( \delta^{ij} - \frac{2 (y -b_1)^i (y -b_1)^j}{|{\un y}  - {\un b}_1|^2} \right)  \, \left[  G^{p \, \textrm{adj}}_{{\un x}, {\un y}} (2 k^- p_1^+) - G^{p \, \textrm{adj}}_{{\un x}, {\un b}_1} (2 k^- p_1^+)   \right] \Bigg\} .
\end{align}

Adding the $U_{\un x}^{\textrm{G} [1]}$-containing terms from Eqs.~\eqref{AiAeik_dip2} and \eqref{BCiFi_dip} to \eq{AB_UiG2_F} and dropping $\int d^2 b_{10} \, \epsilon^{ii'} \, b^{i'}_{10} \, f (b_{10}^2)$ in the latter yields
\begin{align}\label{AB_UiG2_final}
& \frac{d \sigma^{A^i \, A^*_\textrm{eik} + \mbox{c.c.}} }{d^2 k_T \, dy} + \frac{d \sigma^{B \, (C^{i*} + F^{i*}) + \mbox{c.c.}} }{d^2 k_T \, dy} = - \frac{2 \as N_c}{(2 \pi)^4}  \frac{1}{s} \, \int d^2 x \, d^2 y \, d^2 b_1 \, e^{- i {\un k} \cdot ({\un x} - {\un y})}  \\ 
& \times  \, \Bigg\{
\epsilon^{kj} \, \left( \delta^{ik} - \frac{2 (y-b_1)^i (y-b_1)^k}{|{\un y} - {\un b}_1|^2} \right)  \, \frac{({x} - {b}_0)^j}{|{\un x} - {\un b}_0|^2} 
\,  \left[ G^\textrm{adj}_{{\un x}, {\un y}} (2 k^- p_1^+)- G^\textrm{adj}_{{\un x}, {\un b}_0} (2 k^- p_1^+)  \right]   \notag \\ 
& + \epsilon^{kj} \, \left( \delta^{ik} - \frac{2 (x-b_1)^i (x-b_1)^k}{|{\un x} - {\un b}_1|^2} \right)  \, \frac{({y} - {b}_0)^j}{|{\un y} - {\un b}_0|^2}  \, \left[ G^\textrm{adj}_{{\un x}, {\un y}} (2 k^- p_1^+)  - G^\textrm{adj}_{{\un x}, {\un b}_1} (2 k^- p_1^+) \right] \notag \\
& \left[ (\pd_x^p - i \,  k^p )  \left( \delta^{ij} - \frac{2 (x -b_1)^i (x -b_1)^j}{|{\un x}  - {\un b}_1|^2} \right) \right] \, \frac{(y-b_0)^j}{|{\un y} - {\un b}_0|^2} \, \left[ G^{p \, \textrm{adj}}_{{\un x}, {\un y}} (2 k^- p_1^+) - G^{p \, \textrm{adj}}_{{\un x}, {\un b}_0} (2 k^- p_1^+)  \right] \notag \\
& -  \left[ (\pd_x^p - i \,  k^p ) \, \frac{(x-b_0)^j}{|{\un x} - {\un b}_0|^2}  \right] \, \left( \delta^{ij} - \frac{2 (y -b_1)^i (y -b_1)^j}{|{\un y}  - {\un b}_1|^2} \right)  \, \left[  G^{p \, \textrm{adj}}_{{\un x}, {\un y}} (2 k^- p_1^+) - G^{p \, \textrm{adj}}_{{\un x}, {\un b}_1} (2 k^- p_1^+)   \right] \Bigg\} . \notag
\end{align}
Indeed, everything in \eq{AB_UiG2_final} is expressed in terms of the polarized dipole amplitudes $G_T^{\textrm{adj}}$ and $G_{T}^{i \, \textrm{adj}}$.

Lastly, the $(D^i + E^i) A_\textrm{eik}^{*}$ terms depend on the  $\mathscr{D}^p -  \cev{\mathscr{D}}^p$ operator by definition, that is, diagrams $D^i$ and $E^i$ already have the $\mathscr{D}^p -  \cev{\mathscr{D}}^p$ operator in the black circle. Therefore, such contribution depends on $G_{T}^{i \, \textrm{adj}}$. For completeness, let us quote the $(D^i + E^i) A_\textrm{eik}^{*}$ + c.c. contribution, which can be read from Eqs.~\eqref{DAeik_dip} and \eqref{EAeik_dip} by replacing $G^{\textrm{adj}} \to G^{i \, \textrm{adj}}$ in them and dropping the overall factor of $\lambda$. We get
\begin{align}\label{DEiAeik_dip}
& \frac{d \sigma^{D^i \, A^*_\textrm{eik} + \mbox{c.c.}} }{d^2 k_T \, dy}  + \frac{d \sigma^{E^i \, A^*_\textrm{eik} + \mbox{c.c.}} }{d^2 k_T \, dy}  \\
& = - \frac{\as}{4 \pi^4} \frac{1}{s} \, N_c \,   \int d^2 x \, d^2 y \, d^2 b_1 \, e^{- i {\un k} \cdot ({\un x} - {\un y})} \, \left\{  \left( \frac{{\un x} - {\un b}_1}{|{\un x} - {\un b}_1|^2} - \frac{{\un x} - {\un b}_0}{|{\un x} - {\un b}_0|^2} \right) \, \cdot \, \frac{{\un y} - {\un b}_1}{|{\un y} - {\un b}_1|^2} \, G^{ i \, \textrm{adj}}_{{\un b}_1, {\un x}} (2 k^- p_1^+)  \right. \notag \\
& + \frac{{\un x} - {\un b}_1}{|{\un x} - {\un b}_1|^2} \cdot \left( \frac{{\un y} - {\un b}_1}{|{\un y} - {\un b}_1|^2} - \frac{{\un y} - {\un b}_0}{|{\un y} - {\un b}_0|^2} \right) \, G^{i \, \textrm{adj}}_{{\un b}_1, {\un y}} (2 k^- p_1^+) \notag \\
& \left. + \left( \frac{{\un x} - {\un b}_1}{|{\un x} - {\un b}_1|^2} \cdot \frac{{\un y} - {\un b}_0}{|{\un y} - {\un b}_0|^2} + \frac{{\un x} - {\un b}_0}{|{\un x} - {\un b}_0|^2} \cdot \frac{{\un y} - {\un b}_1}{|{\un y} - {\un b}_1|^2} \right) \, G^{i \, \textrm{adj}}_{{\un b}_1, {\un b}_0} (2 k^- p_1^+) \right\}. \notag
\end{align}

Adding Eqs.~\eqref{AB_UiG2_final} and \eqref{DEiAeik_dip} together we arrive at the final result for the polarization-independent dipole producing a gluon at the sub-eikonal order:
\begin{align}\label{cross_sect_i}
& \frac{d \sigma^i}{d^2 k_T \, dy} = - \frac{\as}{4 \pi^4}  \frac{1}{s} \, N_c \, \int d^2 x \, d^2 y \, d^2 b_1 \, e^{- i {\un k} \cdot ({\un x} - {\un y})}  \notag \\ 
& \times \, \Bigg\{  2 \, \epsilon^{kj} \, \left( \delta^{ik} - \frac{2 (y-b_1)^i (y-b_1)^k}{|{\un y} - {\un b}_1|^2} \right)  \, \frac{({x} - {b}_0)^j}{|{\un x} - {\un b}_0|^2} 
\,  \left[ G^\textrm{adj}_{{\un x}, {\un y}} (2 k^- p_1^+) - G^\textrm{adj}_{{\un x}, {\un b}_0} (2 k^- p_1^+)  \right]   \notag \\ 
& + 2 \,\epsilon^{kj} \, \left( \delta^{ik} - \frac{2 (x-b_1)^i (x-b_1)^k}{|{\un x} - {\un b}_1|^2} \right)  \, \frac{({y} - {b}_0)^j}{|{\un y} - {\un b}_0|^2}  \, \left[ G^\textrm{adj}_{{\un x}, {\un y}} (2 k^- p_1^+)  - G^\textrm{adj}_{{\un x}, {\un b}_1} (2 k^- p_1^+) \right] \notag \\
& 2 \, \left[ (\pd_x^p - i \,  k^p )  \left( \delta^{ij} - \frac{2 (x -b_1)^i (x -b_1)^j}{|{\un x}  - {\un b}_1|^2} \right) \right] \, \frac{(y-b_0)^j}{|{\un y} - {\un b}_0|^2} \, \left[ G^{p \, \textrm{adj}}_{{\un x}, {\un y}} (2 k^- p_1^+) - G^{p \, \textrm{adj}}_{{\un x}, {\un b}_0} (2 k^- p_1^+)  \right] \notag \\
& - 2 \, \left[ (\pd_x^p - i \,  k^p ) \, \frac{(x-b_0)^j}{|{\un x} - {\un b}_0|^2}  \right] \, \left( \delta^{ij} - \frac{2 (y -b_1)^i (y -b_1)^j}{|{\un y}  - {\un b}_1|^2} \right)  \, \left[  G^{p \, \textrm{adj}}_{{\un x}, {\un y}} (2 k^- p_1^+) - G^{p \, \textrm{adj}}_{{\un x}, {\un b}_1} (2 k^- p_1^+)   \right] \notag \\
& + \left( \frac{{\un x} - {\un b}_1}{|{\un x} - {\un b}_1|^2} - \frac{{\un x} - {\un b}_0}{|{\un x} - {\un b}_0|^2} \right) \, \cdot \, \frac{{\un y} - {\un b}_1}{|{\un y} - {\un b}_1|^2} \, G^{ i \, \textrm{adj}}_{{\un b}_1, {\un x}} (2 k^- p_1^+)   \notag \\
& + \frac{{\un x} - {\un b}_1}{|{\un x} - {\un b}_1|^2} \cdot \left( \frac{{\un y} - {\un b}_1}{|{\un y} - {\un b}_1|^2} - \frac{{\un y} - {\un b}_0}{|{\un y} - {\un b}_0|^2} \right) \, G^{i \, \textrm{adj}}_{{\un b}_1, {\un y}} (2 k^- p_1^+) \notag \\
& + \left( \frac{{\un x} - {\un b}_1}{|{\un x} - {\un b}_1|^2} \cdot \frac{{\un y} - {\un b}_0}{|{\un y} - {\un b}_0|^2} + \frac{{\un x} - {\un b}_0}{|{\un x} - {\un b}_0|^2} \cdot \frac{{\un y} - {\un b}_1}{|{\un y} - {\un b}_1|^2} \right) \, G^{i \, \textrm{adj}}_{{\un b}_1, {\un b}_0} (2 k^- p_1^+) \Bigg\}. 
\end{align}
Once again, we see that everything is expressed in terms of the ${\cal F}^{12}$ and $\mathscr{D}^p -  \cev{\mathscr{D}}^p$ operators, and, therefore, in terms of $G_T^{\textrm{adj}}$ and $G_{T}^{i \, \textrm{adj}}$. After convolution with \eq{ebf}, we see that only the part of the decomposition \eqref{Gi_decomp} containing $G_{2 T}^{\textrm{adj}}$ survives. 

We have thus shown that the interaction with the target in the cross section of small-$x$ evolved longitudinally polarized projectile on a longitudinally polarized target is expressible in terms of $G_T^{\textrm{adj}} (2 k^- p_1^+)$ and $G_{2 T}^{\textrm{adj}} (2 k^- p_1^+)$, as desired.


\bibliography{references, references2, newrefs}

\providecommand{\href}[2]{#2}\begingroup\raggedright\begin{thebibliography}{100}

\bibitem{Kovchegov:2015pbl}
Y.~V. Kovchegov, D.~Pitonyak and M.~D. Sievert, \emph{{Helicity Evolution at
  Small-x}}, \href{https://doi.org/10.1007/JHEP01(2016)072}{\emph{JHEP}
  {\bfseries 01} (2016) 072},
  [\href{https://arxiv.org/abs/1511.06737}{{\ttfamily 1511.06737}}].

\bibitem{Kovchegov:2016zex}
Y.~V. Kovchegov, D.~Pitonyak and M.~D. Sievert, \emph{{Helicity Evolution at
  Small $x$: Flavor Singlet and Non-Singlet Observables}},
  \href{https://doi.org/10.1103/PhysRevD.95.014033}{\emph{Phys. Rev.}
  {\bfseries D95} (2017) 014033},
  [\href{https://arxiv.org/abs/1610.06197}{{\ttfamily 1610.06197}}].

\bibitem{Kovchegov:2018znm}
Y.~V. Kovchegov and M.~D. Sievert, \emph{{Small-$x$ Helicity Evolution: an
  Operator Treatment}},
  \href{https://doi.org/10.1103/PhysRevD.99.054032}{\emph{Phys. Rev.}
  {\bfseries D99} (2019) 054032},
  [\href{https://arxiv.org/abs/1808.09010}{{\ttfamily 1808.09010}}].

\bibitem{Cougoulic:2022gbk}
F.~Cougoulic, Y.~V. Kovchegov, A.~Tarasov and Y.~Tawabutr, \emph{{Quark and
  gluon helicity evolution at small x: revised and updated}},
  \href{https://doi.org/10.1007/JHEP07(2022)095}{\emph{JHEP} {\bfseries 07}
  (2022) 095}, [\href{https://arxiv.org/abs/2204.11898}{{\ttfamily
  2204.11898}}].

\bibitem{Adamiak:2023yhz}
D.~Adamiak, N.~Baldonado, Y.~V. Kovchegov, W.~Melnitchouk, D.~Pitonyak, N.~Sato
  et~al., \emph{{Global analysis of polarized DIS \& SIDIS data with improved
  small-$x$ helicity evolution}},
  \href{https://arxiv.org/abs/2308.07461}{{\ttfamily 2308.07461}}.

\bibitem{EuropeanMuon:1987isl}
{\scshape European Muon} collaboration, J.~Ashman et~al., \emph{{A Measurement
  of the Spin Asymmetry and Determination of the Structure Function g(1) in
  Deep Inelastic Muon-Proton Scattering}},
  \href{https://doi.org/10.1016/0370-2693(88)91523-7}{\emph{Phys. Lett. B}
  {\bfseries 206} (1988) 364}.

\bibitem{Jaffe:1989jz}
R.~L. Jaffe and A.~Manohar, \emph{{The G(1) Problem: Fact and Fantasy on the
  Spin of the Proton}},
  \href{https://doi.org/10.1016/0550-3213(90)90506-9}{\emph{Nucl. Phys.}
  {\bfseries B337} (1990) 509--546}.

\bibitem{Ji:1996ek}
X.-D. Ji, \emph{{Gauge-Invariant Decomposition of Nucleon Spin}},
  \href{https://doi.org/10.1103/PhysRevLett.78.610}{\emph{Phys. Rev. Lett.}
  {\bfseries 78} (1997) 610--613},
  [\href{https://arxiv.org/abs/hep-ph/9603249}{{\ttfamily hep-ph/9603249}}].

\bibitem{Boer:2011fh}
D.~Boer et~al., \emph{{Gluons and the quark sea at high energies:
  Distributions, polarization, tomography}},
  \href{https://arxiv.org/abs/1108.1713}{{\ttfamily 1108.1713}}.

\bibitem{Aidala:2012mv}
C.~A. Aidala, S.~D. Bass, D.~Hasch and G.~K. Mallot, \emph{{The Spin Structure
  of the Nucleon}}, \href{https://doi.org/10.1103/RevModPhys.85.655}{\emph{Rev.
  Mod. Phys.} {\bfseries 85} (2013) 655--691},
  [\href{https://arxiv.org/abs/1209.2803}{{\ttfamily 1209.2803}}].

\bibitem{Accardi:2012qut}
A.~Accardi et~al., \emph{{Electron Ion Collider: The Next QCD Frontier}},
  \href{https://doi.org/10.1140/epja/i2016-16268-9}{\emph{Eur. Phys. J.}
  {\bfseries A52} (2016) 268},
  [\href{https://arxiv.org/abs/1212.1701}{{\ttfamily 1212.1701}}].

\bibitem{Leader:2013jra}
E.~Leader and C.~Lorcé, \emph{{The angular momentum controversy: What's it all
  about and does it matter?}},
  \href{https://doi.org/10.1016/j.physrep.2014.02.010}{\emph{Phys. Rept.}
  {\bfseries 541} (2014) 163--248},
  [\href{https://arxiv.org/abs/1309.4235}{{\ttfamily 1309.4235}}].

\bibitem{Aschenauer:2013woa}
E.~C. Aschenauer et~al., \emph{{The RHIC Spin Program: Achievements and Future
  Opportunities}},  \href{https://arxiv.org/abs/1304.0079}{{\ttfamily
  1304.0079}}.

\bibitem{Aschenauer:2015eha}
E.-C. Aschenauer et~al., \emph{{The RHIC SPIN Program: Achievements and Future
  Opportunities}},  \href{https://arxiv.org/abs/1501.01220}{{\ttfamily
  1501.01220}}.

\bibitem{Proceedings:2020eah}
A.~Prokudin, Y.~Hatta, Y.~Kovchegov and C.~Marquet, eds., \emph{{Proceedings,
  Probing Nucleons and Nuclei in High Energy Collisions: Dedicated to the
  Physics of the Electron Ion Collider}: {Seattle (WA), United States, October
  1 - November 16, 2018}}, WSP, 2020.
\newblock 10.1142/11684.

\bibitem{Ji:2020ena}
X.~Ji, F.~Yuan and Y.~Zhao, \emph{{What we know and what we
  don\textquoteright{}t know about the proton spin after 30 years}},
  \href{https://doi.org/10.1038/s42254-020-00248-4}{\emph{Nature Rev. Phys.}
  {\bfseries 3} (2021) 27--38},
  [\href{https://arxiv.org/abs/2009.01291}{{\ttfamily 2009.01291}}].

\bibitem{AbdulKhalek:2021gbh}
R.~Abdul~Khalek et~al., \emph{{Science Requirements and Detector Concepts for
  the Electron-Ion Collider: EIC Yellow Report}},
  \href{https://arxiv.org/abs/2103.05419}{{\ttfamily 2103.05419}}.

\bibitem{Kuraev:1977fs}
E.~A. Kuraev, L.~N. Lipatov and V.~S. Fadin, \emph{{The Pomeranchuk
  singlularity in non-Abelian gauge theories}}, {\emph{Sov. Phys. JETP}
  {\bfseries 45} (1977) 199--204}.

\bibitem{Balitsky:1978ic}
I.~Balitsky and L.~Lipatov, \emph{{The Pomeranchuk Singularity in Quantum
  Chromodynamics}}, {\emph{Sov.J.Nucl.Phys.} {\bfseries 28} (1978) 822--829}.

\bibitem{Gribov:1981ac}
L.~V. Gribov, E.~M. Levin and M.~G. Ryskin, \emph{Singlet structure function at
  small x: Unitarization of gluon ladders}, {\emph{Nucl. Phys.} {\bfseries
  B188} (1981) 555--576}.

\bibitem{Balitsky:1995ub}
I.~Balitsky, \emph{{Operator expansion for high-energy scattering}},
  \href{https://doi.org/10.1016/0550-3213(95)00638-9}{\emph{Nucl. Phys.}
  {\bfseries B463} (1996) 99--160},
  [\href{https://arxiv.org/abs/hep-ph/9509348}{{\ttfamily hep-ph/9509348}}].

\bibitem{Balitsky:1998ya}
I.~Balitsky, \emph{Factorization and high-energy effective action},
  {\emph{Phys. Rev.} {\bfseries D60} (1999) 014020},
  [\href{https://arxiv.org/abs/hep-ph/9812311}{{\ttfamily hep-ph/9812311}}].

\bibitem{Kovchegov:1999yj}
Y.~V. Kovchegov, \emph{Small-x {$F_2$} structure function of a nucleus
  including multiple pomeron exchanges}, {\emph{Phys. Rev.} {\bfseries D60}
  (1999) 034008}, [\href{https://arxiv.org/abs/hep-ph/9901281}{{\ttfamily
  hep-ph/9901281}}].

\bibitem{Kovchegov:1999ua}
Y.~V. Kovchegov, \emph{Unitarization of the {BFKL} pomeron on a nucleus},
  {\emph{Phys. Rev.} {\bfseries D61} (2000) 074018},
  [\href{https://arxiv.org/abs/hep-ph/9905214}{{\ttfamily hep-ph/9905214}}].

\bibitem{Jalilian-Marian:1997dw}
J.~Jalilian-Marian, A.~Kovner and H.~Weigert, \emph{The {Wilson}
  renormalization group for low x physics: Gluon evolution at finite parton
  density}, {\emph{Phys. Rev.} {\bfseries D59} (1998) 014015},
  [\href{https://arxiv.org/abs/hep-ph/9709432}{{\ttfamily hep-ph/9709432}}].

\bibitem{Jalilian-Marian:1997gr}
J.~Jalilian-Marian, A.~Kovner, A.~Leonidov and H.~Weigert, \emph{The {Wilson}
  renormalization group for low x physics: Towards the high density regime},
  {\emph{Phys. Rev.} {\bfseries D59} (1998) 014014},
  [\href{https://arxiv.org/abs/hep-ph/9706377}{{\ttfamily hep-ph/9706377}}].

\bibitem{Weigert:2000gi}
H.~Weigert, \emph{Unitarity at small {B}jorken x}, {\emph{Nucl. Phys.}
  {\bfseries A703} (2002) 823--860},
  [\href{https://arxiv.org/abs/hep-ph/0004044}{{\ttfamily hep-ph/0004044}}].

\bibitem{Iancu:2001ad}
E.~Iancu, A.~Leonidov and L.~D. McLerran, \emph{{The renormalization group
  equation for the color glass condensate}},
  \href{https://doi.org/10.1016/S0370-2693(01)00524-X}{\emph{Phys. Lett.}
  {\bfseries B510} (2001) 133--144}.

\bibitem{Iancu:2000hn}
E.~Iancu, A.~Leonidov and L.~D. McLerran, \emph{Nonlinear gluon evolution in
  the color glass condensate. {I}}, {\emph{Nucl. Phys.} {\bfseries A692} (2001)
  583--645}, [\href{https://arxiv.org/abs/hep-ph/0011241}{{\ttfamily
  hep-ph/0011241}}].

\bibitem{Ferreiro:2001qy}
E.~Ferreiro, E.~Iancu, A.~Leonidov and L.~McLerran, \emph{Nonlinear gluon
  evolution in the color glass condensate. {II}}, {\emph{Nucl. Phys.}
  {\bfseries A703} (2002) 489--538},
  [\href{https://arxiv.org/abs/hep-ph/0109115}{{\ttfamily hep-ph/0109115}}].

\bibitem{Gribov:1984tu}
L.~V. Gribov, E.~M. Levin and M.~G. Ryskin, \emph{{Semihard Processes in QCD}},
  {\emph{Phys. Rept.} {\bfseries 100} (1983) 1--150}.

\bibitem{Iancu:2003xm}
E.~Iancu and R.~Venugopalan, \emph{{The Color glass condensate and high-energy
  scattering in QCD}}, pp.~249--3363.
\newblock 3, 2003.
\newblock \href{https://arxiv.org/abs/hep-ph/0303204}{{\ttfamily
  hep-ph/0303204}}.
\newblock 10.1142/9789812795533$\_$0005.

\bibitem{Weigert:2005us}
H.~Weigert, \emph{Evolution at small {$x_{bj}$: The Color Glass Condensate}},
  {\emph{Prog. Part. Nucl. Phys.} {\bfseries 55} (2005) 461--565},
  [\href{https://arxiv.org/abs/hep-ph/0501087}{{\ttfamily hep-ph/0501087}}].

\bibitem{JalilianMarian:2005jf}
J.~Jalilian-Marian and Y.~V. Kovchegov, \emph{{Saturation physics and
  deuteron-Gold collisions at RHIC}},
  \href{https://doi.org/10.1016/j.ppnp.2005.07.002}{\emph{Prog. Part. Nucl.
  Phys.} {\bfseries 56} (2006) 104--231},
  [\href{https://arxiv.org/abs/hep-ph/0505052}{{\ttfamily hep-ph/0505052}}].

\bibitem{Gelis:2010nm}
F.~Gelis, E.~Iancu, J.~Jalilian-Marian and R.~Venugopalan, \emph{{The Color
  Glass Condensate}},
  \href{https://doi.org/10.1146/annurev.nucl.010909.083629}{\emph{Ann.Rev.Nucl.Part.Sci.}
  {\bfseries 60} (2010) 463--489},
  [\href{https://arxiv.org/abs/1002.0333}{{\ttfamily 1002.0333}}].

\bibitem{Albacete:2014fwa}
J.~L. Albacete and C.~Marquet, \emph{{Gluon saturation and initial conditions
  for relativistic heavy ion collisions}},
  \href{https://doi.org/10.1016/j.ppnp.2014.01.004}{\emph{Prog.Part.Nucl.Phys.}
  {\bfseries 76} (2014) 1--42},
  [\href{https://arxiv.org/abs/1401.4866}{{\ttfamily 1401.4866}}].

\bibitem{Kovchegov:2012mbw}
Y.~V. Kovchegov and E.~Levin, \emph{{Quantum chromodynamics at high energy}},
  vol.~33.
\newblock Cambridge University Press, 2012.

\bibitem{Morreale:2021pnn}
A.~Morreale and F.~Salazar, \emph{{Mining for Gluon Saturation at Colliders}},
  \href{https://doi.org/10.3390/universe7080312}{\emph{Universe} {\bfseries 7}
  (2021) 312}, [\href{https://arxiv.org/abs/2108.08254}{{\ttfamily
  2108.08254}}].

\bibitem{Hatta:2016aoc}
Y.~Hatta, Y.~Nakagawa, F.~Yuan, Y.~Zhao and B.~Xiao, \emph{{Gluon orbital
  angular momentum at small-$x$}},
  \href{https://doi.org/10.1103/PhysRevD.95.114032}{\emph{Phys. Rev.}
  {\bfseries D95} (2017) 114032},
  [\href{https://arxiv.org/abs/1612.02445}{{\ttfamily 1612.02445}}].

\bibitem{Kovchegov:2016weo}
Y.~V. Kovchegov, D.~Pitonyak and M.~D. Sievert, \emph{{Small-$x$ asymptotics of
  the quark helicity distribution}},
  \href{https://doi.org/10.1103/PhysRevLett.118.052001}{\emph{Phys. Rev. Lett.}
  {\bfseries 118} (2017) 052001},
  [\href{https://arxiv.org/abs/1610.06188}{{\ttfamily 1610.06188}}].

\bibitem{Kovchegov:2017jxc}
Y.~V. Kovchegov, D.~Pitonyak and M.~D. Sievert, \emph{{Small-$x$ Asymptotics of
  the Quark Helicity Distribution: Analytic Results}},
  \href{https://doi.org/10.1016/j.physletb.2017.06.032}{\emph{Phys. Lett.}
  {\bfseries B772} (2017) 136--140},
  [\href{https://arxiv.org/abs/1703.05809}{{\ttfamily 1703.05809}}].

\bibitem{Kovchegov:2017lsr}
Y.~V. Kovchegov, D.~Pitonyak and M.~D. Sievert, \emph{{Small-$x$ Asymptotics of
  the Gluon Helicity Distribution}},
  \href{https://doi.org/10.1007/JHEP10(2017)198}{\emph{JHEP} {\bfseries 10}
  (2017) 198}, [\href{https://arxiv.org/abs/1706.04236}{{\ttfamily
  1706.04236}}].

\bibitem{Kovchegov:2019rrz}
Y.~V. Kovchegov, \emph{{Orbital Angular Momentum at Small $x$}},
  \href{https://doi.org/10.1007/JHEP03(2019)174}{\emph{JHEP} {\bfseries 03}
  (2019) 174}, [\href{https://arxiv.org/abs/1901.07453}{{\ttfamily
  1901.07453}}].

\bibitem{Cougoulic:2019aja}
F.~Cougoulic and Y.~V. Kovchegov, \emph{{Helicity-dependent generalization of
  the JIMWLK evolution}},
  \href{https://doi.org/10.1103/PhysRevD.100.114020}{\emph{Phys. Rev.}
  {\bfseries D100} (2019) 114020},
  [\href{https://arxiv.org/abs/1910.04268}{{\ttfamily 1910.04268}}].

\bibitem{Kovchegov:2020hgb}
Y.~V. Kovchegov and Y.~Tawabutr, \emph{{Helicity at Small $x$: Oscillations
  Generated by Bringing Back the Quarks}},
  \href{https://doi.org/10.1007/JHEP08(2020)014}{\emph{JHEP} {\bfseries 08}
  (2020) 014}, [\href{https://arxiv.org/abs/2005.07285}{{\ttfamily
  2005.07285}}].

\bibitem{Cougoulic:2020tbc}
F.~Cougoulic and Y.~V. Kovchegov, \emph{{Helicity-dependent extension of the
  McLerran-Venugopalan model}},
  \href{https://doi.org/10.1016/j.nuclphysa.2020.122051}{\emph{Nucl. Phys. A}
  {\bfseries 1004} (2020) 122051},
  [\href{https://arxiv.org/abs/2005.14688}{{\ttfamily 2005.14688}}].

\bibitem{Chirilli:2021lif}
G.~A. Chirilli, \emph{{High-energy operator product expansion at sub-eikonal
  level}}, \href{https://doi.org/10.1007/JHEP06(2021)096}{\emph{JHEP}
  {\bfseries 06} (2021) 096},
  [\href{https://arxiv.org/abs/2101.12744}{{\ttfamily 2101.12744}}].

\bibitem{Kovchegov:2021lvz}
Y.~V. Kovchegov, A.~Tarasov and Y.~Tawabutr, \emph{{Helicity evolution at small
  x: the single-logarithmic contribution}},
  \href{https://doi.org/10.1007/JHEP03(2022)184}{\emph{JHEP} {\bfseries 03}
  (2022) 184}, [\href{https://arxiv.org/abs/2104.11765}{{\ttfamily
  2104.11765}}].

\bibitem{Borden:2023ugd}
J.~Borden and Y.~V. Kovchegov, \emph{{Analytic solution for the revised
  helicity evolution at small x and large Nc: New resummed gluon-gluon
  polarized anomalous dimension and intercept}},
  \href{https://doi.org/10.1103/PhysRevD.108.014001}{\emph{Phys. Rev. D}
  {\bfseries 108} (2023) 014001},
  [\href{https://arxiv.org/abs/2304.06161}{{\ttfamily 2304.06161}}].

\bibitem{Adamiak:2023okq}
D.~Adamiak, Y.~V. Kovchegov and Y.~Tawabutr, \emph{{Helicity evolution at small
  x: Revised asymptotic results at large Nc and Nf}},
  \href{https://doi.org/10.1103/PhysRevD.108.054005}{\emph{Phys. Rev. D}
  {\bfseries 108} (2023) 054005},
  [\href{https://arxiv.org/abs/2306.01651}{{\ttfamily 2306.01651}}].

\bibitem{Mueller:1994rr}
A.~H. Mueller, \emph{Soft gluons in the infinite momentum wave function and the
  {BFKL} pomeron}, {\emph{Nucl. Phys.} {\bfseries B415} (1994) 373--385}.

\bibitem{Mueller:1994jq}
A.~H. Mueller and B.~Patel, \emph{Single and double {BFKL} pomeron exchange and
  a dipole picture of high-energy hard processes}, {\emph{Nucl. Phys.}
  {\bfseries B425} (1994) 471--488},
  [\href{https://arxiv.org/abs/hep-ph/9403256}{{\ttfamily hep-ph/9403256}}].

\bibitem{Mueller:1995gb}
A.~H. Mueller, \emph{Unitarity and the {BFKL} pomeron}, {\emph{Nucl. Phys.}
  {\bfseries B437} (1995) 107--126},
  [\href{https://arxiv.org/abs/hep-ph/9408245}{{\ttfamily hep-ph/9408245}}].

\bibitem{Altinoluk:2014oxa}
T.~Altinoluk, N.~Armesto, G.~Beuf, M.~Martinez and C.~A. Salgado,
  \emph{{Next-to-eikonal corrections in the CGC: gluon production and spin
  asymmetries in pA collisions}},
  \href{https://doi.org/10.1007/JHEP07(2014)068}{\emph{JHEP} {\bfseries 07}
  (2014) 068}, [\href{https://arxiv.org/abs/1404.2219}{{\ttfamily 1404.2219}}].

\bibitem{Balitsky:2015qba}
I.~Balitsky and A.~Tarasov, \emph{{Rapidity evolution of gluon TMD from low to
  moderate x}}, \href{https://doi.org/10.1007/JHEP10(2015)017}{\emph{JHEP}
  {\bfseries 10} (2015) 017},
  [\href{https://arxiv.org/abs/1505.02151}{{\ttfamily 1505.02151}}].

\bibitem{Balitsky:2016dgz}
I.~Balitsky and A.~Tarasov, \emph{{Gluon TMD in particle production from low to
  moderate x}}, \href{https://doi.org/10.1007/JHEP06(2016)164}{\emph{JHEP}
  {\bfseries 06} (2016) 164},
  [\href{https://arxiv.org/abs/1603.06548}{{\ttfamily 1603.06548}}].

\bibitem{Chirilli:2018kkw}
G.~A. Chirilli, \emph{{Sub-eikonal corrections to scattering amplitudes at high
  energy}}, \href{https://doi.org/10.1007/JHEP01(2019)118}{\emph{JHEP}
  {\bfseries 01} (2019) 118},
  [\href{https://arxiv.org/abs/1807.11435}{{\ttfamily 1807.11435}}].

\bibitem{Jalilian-Marian:2018iui}
J.~Jalilian-Marian, \emph{{Quark jets scattering from a gluon field: from
  saturation to high $p_t$}},
  \href{https://doi.org/10.1103/PhysRevD.99.014043}{\emph{Phys. Rev.}
  {\bfseries D99} (2019) 014043},
  [\href{https://arxiv.org/abs/1809.04625}{{\ttfamily 1809.04625}}].

\bibitem{Jalilian-Marian:2019kaf}
J.~Jalilian-Marian, \emph{{Rapidity loss, spin and angular asymmetries in
  scattering of a quark from color field of a proton (nucleus)}},
  \href{https://arxiv.org/abs/1912.08878}{{\ttfamily 1912.08878}}.

\bibitem{Altinoluk:2020oyd}
T.~Altinoluk, G.~Beuf, A.~Czajka and A.~Tymowska, \emph{{Quarks at
  next-to-eikonal accuracy in the CGC: Forward quark-nucleus scattering}},
  \href{https://doi.org/10.1103/PhysRevD.104.014019}{\emph{Phys. Rev. D}
  {\bfseries 104} (2021) 014019},
  [\href{https://arxiv.org/abs/2012.03886}{{\ttfamily 2012.03886}}].

\bibitem{Kovchegov:2021iyc}
Y.~V. Kovchegov and M.~G. Santiago, \emph{{Quark sivers function at small x:
  spin-dependent odderon and the sub-eikonal evolution}},
  \href{https://doi.org/10.1007/JHEP11(2021)200}{\emph{JHEP} {\bfseries 11}
  (2021) 200}, [\href{https://arxiv.org/abs/2108.03667}{{\ttfamily
  2108.03667}}].

\bibitem{Altinoluk:2021lvu}
T.~Altinoluk and G.~Beuf, \emph{{Quark and scalar propagators at
  next-to-eikonal accuracy in the CGC through a dynamical background gluon
  field}}, \href{https://doi.org/10.1103/PhysRevD.105.074026}{\emph{Phys. Rev.
  D} {\bfseries 105} (2022) 074026},
  [\href{https://arxiv.org/abs/2109.01620}{{\ttfamily 2109.01620}}].

\bibitem{Kovchegov:2022kyy}
Y.~V. Kovchegov and M.~G. Santiago, \emph{{T-odd leading-twist quark TMDs at
  small x}}, \href{https://doi.org/10.1007/JHEP11(2022)098}{\emph{JHEP}
  {\bfseries 11} (2022) 098},
  [\href{https://arxiv.org/abs/2209.03538}{{\ttfamily 2209.03538}}].

\bibitem{Altinoluk:2022jkk}
T.~Altinoluk, G.~Beuf, A.~Czajka and A.~Tymowska, \emph{{DIS dijet production
  at next-to-eikonal accuracy in the CGC}},
  \href{https://doi.org/10.1103/PhysRevD.107.074016}{\emph{Phys. Rev. D}
  {\bfseries 107} (2023) 074016},
  [\href{https://arxiv.org/abs/2212.10484}{{\ttfamily 2212.10484}}].

\bibitem{Altinoluk:2023qfr}
T.~Altinoluk, N.~Armesto and G.~Beuf, \emph{{Probing quark transverse momentum
  distributions in the color glass condensate: Quark-gluon dijets in deep
  inelastic scattering at next-to-eikonal accuracy}},
  \href{https://doi.org/10.1103/PhysRevD.108.074023}{\emph{Phys. Rev. D}
  {\bfseries 108} (2023) 074023},
  [\href{https://arxiv.org/abs/2303.12691}{{\ttfamily 2303.12691}}].

\bibitem{Altinoluk:2023dww}
T.~Altinoluk, G.~Beuf and J.~Jalilian-Marian, \emph{{Renormalization of the
  gluon distribution function in the background field formalism}},
  \href{https://arxiv.org/abs/2305.11079}{{\ttfamily 2305.11079}}.

\bibitem{Li:2023tlw}
M.~Li, \emph{{Small x physics beyond eikonal approximation: an effective
  Hamiltonian approach}},
  \href{https://doi.org/10.1007/JHEP07(2023)158}{\emph{JHEP} {\bfseries 07}
  (2023) 158}, [\href{https://arxiv.org/abs/2304.12842}{{\ttfamily
  2304.12842}}].

\bibitem{tHooft:1973alw}
G.~'t~Hooft, \emph{{A Planar Diagram Theory for Strong Interactions}},
  \href{https://doi.org/10.1016/0550-3213(74)90154-0}{\emph{Nucl. Phys. B}
  {\bfseries 72} (1974) 461}.

\bibitem{Veneziano:1976wm}
G.~Veneziano, \emph{{Some Aspects of a Unified Approach to Gauge, Dual and
  Gribov Theories}},
  \href{https://doi.org/10.1016/0550-3213(76)90412-0}{\emph{Nucl. Phys. B}
  {\bfseries 117} (1976) 519--545}.

\bibitem{Bartels:1995iu}
J.~Bartels, B.~Ermolaev and M.~Ryskin, \emph{{Nonsinglet contributions to the
  structure function g1 at small x}}, {\emph{Z.Phys.} {\bfseries C70} (1996)
  273--280}, [\href{https://arxiv.org/abs/hep-ph/9507271}{{\ttfamily
  hep-ph/9507271}}].

\bibitem{Bartels:1996wc}
J.~Bartels, B.~Ermolaev and M.~Ryskin, \emph{{Flavor singlet contribution to
  the structure function G(1) at small x}},
  \href{https://doi.org/10.1007/s002880050285}{\emph{Z.Phys.} {\bfseries C72}
  (1996) 627--635}, [\href{https://arxiv.org/abs/hep-ph/9603204}{{\ttfamily
  hep-ph/9603204}}].

\bibitem{Gorshkov:1966ht}
V.~G. Gorshkov, V.~N. Gribov, L.~N. Lipatov and G.~V. Frolov, \emph{{Doubly
  logarithmic asymptotic behavior in quantum electrodynamics}}, {\emph{Sov. J.
  Nucl. Phys.} {\bfseries 6} (1968) 95}.

\bibitem{Kirschner:1983di}
R.~Kirschner and L.~Lipatov, \emph{{Double Logarithmic Asymptotics and Regge
  Singularities of Quark Amplitudes with Flavor Exchange}},
  \href{https://doi.org/10.1016/0550-3213(83)90178-5}{\emph{Nucl.Phys.}
  {\bfseries B213} (1983) 122--148}.

\bibitem{Kirschner:1994rq}
R.~Kirschner, \emph{{Reggeon interactions in perturbative QCD}},
  \href{https://doi.org/10.1007/BF01556138}{\emph{Z.Phys.} {\bfseries C65}
  (1995) 505--510}, [\href{https://arxiv.org/abs/hep-th/9407085}{{\ttfamily
  hep-th/9407085}}].

\bibitem{Kirschner:1994vc}
R.~Kirschner, \emph{{Regge asymptotics of scattering with flavor exchange in
  QCD}}, \href{https://doi.org/10.1007/BF01624588}{\emph{Z.Phys.} {\bfseries
  C67} (1995) 459--466},
  [\href{https://arxiv.org/abs/hep-th/9404158}{{\ttfamily hep-th/9404158}}].

\bibitem{Blumlein:1995jp}
J.~Blumlein and A.~Vogt, \emph{{On the behavior of nonsinglet structure
  functions at small x}},
  \href{https://doi.org/10.1016/0370-2693(95)01568-X}{\emph{Phys. Lett. B}
  {\bfseries 370} (1996) 149--155},
  [\href{https://arxiv.org/abs/hep-ph/9510410}{{\ttfamily hep-ph/9510410}}].

\bibitem{Griffiths:1999dj}
S.~Griffiths and D.~Ross, \emph{{Studying the perturbative Reggeon}},
  \href{https://doi.org/10.1007/s100529900240}{\emph{Eur.Phys.J.} {\bfseries
  C12} (2000) 277--286},
  [\href{https://arxiv.org/abs/hep-ph/9906550}{{\ttfamily hep-ph/9906550}}].

\bibitem{Moch:2014sna}
S.~Moch, J.~A.~M. Vermaseren and A.~Vogt, \emph{{The Three-Loop Splitting
  Functions in QCD: The Helicity-Dependent Case}},
  \href{https://doi.org/10.1016/j.nuclphysb.2014.10.016}{\emph{Nucl. Phys. B}
  {\bfseries 889} (2014) 351--400},
  [\href{https://arxiv.org/abs/1409.5131}{{\ttfamily 1409.5131}}].

\bibitem{Blumlein:2021ryt}
J.~Bl\"umlein, P.~Marquard, C.~Schneider and K.~Sch\"onwald, \emph{{The
  three-loop polarized singlet anomalous dimensions from off-shell operator
  matrix elements}}, \href{https://doi.org/10.1007/JHEP01(2022)193}{\emph{JHEP}
  {\bfseries 01} (2022) 193},
  [\href{https://arxiv.org/abs/2111.12401}{{\ttfamily 2111.12401}}].

\bibitem{Blumlein:1996hb}
J.~Bl{\"u}mlein and A.~Vogt, \emph{{The Singlet contribution to the structure
  function g1 (x, Q**2) at small x}},
  \href{https://doi.org/10.1016/0370-2693(96)00958-6}{\emph{Phys. Lett. B}
  {\bfseries 386} (1996) 350--358},
  [\href{https://arxiv.org/abs/hep-ph/9606254}{{\ttfamily hep-ph/9606254}}].

\bibitem{Ermolaev:1999jx}
B.~I. Ermolaev, M.~Greco and S.~I. Troian, \emph{{QCD running coupling effects
  for the nonsinglet structure function at small $x$}},
  \href{https://doi.org/10.1016/S0550-3213(99)00812-3}{\emph{Nucl. Phys.}
  {\bfseries B571} (2000) 137--150},
  [\href{https://arxiv.org/abs/hep-ph/9906276}{{\ttfamily hep-ph/9906276}}].

\bibitem{Ermolaev:2000sg}
B.~I. Ermolaev, M.~Greco and S.~I. Troyan, \emph{{Intercepts of the nonsinglet
  structure functions}},
  \href{https://doi.org/10.1016/S0550-3213(00)00647-7}{\emph{Nucl. Phys.}
  {\bfseries B594} (2001) 71--88},
  [\href{https://arxiv.org/abs/hep-ph/0009037}{{\ttfamily hep-ph/0009037}}].

\bibitem{Ermolaev:2003zx}
B.~I. Ermolaev, M.~Greco and S.~I. Troyan, \emph{{Running coupling effects for
  the singlet structure function $g_1$ at small $x$}},
  \href{https://doi.org/10.1016/j.physletb.2003.11.016}{\emph{Phys. Lett.}
  {\bfseries B579} (2004) 321--330},
  [\href{https://arxiv.org/abs/hep-ph/0307128}{{\ttfamily hep-ph/0307128}}].

\bibitem{Ermolaev:2009cq}
B.~I. Ermolaev, M.~Greco and S.~I. Troyan, \emph{{Overview of the spin
  structure function $g_1$ at arbitrary $x$ and $Q^2$}},
  \href{https://doi.org/10.1393/ncr/i2010-10052-3}{\emph{Riv. Nuovo Cim.}
  {\bfseries 33} (2010) 57--122},
  [\href{https://arxiv.org/abs/0905.2841}{{\ttfamily 0905.2841}}].

\bibitem{Adamiak:2021ppq}
{\scshape Jefferson Lab Angular Momentum} collaboration, D.~Adamiak, Y.~V.
  Kovchegov, W.~Melnitchouk, D.~Pitonyak, N.~Sato and M.~D. Sievert,
  \emph{{First analysis of world polarized DIS data with small-x helicity
  evolution}}, \href{https://doi.org/10.1103/PhysRevD.104.L031501}{\emph{Phys.
  Rev. D} {\bfseries 104} (2021) L031501},
  [\href{https://arxiv.org/abs/2102.06159}{{\ttfamily 2102.06159}}].

\bibitem{Gluck:2000dy}
M.~Gluck, E.~Reya, M.~Stratmann and W.~Vogelsang, \emph{{Models for the
  polarized parton distributions of the nucleon}},
  \href{https://doi.org/10.1103/PhysRevD.63.094005}{\emph{Phys. Rev. D}
  {\bfseries 63} (2001) 094005},
  [\href{https://arxiv.org/abs/hep-ph/0011215}{{\ttfamily hep-ph/0011215}}].

\bibitem{Leader:2005ci}
E.~Leader, A.~V. Sidorov and D.~B. Stamenov, \emph{{Longitudinal polarized
  parton densities updated}},
  \href{https://doi.org/10.1103/PhysRevD.73.034023}{\emph{Phys. Rev.}
  {\bfseries D73} (2006) 034023},
  [\href{https://arxiv.org/abs/hep-ph/0512114}{{\ttfamily hep-ph/0512114}}].

\bibitem{deFlorian:2009vb}
D.~de~Florian, R.~Sassot, M.~Stratmann and W.~Vogelsang, \emph{{Extraction of
  Spin-Dependent Parton Densities and Their Uncertainties}},
  \href{https://doi.org/10.1103/PhysRevD.80.034030}{\emph{Phys. Rev.}
  {\bfseries D80} (2009) 034030},
  [\href{https://arxiv.org/abs/0904.3821}{{\ttfamily 0904.3821}}].

\bibitem{Leader:2010rb}
E.~Leader, A.~V. Sidorov and D.~B. Stamenov, \emph{{Determination of Polarized
  PDFs from a QCD Analysis of Inclusive and Semi-inclusive Deep Inelastic
  Scattering Data}},
  \href{https://doi.org/10.1103/PhysRevD.82.114018}{\emph{Phys. Rev.}
  {\bfseries D82} (2010) 114018},
  [\href{https://arxiv.org/abs/1010.0574}{{\ttfamily 1010.0574}}].

\bibitem{Jimenez-Delgado:2013boa}
P.~Jimenez-Delgado, A.~Accardi and W.~Melnitchouk, \emph{{Impact of hadronic
  and nuclear corrections on global analysis of spin-dependent parton
  distributions}},
  \href{https://doi.org/10.1103/PhysRevD.89.034025}{\emph{Phys. Rev.}
  {\bfseries D89} (2014) 034025},
  [\href{https://arxiv.org/abs/1310.3734}{{\ttfamily 1310.3734}}].

\bibitem{Ball:2013lla}
{\scshape NNPDF} collaboration, R.~D. Ball, S.~Forte, A.~Guffanti, E.~R.
  Nocera, G.~Ridolfi and J.~Rojo, \emph{{Unbiased determination of polarized
  parton distributions and their uncertainties}},
  \href{https://doi.org/10.1016/j.nuclphysb.2013.05.007}{\emph{Nucl. Phys.}
  {\bfseries B874} (2013) 36--84},
  [\href{https://arxiv.org/abs/1303.7236}{{\ttfamily 1303.7236}}].

\bibitem{Nocera:2014gqa}
{\scshape NNPDF} collaboration, E.~R. Nocera, R.~D. Ball, S.~Forte, G.~Ridolfi
  and J.~Rojo, \emph{{A first unbiased global determination of polarized PDFs
  and their uncertainties}},
  \href{https://doi.org/10.1016/j.nuclphysb.2014.08.008}{\emph{Nucl. Phys.}
  {\bfseries B887} (2014) 276--308},
  [\href{https://arxiv.org/abs/1406.5539}{{\ttfamily 1406.5539}}].

\bibitem{deFlorian:2014yva}
D.~de~Florian, R.~Sassot, M.~Stratmann and W.~Vogelsang, \emph{{Evidence for
  polarization of gluons in the proton}},
  \href{https://doi.org/10.1103/PhysRevLett.113.012001}{\emph{Phys. Rev. Lett.}
  {\bfseries 113} (2014) 012001},
  [\href{https://arxiv.org/abs/1404.4293}{{\ttfamily 1404.4293}}].

\bibitem{Leader:2014uua}
E.~Leader, A.~V. Sidorov and D.~B. Stamenov, \emph{{New analysis concerning the
  strange quark polarization puzzle}},
  \href{https://doi.org/10.1103/PhysRevD.91.054017}{\emph{Phys. Rev.}
  {\bfseries D91} (2015) 054017},
  [\href{https://arxiv.org/abs/1410.1657}{{\ttfamily 1410.1657}}].

\bibitem{Sato:2016tuz}
{\scshape Jefferson Lab Angular Momentum} collaboration, N.~Sato,
  W.~Melnitchouk, S.~E. Kuhn, J.~J. Ethier and A.~Accardi, \emph{{Iterative
  Monte Carlo analysis of spin-dependent parton distributions}},
  \href{https://doi.org/10.1103/PhysRevD.93.074005}{\emph{Phys. Rev.}
  {\bfseries D93} (2016) 074005},
  [\href{https://arxiv.org/abs/1601.07782}{{\ttfamily 1601.07782}}].

\bibitem{Ethier:2017zbq}
J.~Ethier, N.~Sato and W.~Melnitchouk, \emph{{First simultaneous extraction of
  spin-dependent parton distributions and fragmentation functions from a global
  QCD analysis}},
  \href{https://doi.org/10.1103/PhysRevLett.119.132001}{\emph{Phys.\ Rev.\
  Lett.} {\bfseries 119} (2017) 132001},
  [\href{https://arxiv.org/abs/1705.05889}{{\ttfamily 1705.05889}}].

\bibitem{DeFlorian:2019xxt}
D.~De~Florian, G.~A. Lucero, R.~Sassot, M.~Stratmann and W.~Vogelsang,
  \emph{{Monte Carlo sampling variant of the DSSV14 set of helicity parton
  densities}}, \href{https://doi.org/10.1103/PhysRevD.100.114027}{\emph{Phys.
  Rev. D} {\bfseries 100} (2019) 114027},
  [\href{https://arxiv.org/abs/1902.10548}{{\ttfamily 1902.10548}}].

\bibitem{Borsa:2020lsz}
I.~Borsa, G.~Lucero, R.~Sassot, E.~C. Aschenauer and A.~S. Nunes,
  \emph{{Revisiting helicity parton distributions at a future electron-ion
  collider}}, \href{https://doi.org/10.1103/PhysRevD.102.094018}{\emph{Phys.
  Rev. D} {\bfseries 102} (2020) 094018},
  [\href{https://arxiv.org/abs/2007.08300}{{\ttfamily 2007.08300}}].

\bibitem{Zhou:2022wzm}
{\scshape JAM Collaboration} collaboration, Y.~Zhou, N.~Sato and
  W.~Melnitchouk, \emph{{How well do we know the gluon polarization in the
  proton?}}, \href{https://doi.org/10.1103/PhysRevD.105.074022}{\emph{Phys.
  Rev. D} {\bfseries 105} (2022) 074022},
  [\href{https://arxiv.org/abs/2201.02075}{{\ttfamily 2201.02075}}].

\bibitem{Cocuzza:2022jye}
{\scshape JAM Collaboration} collaboration, C.~Cocuzza, W.~Melnitchouk, A.~Metz
  and N.~Sato, \emph{{Polarized antimatter in the proton from a global QCD
  analysis}}, \href{https://doi.org/10.1103/PhysRevD.106.L031502}{\emph{Phys.
  Rev. D} {\bfseries 106} (2022) L031502},
  [\href{https://arxiv.org/abs/2202.03372}{{\ttfamily 2202.03372}}].

\bibitem{Gribov:1972ri}
V.~N. Gribov and L.~N. Lipatov, \emph{{Deep inelastic e p scattering in
  perturbation theory}}, {\emph{Sov. J. Nucl. Phys.} {\bfseries 15} (1972)
  438--450}.

\bibitem{Altarelli:1977zs}
G.~Altarelli and G.~Parisi, \emph{{Asymptotic Freedom in Parton Language}},
  \href{https://doi.org/10.1016/0550-3213(77)90384-4}{\emph{Nucl. Phys.}
  {\bfseries B126} (1977) 298}.

\bibitem{Dokshitzer:1977sg}
Y.~L. Dokshitzer, \emph{{Calculation of the Structure Functions for Deep
  Inelastic Scattering and $e^+ e^-$ Annihilation by Perturbation Theory in
  Quantum Chromodynamics}}, {\emph{Sov. Phys. JETP} {\bfseries 46} (1977)
  641--653}.

\bibitem{STAR:2014wox}
{\scshape STAR} collaboration, L.~Adamczyk et~al., \emph{{Precision Measurement
  of the Longitudinal Double-spin Asymmetry for Inclusive Jet Production in
  Polarized Proton Collisions at $\sqrt{s}=200$ GeV}},
  \href{https://doi.org/10.1103/PhysRevLett.115.092002}{\emph{Phys. Rev. Lett.}
  {\bfseries 115} (2015) 092002},
  [\href{https://arxiv.org/abs/1405.5134}{{\ttfamily 1405.5134}}].

\bibitem{PHENIX:2015fxo}
{\scshape PHENIX} collaboration, A.~Adare et~al., \emph{{Inclusive cross
  section and double-helicity asymmetry for $\pi^{0}$ production at midrapidity
  in $p+p$ collisions at $\sqrt{s}=510$ GeV}},
  \href{https://doi.org/10.1103/PhysRevD.93.011501}{\emph{Phys. Rev. D}
  {\bfseries 93} (2016) 011501},
  [\href{https://arxiv.org/abs/1510.02317}{{\ttfamily 1510.02317}}].

\bibitem{Kovchegov:1998bi}
Y.~V. Kovchegov and A.~H. Mueller, \emph{Gluon production in current nucleus
  and nucleon nucleus collisions in a quasi-classical approximation},
  {\emph{Nucl. Phys.} {\bfseries B529} (1998) 451--479},
  [\href{https://arxiv.org/abs/hep-ph/9802440}{{\ttfamily hep-ph/9802440}}].

\bibitem{Kopeliovich:1998nw}
B.~Z. Kopeliovich, A.~V. Tarasov and A.~Schafer, \emph{Bremsstrahlung of a
  quark propagating through a nucleus}, {\emph{Phys. Rev.} {\bfseries C59}
  (1999) 1609--1619}, [\href{https://arxiv.org/abs/hep-ph/9808378}{{\ttfamily
  hep-ph/9808378}}].

\bibitem{Braun:2000bh}
M.~A. Braun, \emph{{Inclusive jet production on the nucleus in the perturbative
  QCD with $N_c \rightarrow \infty$}},
  \href{https://doi.org/10.1016/S0370-2693(00)00570-0}{\emph{Phys. Lett.}
  {\bfseries B483} (2000) 105--114},
  [\href{https://arxiv.org/abs/hep-ph/0003003}{{\ttfamily hep-ph/0003003}}].

\bibitem{Dumitru:2001ux}
A.~Dumitru and L.~D. McLerran, \emph{How protons shatter colored glass},
  {\emph{Nucl. Phys.} {\bfseries A700} (2002) 492--508},
  [\href{https://arxiv.org/abs/hep-ph/0105268}{{\ttfamily hep-ph/0105268}}].

\bibitem{Kovchegov:2001sc}
Y.~V. Kovchegov and K.~Tuchin, \emph{Inclusive gluon production in dis at high
  parton density}, {\emph{Phys. Rev.} {\bfseries D65} (2002) 074026},
  [\href{https://arxiv.org/abs/hep-ph/0111362}{{\ttfamily hep-ph/0111362}}].

\bibitem{Kharzeev:2003wz}
D.~Kharzeev, Y.~V. Kovchegov and K.~Tuchin, \emph{Cronin effect and high-p(t)
  suppression in p a collisions}, {\emph{Phys. Rev.} {\bfseries D68} (2003)
  094013}, [\href{https://arxiv.org/abs/hep-ph/0307037}{{\ttfamily
  hep-ph/0307037}}].

\bibitem{Balitsky:2004rr}
I.~Balitsky, \emph{{Scattering of shock waves in QCD}},
  \href{https://doi.org/10.1103/PhysRevD.70.114030}{\emph{Phys. Rev.}
  {\bfseries D70} (2004) 114030},
  [\href{https://arxiv.org/abs/hep-ph/0409314}{{\ttfamily hep-ph/0409314}}].

\bibitem{Chirilli:2015tea}
G.~A. Chirilli, Y.~V. Kovchegov and D.~E. Wertepny, \emph{{Classical Gluon
  Production Amplitude for Nucleus-Nucleus Collisions: First Saturation
  Correction in the Projectile}},
  \href{https://doi.org/10.1007/JHEP03(2015)015}{\emph{JHEP} {\bfseries 03}
  (2015) 015}, [\href{https://arxiv.org/abs/1501.03106}{{\ttfamily
  1501.03106}}].

\bibitem{Li:2021zmf}
M.~Li and V.~V. Skokov, \emph{{First saturation correction in high energy
  proton-nucleus collisions. Part I. Time evolution of classical Yang-Mills
  fields beyond leading order}},
  \href{https://doi.org/10.1007/JHEP06(2021)140}{\emph{JHEP} {\bfseries 06}
  (2021) 140}, [\href{https://arxiv.org/abs/2102.01594}{{\ttfamily
  2102.01594}}].

\bibitem{Li:2021yiv}
M.~Li and V.~V. Skokov, \emph{{First saturation correction in high energy
  proton-nucleus collisions. Part II. Single inclusive semi-hard gluon
  production}}, \href{https://doi.org/10.1007/JHEP06(2021)141}{\emph{JHEP}
  {\bfseries 06} (2021) 141},
  [\href{https://arxiv.org/abs/2104.01879}{{\ttfamily 2104.01879}}].

\bibitem{Li:2021ntt}
M.~Li and V.~V. Skokov, \emph{{First saturation correction in high energy
  proton-nucleus collisions. Part III. Ensemble averaging}},
  \href{https://doi.org/10.1007/JHEP01(2022)160}{\emph{JHEP} {\bfseries 01}
  (2022) 160}, [\href{https://arxiv.org/abs/2111.05304}{{\ttfamily
  2111.05304}}].

\bibitem{Mueller:1989st}
A.~H. Mueller, \emph{{Small x Behavior and Parton Saturation: A QCD Model}},
  {\emph{Nucl. Phys.} {\bfseries B335} (1990) 115}.

\bibitem{McLerran:1993ni}
L.~D. McLerran and R.~Venugopalan, \emph{Computing quark and gluon distribution
  functions for very large nuclei}, {\emph{Phys. Rev.} {\bfseries D49} (1994)
  2233--2241}, [\href{https://arxiv.org/abs/hep-ph/9309289}{{\ttfamily
  hep-ph/9309289}}].

\bibitem{McLerran:1993ka}
L.~D. McLerran and R.~Venugopalan, \emph{Gluon distribution functions for very
  large nuclei at small transverse momentum}, {\emph{Phys. Rev.} {\bfseries
  D49} (1994) 3352--3355},
  [\href{https://arxiv.org/abs/hep-ph/9311205}{{\ttfamily hep-ph/9311205}}].

\bibitem{McLerran:1994vd}
L.~D. McLerran and R.~Venugopalan, \emph{Green's functions in the color field
  of a large nucleus}, {\emph{Phys. Rev.} {\bfseries D50} (1994) 2225--2233},
  [\href{https://arxiv.org/abs/hep-ph/9402335}{{\ttfamily hep-ph/9402335}}].

\bibitem{Lepage:1980fj}
G.~P. Lepage and S.~J. Brodsky, \emph{Exclusive processes in perturbative
  quantum chromodynamics}, {\emph{Phys. Rev.} {\bfseries D22} (1980) 2157}.

\bibitem{Kovchegov:2012ga}
Y.~V. Kovchegov and M.~D. Sievert, \emph{{A New Mechanism for Generating a
  Single Transverse Spin Asymmetry}},
  \href{https://doi.org/10.1103/PhysRevD.86.034028}{\emph{Phys.Rev.} {\bfseries
  D86} (2012) 034028}, [\href{https://arxiv.org/abs/1201.5890}{{\ttfamily
  1201.5890}}].

\bibitem{Itakura:2003jp}
K.~Itakura, Y.~V. Kovchegov, L.~McLerran and D.~Teaney, \emph{{Baryon stopping
  and valence quark distribution at small x}},
  \href{https://doi.org/10.1016/j.nuclphysa.2003.10.016}{\emph{Nucl. Phys.}
  {\bfseries A730} (2004) 160--190},
  [\href{https://arxiv.org/abs/hep-ph/0305332}{{\ttfamily hep-ph/0305332}}].

\bibitem{Mulders:2000sh}
P.~J. Mulders and J.~Rodrigues, \emph{{Transverse momentum dependence in gluon
  distribution and fragmentation functions}},
  \href{https://doi.org/10.1103/PhysRevD.63.094021}{\emph{Phys. Rev.}
  {\bfseries D63} (2001) 094021},
  [\href{https://arxiv.org/abs/hep-ph/0009343}{{\ttfamily hep-ph/0009343}}].

\bibitem{PHENIX:2014gbf}
{\scshape PHENIX} collaboration, A.~Adare et~al., \emph{{Inclusive
  double-helicity asymmetries in neutral-pion and eta-meson production in
  $\vec{p}+\vec{p}$ collisions at $\sqrt{s}=200$ GeV}},
  \href{https://doi.org/10.1103/PhysRevD.90.012007}{\emph{Phys. Rev. D}
  {\bfseries 90} (2014) 012007},
  [\href{https://arxiv.org/abs/1402.6296}{{\ttfamily 1402.6296}}].

\bibitem{STAR:2021mqa}
{\scshape STAR} collaboration, M.~S. Abdallah et~al., \emph{{Longitudinal
  double-spin asymmetry for inclusive jet and dijet production in polarized
  proton collisions at $\sqrt{s}=510$ GeV}},
  \href{https://doi.org/10.1103/PhysRevD.105.092011}{\emph{Phys. Rev. D}
  {\bfseries 105} (2022) 092011},
  [\href{https://arxiv.org/abs/2110.11020}{{\ttfamily 2110.11020}}].

\bibitem{Kovchegov:2023yzd}
Y.~V. Kovchegov and B.~Manley, \emph{{Orbital angular momentum at small x
  revisited}}, \href{https://doi.org/10.1007/JHEP02(2024)060}{\emph{JHEP}
  {\bfseries 02} (2024) 060},
  [\href{https://arxiv.org/abs/2310.18404}{{\ttfamily 2310.18404}}].

\end{thebibliography}\endgroup
\bibliographystyle{JHEP}

\end{document}